\documentclass[preprint]{aastex}

\newcommand{\be}{\begin{equation}}
\newcommand{\ee}{\end{equation}}

\begin{document}

\title{A test for the universality of extinction curve shape}

\author{Anna Geminale\altaffilmark{1} and Piotr Popowski}
\affil{Max-Planck-Institut f\"{u}r Astrophysik,
Karl-Schwarzschild-Str.\ 1, Postfach 1317, 85741 Garching bei
M\"{u}nchen, Germany\\
E-mail: {\tt geminale@pd.astro.it, popowski@mpa-garching.mpg.de}}
\altaffiltext{1}{Ph.D. student, Department of Astronomy, 
Vicolo dell'Osservatorio, 2, I-35122 Padova, Italy} 

\begin{abstract} 
We present an analysis of 436 lines of sight with
extinction data covering wavelength range from near-infrared
(NIR) to ultraviolet (UV). We use $J,H,K$ photometry from 2MASS
database, the IR intrinsic colors from \citet{b15}, and UV extinction
data from Wegner (2002). We exclude 19 lines of sight (4\%) from the original
sample because of suspected photometric problems. 
We derive total to selective extinction
ratios ($R_V$) based on the Cardelli, Clayton, \& Mathis (1989; CCM)
law, which is typically used to fit the extinction data for both
diffuse and dense interstellar medium. We conclude that CCM law is
able to fit well most of the extinction curves in our sample (71\%),
and we present a catalog of $R_V$ and visual extinction ($A_V$) values
for those cases. We divide the
remaining lines of sight with peculiar extinction into two groups
according to two main behaviors: a) the NIR or/and UV wavelength
regions cannot be reproduced by CCM formula (14\% of the entire
sample), b) the NIR and UV extinction data taken separately are best
fit by CCM laws with significantly different values of $R_V$ (10\% of
the entire sample). We present examples of such curves. We caution
that some peculiarities of the extinction curves may not be intrinsic
but simply caused by faulty data. The study of the intrinsically
peculiar cases could help us to learn about the physical processes
that affect dust in the interstellar medium, e.g., formation of
mantles on the surface of grains, evaporation, growing or shattering.

\end{abstract}

\section{Introduction}
\label{introduction}
The interstellar grains affect starlight which passes through them by 
absorbing and scattering photons. These two physical processes
produce interstellar extinction which depends on the properties of 
dust grains, e.g., size distribution and composition. The
extinction curve shows the extinction as a function of wavelength
from the infrared to ultraviolet. \citet{b13} presented the average
extinction curve of our Galaxy at various wavelengths. It shows some
evident features: it rises in the infrared, it shows a slight knee in
the optical, it is characterized by a bump at 2175$\AA$, and it rises
in the far-ultraviolet. These features are common between different 
environments. The interstellar grain properties are different in
diffuse and dense interstellar medium and thus also the
extinction changes. Cardelli, Clayton and Mathis (1989; hereafter CCM)
found an average extinction law valid over the wavelength range
$0.125\mu m \leq \lambda \leq 3.5\mu m$, which is applicable to both
diffuse and dense regions of the interstellar medium. This extinction
law depends on only one parameter, $R_V=A_V/E(B-V)$, where $A_V$ is the
visual extinction and $E(B-V)$ is the color excess or reddening. The
$R_V$ parameter ranges from about 2.0 to about 5.5 (with a typical
value of 3.1) when one goes from diffuse to dense interstellar medium
and thus $R_V$ characterizes the region that produces the extinction.

If one knows the value of $R_V$ along a particular line of sight, one can
obtain the extinction curve at any wavelength $\lambda$ from the
infrared to ultraviolet using CCM law:
\begin{equation}
\frac{A_{\lambda}}{A_V}=a(x)+b(x) \cdot R_V^{-1},
\label{CCMlaw}
\end{equation}
where $x=1/\lambda$, and $a(x)$ and $b(x)$ are the
wavelength-dependent coefficients.

There are different ways to obtain $R_V$ using optical/NIR or UV
extinction data. \citet{b17} computed $R_V$ values for the sample of
597 OB stars for which the optical/NIR magnitudes are known. He
assumed that, in the infrared spectral region, the normalized
extinction curve is proportional to $\lambda^{-3}$ or $\lambda^{-4}$. 
Extrapolating the IR interstellar extinction curve to $1/\lambda=0$ he
derived $R_V$ as:
\begin{equation}
R_V=- \left [ \frac{E(\lambda-V)}{E(B-V)} \right] _{\lambda
\longrightarrow \infty}.
\label{RvWegner}
\end{equation}
\citet{b8} applied the $\chi^2$ minimization method to compute
$R_V$ values for a sample of ultraviolet (UV) extinction data using
the linear relation (\ref{CCMlaw}). \citet{b4} extended the analysis
of \citet{b8} by using non-equal weights derived from observational
errors to determine $R_V$ and $A_V$ values toward the sample of 782
stars with known ultraviolet color excesses.

In this paper we use both near-infrared ($JHK$) and UV
photometry to obtain $R_V$ values for a sample of 436 lines of
sight. We arrive at two main conclusions: (i) there are lines of sight
with extinction in the NIR or/and UV which generically don't
follow CCM law, and (ii) there are lines of sight which show an
extinction curve that cannot be reproduced with a single $R_V$ value
in the whole wavelength range. 

The structure of this paper is the following. In \S \ref{Th.Basis} we
discuss the theoretical basis of the $\chi^2$ minimization method we
use to derive $R_V$ values. In \S \ref{Data} we describe our data
sources. In \S \ref{Results} we present the results and show two main 
peculiar classes of extinction curves present in our sample. Finally
in \S \ref{Conclusion} we summarize our results.

\section{Theoretical Considerations}
\label{Th.Basis}

We normalize extinction in a standard way: 
\begin{equation}
\epsilon(\lambda-V)=\frac{E(\lambda-V)}{E(B-V)}.
\end{equation}
The absolute extinction may be deduced from the relative extinction by
using total-to-selective extinction ratio $R_V$:
\begin{equation}
R_V=\frac{A_V}{E(B-V)},
\end{equation}
because:
\begin{equation}
\epsilon(\lambda-V)=\frac{E(\lambda-V)}{E(B-V)}=
\frac{A_{\lambda}-A_{V}}{E(B-V)} 
= R_V \left \{\frac{A_{\lambda}}{A_{V}}-1 \right \}.
\label{Eps}
\end{equation}
For each individual band, equation (\ref{CCMlaw}) and (\ref{Eps}) can
be combined to derive an $R_V$ value. More generally, the $\chi^2$ 
minimization can be used to obtain the $R_V$ value that provides the
best CCM fit to all observed extinction data. Here
we follow our previous method \citep{b4} and use a weighted
formula to find $R_V$ by minimizing the following $\chi^2$:
\begin{equation}
\chi^2=\sum_{i=1}^{N_{\rm bands}} w_{\lambda_i} \{
\epsilon(\lambda_i-V) - [R_V(a(x_i) -1)+
b(x_i)] \} ^2 ~E^2(B-V),
\label{chiweighted}
\end{equation}
where $\omega_i \equiv 1/\sigma_i^2$ are the weights associated with
each band, and $a(x_i)$ and $b(x_i)$ are the coefficients of CCM law.

Minimizing equation (\ref{chiweighted}) with respect to $R_V$, yields:
\begin{equation}
R_V=\frac{\sum_{i=1}^{N_{\rm bands}} \{ (a(x_i)-1)\cdot
(\epsilon(\lambda_i-V)-b(x_i))/ \sigma_i^2
\}}{\sum_{i=1}^{N_{\rm bands}} \{ (a(x_i)-1)^2/
\sigma_i^2 \} },
\label{Rvweighted}
\end{equation}
where in the current work $\sigma_i$ values are taken from \citet{b16}
and they were computed according to: 
\begin{equation}
\sigma_i^2 \equiv  \sigma^2 [\epsilon(\lambda_i-V)] =\left[
\frac{1}{E(B-V)} \right]^2
\, \left \{\sigma^2[E(B-V)] + [E(\lambda_i-V)]^2 \,
\sigma^2[E(\lambda_i-V)] \right \}
\label{sigma}
\end{equation}
We compute the error in $R_V$ as\footnote{See \citet{b4} for further
details.}: 
\begin{equation}
\sigma(R_V) \equiv \sum_{j=1}^{N_{\rm bands}} \left
|\frac{\partial R_V}{\partial
\epsilon(\lambda_j-V)} \right| \cdot \sigma_j
= \frac{1}{\sum_{i=1}^{N_{\rm bands}}
(a(x_i)-1)^2/\sigma_i^2} \cdot \sum_{j=1}^{N_{\rm bands}} \left | 
\frac{a(x_j)-1}{\sigma_j} \right|
\label{errabs}
\end{equation}

\section{Data}
\label{Data}
\begin{figure}
\includegraphics[width=150mm]{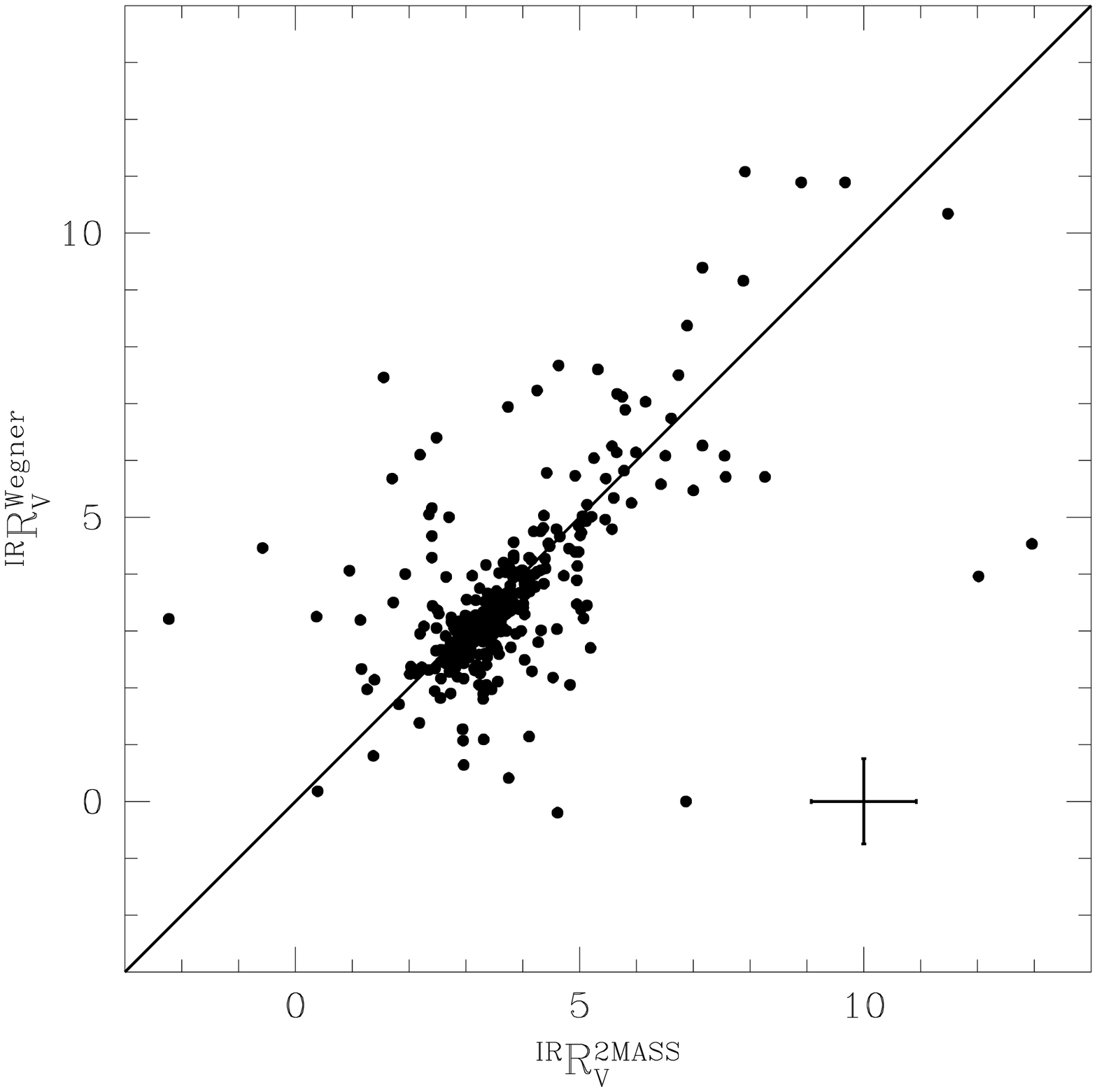}
\caption{Comparison between $^{\rm{IR}}\!R_V$ values obtained using
Wegner's (2002) data with the ones computed from 2MASS' photometry. The
line represents the 1-to-1 relation and the mean errors in both
coordinates are shown in the lower right corner of the plot.}
\label{figure1}
\end{figure}

The data used for the determination of extinction curve shapes have
two essential ingredients: 1) photometric measurements, 2) intrinsic
colors for a star with a given spectral type and luminosity
class. Since \citet{b16} provided a self-consistent set of intrinsic
colors over the entire wavelength range for 436 lines of sight, we use
his sample of stars in our analysis. Wegner's (2002) catalog provides
both intrinsic colors \citep{b15} and photometric measurements in
$UBVRIJHKLM$ bands. In our preliminary analysis \citep{b5} we use
Wegner's (2002) colors as our only data source. However, Wegner's
(2002) photometric data
in the infrared come from several sources and as a result are rather 
inhomogeneous\footnote{\citet{b16} takes infrared magnitudes in
$J,H,K,L,M$ filters mostly from the catalog of
Gezari, Schmitz, \& Mead (1984) and \citet{b7}. The catalog of Gezari
et al.\ (1984; 1993) contains infrared observations published in the
scientific literature from 1965 through 1990. Wegner's (2002) $R$ and $I$
magnitudes with accuracy of $\pm 0.01$ mag originate from \citet{b10}
and \citet{b3}. The spectral classification and $UBV$ data with
accuracy of $\pm 0.01$ mag are quoted after SIMBAD database.
The effective wavelengths of the optical/NIR bands are:
$\lambda_U=0.36 \mu m$, $\lambda_R=0.71 \mu m$, $\lambda_I=0.97\mu m$,
$\lambda_J=1.25 \mu m$, $\lambda_H=1.65 \mu m$, $\lambda_K=2.2 \mu m$,
$\lambda_L=3.5 \mu m$, $\lambda_M=4.8 \mu m$.}.
Therefore, here we opt for using the Two Micron All Sky Survey 
(2MASS) database \citep{b9} as our only input outside of the UV
spectral range. It is a very homogeneous database covering the entire
sky. Complete 2MASS photometry with unflagged errors is available 
for all stars (except
for HD7252, HD34087, HD183143) in Wegner's (2002) sample. Since
2MASS catalog includes only $J,H,K_S$ bands ($\lambda_J=1.235\mu m$, 
$\lambda_H=1.662 \mu m$, $\lambda_{K_S}=2.159 \mu m$) and so
measurements span a
relatively narrow wavelength range we test for possible systematic
problems by comparing $R_V$ values obtained using 2MASS data
($^{\rm{IR}}\!R_V^{\rm{2MASS}}$) with the ones computed using Wegner's
(2002) optical/NIR data ($^{\rm{IR}}\!R_V^{\rm{Wegner}}$). The results
are shown in Figure \ref{figure1}. Only 4\% of lines of sight have  
$^{\rm{IR}}\!R_V$ values strongly deviating from the 1-to-1
relationship and those are listed in Table \ref{table1}. In
the first column we give stellar IDs; 
in the second column we report $^{\rm{IR}}\!R_V^{\rm{Wegner}}$ values computed using Wegner's (2002)
data in UV; the third column shows the error in
$^{\rm{IR}}\!R_V^{\rm{Wegner}}$; 
the fourth column lists
$^{\rm{IR}}\!R_V^{\rm{2MASS}}$ values obtained using $J,H,K$ bands
from 2MASS; errors in $^{\rm{IR}}\!R_V^{\rm{2MASS}}$
are given in the fifth column. The sixth
column shows the difference in $R_V$ obtained by comparing columns
two and four normalized to the combined error. We remove these
19 suspicious lines of sight and advance 414 stars to further analysis.

\placefigure{figure1}

In summary, 2MASS is our only source of IR photometry and we adopt
intrinsic colors used by \citet{b16}, which originate from his previous
work \citep{b15}. Since
the effective wavelengths in IR are different between 2MASS and
\citet{b15} we discuss whether any corrections are necessary to put
them on the same system (Appendix \ref{appendix}). 

The ultraviolet photometry is taken from \citet{b19} and based on 
\emph{Astronomical Netherlands Satellite} (ANS); the effective wavelengths of 
the $UV$ bands are: 0.1549, 0.1799, 0.2200, 0.2493, and 0.3294 $\mu m$. 

\placetable{table1}

\section{Results}
\label{Results}
In this section we present the results of our analysis. We analyze the
goodness of CCM fit for our sample (\S \ref{Chitest}) and test the 
universality of CCM law as a function of wavelength (\S \ref{univtest}).

\subsection{$\chi^2$ test}
\label{Chitest}

\begin{figure}
\begin{center}
\includegraphics[width=80mm]{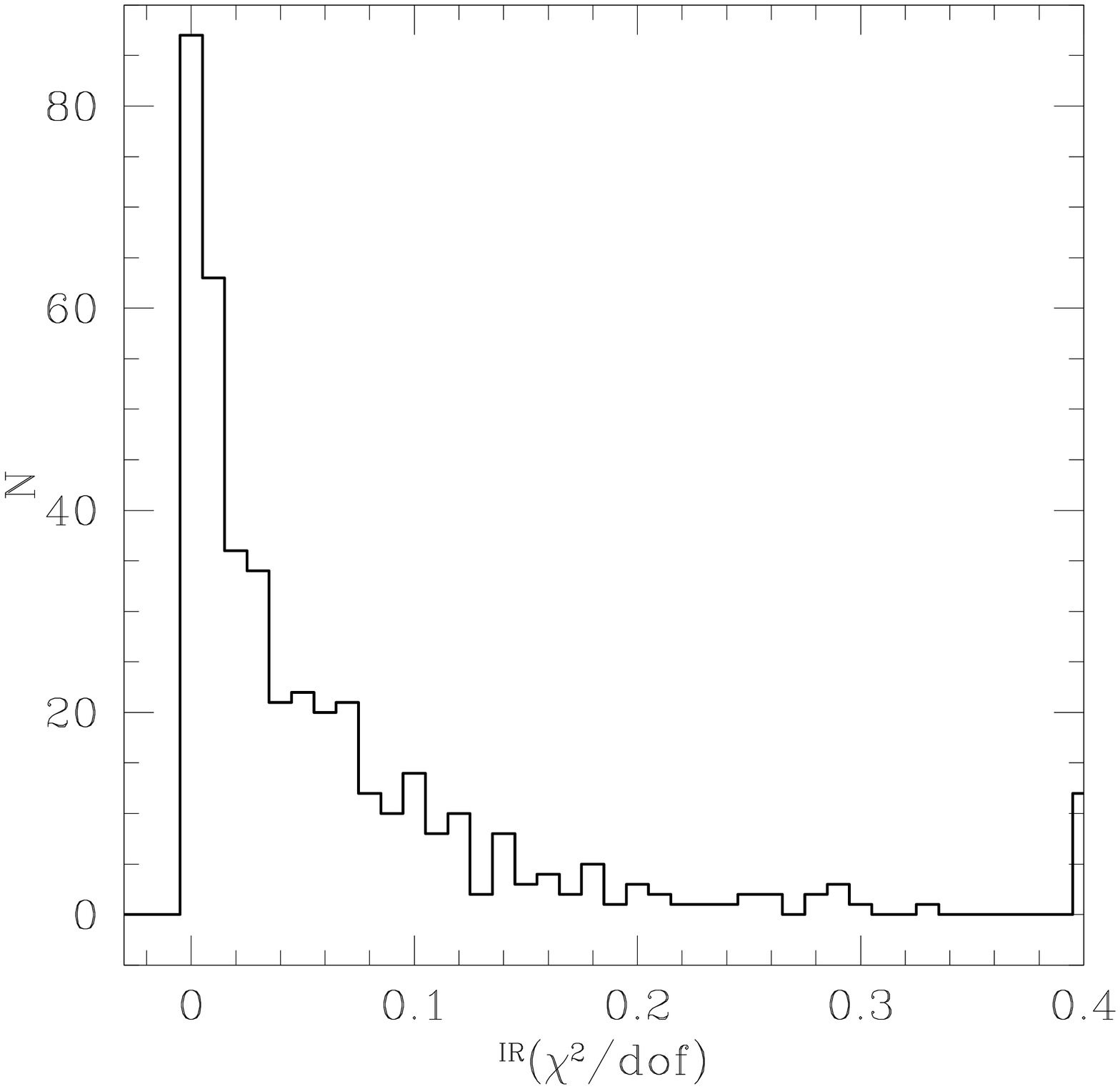}
\includegraphics[width=80mm]{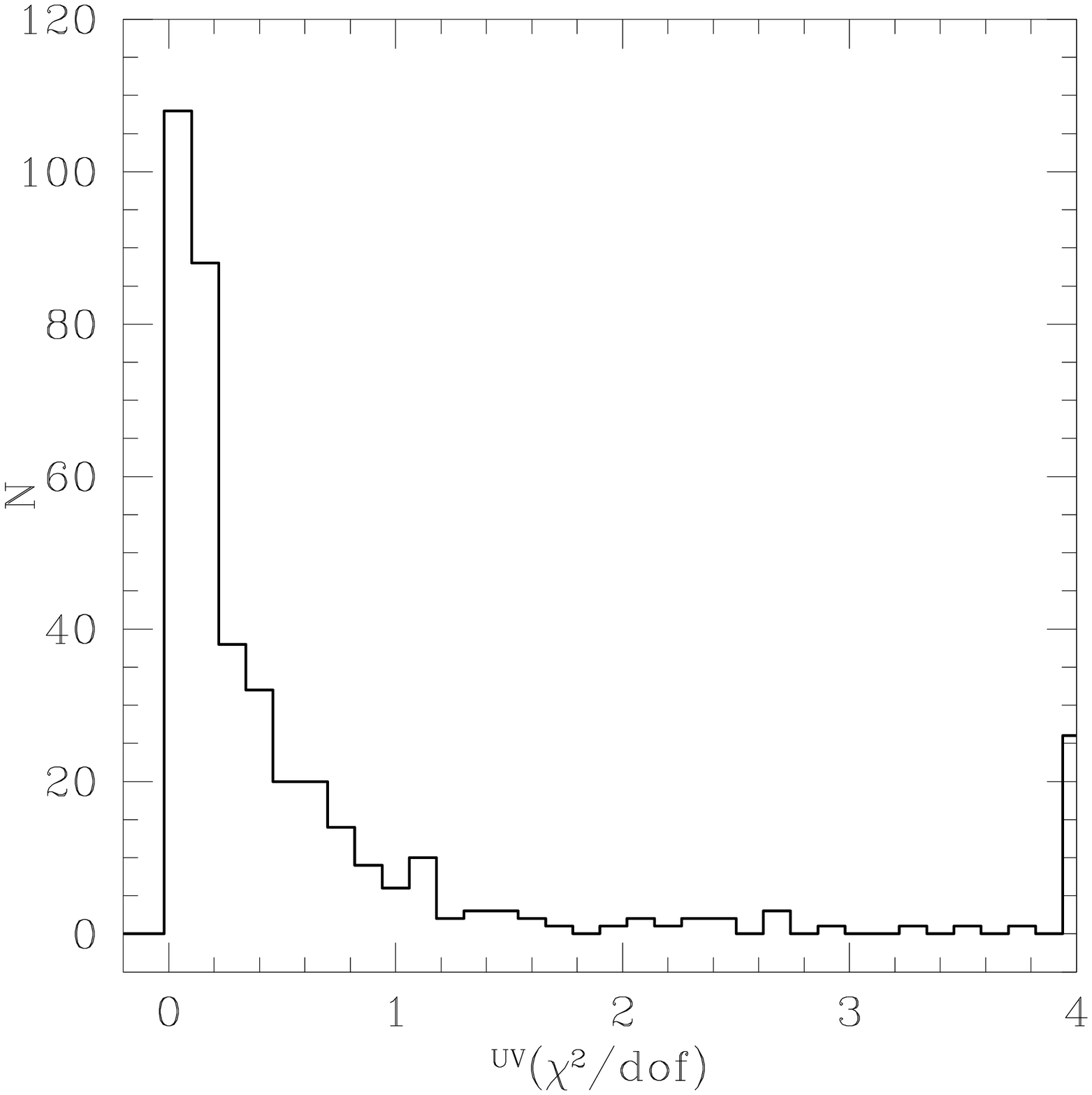}
\caption{Histograms of $\chi^2/dof$ for the CCM fits in IR (left) and
UV (right).}
\label{figure2}
\end{center}
\end{figure}

\placefigure{figure2}

A useful method to test if all extinction data points are well
fitted by CCM law is to compute $\chi^2$ per degree of freedom
($\chi^2/dof$) based on equation (\ref{chiweighted}). The number of
degrees of freedom is the number of points minus the number of fitted
parameters (in our case the only parameter is $R_V$). We note that our
$\chi^2/dof$ have a mean value less than one in both wavelength
ranges, indicating that the errors might have been
overestimated. There are two major contributors to errors: 1)
photometric uncertainties from 2MASS in IR and from \citet{b16} in UV,
2) intrinsic color uncertainties set by \citet{b16}, and it is not
obvious how to scale the errors down. Therefore, 
we do not renormalize our errors requesting $<\chi^2/dof>=1$. 
However, we believe that it is safer to treat
our $\chi^2/dof$ as a measure of a relative rather than absolute
quality of the CCM fit. We remove the lines of sight
with extinction curves not well fitted by CCM law. We define outliers 
based on the tail of our $\chi^2/dof$ distributions: $\chi^2/dof >
0.28$ for the IR data, and $\chi^2/dof > 1.6$ for the UV data (Figure 
\ref{figure2}). We find that 60 lines of sight (14\% of the entire
sample) disagree with CCM law at this level. They are listed in Table
\ref{table2}. In the first column we give stellar IDs; in the
second column we report $^{\rm{IR}}\!R_V^{\rm{2MASS}}$ values obtained using 
$J,H,K$ bands from 2MASS; the third column shows the error in
$^{\rm{IR}}\!R_V^{\rm{2MASS}}$; the fourth column lists $\chi^2/dof$
values for the CCM fit in IR; in the fifth column we give
$^{\rm{UV}}\!R_V^{\rm{Wegner}}$ values computed using Wegner's (2002)
data in UV, and in the sixth column the error in
$^{\rm{UV}}\!R_V^{\rm{Wegner}}$. Finally in the seventh column there are
$\chi^2/dof$ values for the CCM fit in UV.

\placetable{table2}

\subsection{Test for the universality of CCM law}
\label{univtest}

\begin{figure}
\begin{center}
\includegraphics[width=80mm]{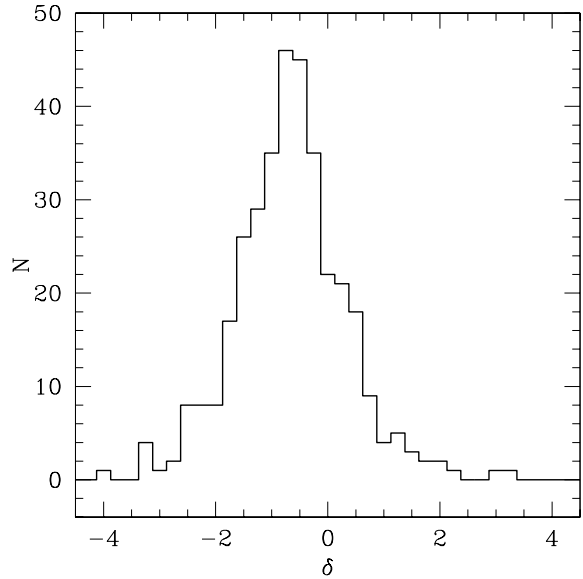}
\caption{Histogram of $\delta$ computed according to equation
(\ref{deviation}). The median of the
distribution is at $-0.64$. We define outliers as the ones with
$\delta \leq -2.0$ or $\delta \geq 1.0$.}
\label{figure3}
\end{center}
\end{figure}

\placefigure{figure3}

We further analyze the 354 lines of sight with $\chi^2/dof$ below the
limits set in \S \ref{Chitest} for which we
expect that CCM law fits well all observed extinction data from
NIR to UV. The usual assumption is that the knowledge of the
$R_V$ value obtained from the optical/NIR part of the extinction
curve may be used to obtain the entire extinction curve by using 
CCM law. We critically test this assumption using two sets of $R_V$
values for each line of sight: the NIR $R_V$
($^{\rm{IR}}\!R_{V}$) and ultraviolet $R_V$ ($^{\rm{UV}}\!R_{V}$). We
compute the following statistic:

\begin{equation}
\delta=
\frac{^{\rm{UV}}\!R_{V}^{\rm{Wegner}}-^{\rm{IR}}\!\!R_{V}^{\rm{2MASS}}}
{\sqrt{\sigma^2[^{\rm{IR}}\!R_{V}^{\rm{2MASS}}]+
\sigma^2[^{\rm{UV}}\!R_{V}^{\rm{Wegner}}]}}
\label{deviation}
\end{equation}

Figure \ref{figure3} shows the histogram of $\delta$. The peak of the 
histogram around a median of $-0.64$ indicates that the number of
extinction curves with $^{\rm{UV}}\!R_V < ~^{\rm{IR}}\!R_V$ is higher
than the one with $^{\rm{UV}}\!R_V > ~^{\rm{IR}}\!R_V$ or equivalently
that $R_V$ derived from UV spectral region tends to be smaller on
average. This is an
important systematic effect that deserves further study but is beyond
the scope of this paper. We choose to define outliers considering 
the symmetric distribution around the peak: specifically, we select
the outliers as the lines of sight characterized by $\delta
\leq -2.0$ or $\delta \geq 1.0$. We classify them as peculiar in the
sense that CCM law is not able to reproduce the whole extinction
curve with a single value of $R_V$. Figure \ref{figure4} shows these 
peculiar extinction curves. All 45 of them (10\% of the entire sample)
are listed in Table \ref{table3}. In
the first column we give stellar IDs; in the second column we report
$^{\rm{IR}}\!R_V^{\rm{2MASS}}$ values obtained using $J,H,K$ bands
from 2MASS; the third column shows the error in 
$^{\rm{IR}}\!R_V^{\rm{2MASS}}$; the fourth column lists 
$^{\rm{UV}}\!R_V^{\rm{Wegner}}$ values computed using Wegner's (2002)
data in UV; the errors in $^{\rm{UV}}\!R_V^{\rm{Wegner}}$ are given in
the fifth column. The sixth
column shows $\delta$ values obtained by comparing columns
two and four.

\placefigure{figure4}
\placetable{table3}

Figure \ref{figure5} presents the comparison between the $R_V$ values
obtained from NIR and UV extinction data. There is a large
scatter around the 1-to-1 relationship; however, only the 45 points with 
error bars shown are those which deviate from this relation
significantly (according to our $\delta$ cut). Valencic, Clayton \&
Gordon (2004) made a similar comparison for their sample. They compared
the $R_V$ values obtained using 2MASS photometry in IR with the $R_V$
values in UV obtained using {\it International Ultraviolet Explorer}
(IUE)\footnote{We remind the reader that in contrast to \citet{b14} we
use UV photometry from ANS satellite.}. They found that 93\% of their
sample shows a good agreement
(within 3$\sigma$) between the two values, which is consistent with
our result. As opposed to \citet{b14} however, we don't assume that our
outliers must result from faulty data. We will investigate this issue
 in the future to see to what extent the peculiar properties come from
systematic problem with the data and to what degree they are
 due to peculiar grain properties.

\placefigure{figure5}

Figure \ref{figure6} shows two examples of extinction curves spanning
wavelength range from NIR to UV which are
well fitted by the CCM law with a single $R_V$ value. Table
\ref{table4} lists $R_V$ and $A_V$ values for all 309 of such lines
of sight. Here we present only the first
15 objects. The complete table is available in the electronic version
of the Journal and on the World Wide Web\footnote{See {\tt
http://dipastro.pd.astro.it/geminale}.}. In the first column we list
the stellar ID, and in the second and third columns we give the
coordinates. 
The fourth column contains $E(B-V)$ values taken from
\citet{b16}; in the fifth and sixth columns we list $R_V$
values (obtained using all data from infrared to UV) and their errors;
in the seventh column we provide $\chi^2/dof$ of the CCM fit; in the
eighth and ninth columns we list
$A_V$ values with their errors\footnote{See equation (22) and (23) in
\citet{b4}.}. There are a lot of cases for which the $\chi^2/dof$ is
substantially less than one, which means that the errors of
the individual points might have been overestimated. For this reason
the $\chi^2/dof$ listed in Table \ref{table4} is probably a better
measure of a relative rather than absolute quality of the CCM fit.

\placefigure{figure6}
\placetable{table4}

\section{Conclusion}
\label{Conclusion}

We use a $\chi^2$ minimization method to compute $R_V$ values for
a sample of 436 lines of sight. We compare $^{\rm{IR}}\!R_V$ values
obtained using 2MASS data with the ones computed using Wegner's (2002)
data in optical/IR. We conclude that 414 lines of sight have $^{\rm{IR}}\!R_V$
that agree using different datasets. For this final sample of 414 lines
of sight we derive our $R_V$ values from IR and UV data assuming CCM
law. We analyze the
goodness of CCM fit for all lines of sight
using the $\chi^2/dof$ statistic and test the universality of CCM
law by computing $R_V$ values separately for the NIR and UV
part of the extinction curve. We
construct a catalog of $R_V$ and $A_V$ values for the 309 extinction
curves with good fits to the CCM law. We divided the remaining 24\% of
cases into two groups, according to two
main peculiarities: a) the NIR and UV extinction data points
cannot be fitted well by the CCM law (14\% of the entire sample), b) $R_V$
values are significantly different for the two spectral regions,
NIR and UV (10\% of the entire sample). 
Unless caused by faulty data, these peculiar extinction curves might
come from unusual properties of dust grains. 
Therefore, theoretical modeling of these extinction
curves (e.g., Mishchenko 1989; Saija et al.\ 2001; Weingartner \&
Draine 2001) may help us to
understand the processes that modify the properties of interstellar
grains. 

\acknowledgments
AG acknowledges the financial support from EARASTARGAL fellowship
at Max-Planck-Institute for Astrophysics, where this work has
been completed.

\appendix

\section{Transformation formulae for intrinsic colors}
\label{appendix}

2MASS catalog is our primary source of IR
photometry. However, since we want to use a self-consistent set of
intrinsic colors over the entire wavelength range, we take the intrinsic
colors from \citet{b15} to obtain the IR color excesses. As reported in 
\S \ref{Data}, the effective wavelengths of the near-infrared bands
used by \citet{b15} are 
$\lambda_J=1.25 \mu m$, $\lambda_H=1.65 \mu m$, $\lambda_K=2.2 \mu m$
and are somewhat different from the 2MASS ones:
$\lambda_J=1.235 \mu m$; $\lambda_H=1.662 \mu m$; $\lambda_K=2.159 \mu
m$. Well-defined color excesses come from photometry and intrinsic colors
corresponding to the same wavelength. The first possibility to deal
with this mismatch is to
convert our adopted intrinsic colors to the
2MASS' photometric system. To this aim, we use linear interpolation
formulae: 
\begin{eqnarray}
(J_{2M}-V)_0&=&(H-V)_{0W}+
\frac{\lambda_{J2M}-\lambda_{HW}}{\lambda_{JW}-\lambda_{HW}}
[(J-V)_{0W}-(H-V)_{0W}],\\
(H_{2M}-V)_0&=&(H-V)_{0W}+
\frac{\lambda_{H2M}-\lambda_{HW}}{\lambda_{KW}-\lambda_{HW}}
[(K-V)_{0W}-(H-V)_{0W}],\\
(K_{2M}-V)_0&=&(H-V)_{0W}+
\frac{\lambda_{K2M}-\lambda_{HW}}{\lambda_{KW}-\lambda_{HW}}
[(K-V)_{0W}-(H-V)_{0W}],
\end{eqnarray}
where the subscripts $2M$ indicate 2MASS, $W$ mean Wegner, and $0$
refer to the intrinsic colors. For the specific wavelengths used here
we obtain:
\begin{eqnarray}
(J_{2M}-V)_0&=&(J-V)_{0W}+0.025 \cdot [(J-V)_{0W}-(H-V)_{0W}],\\
(H_{2M}-V)_0&=&(H-V)_{0W}+0.018 \cdot [(K-V)_{0W}-(H-V)_{0W}],\\
(K_{2M}-V)_0&=&(K-V)_{0W}+0.073 \cdot [(H-V)_{0W}-(K-V)_{0W}].
\end{eqnarray}
For example, if we take the spectral type B1V, the color
transformation will take the following form:
\begin{eqnarray}
(J_{2M}-V)_0&=&(J-V)_{0W}-0.00225,\\
(H_{2M}-V)_0&=&(H-V)_{0W}+0.00108,\\
(K_{2M}-V)_0&=&(K-V)_{0W}-0.0044.
\end{eqnarray}
For all relevant spectral types, we find that the color adjustments are
extremely small, between 0.0001 and 0.006. Since the errors of the
intrinsic colors are much bigger than the above color shifts, we
neglect these corrections and use the intrinsic colors given by
\citet{b15} when analyzing 2MASS data.

It is possible to use a complementary approach and instead of
adjusting intrinsic colors make a photometric transformation of 2MASS
magnitudes to the Johnson's (1966) system adopted by Wegner
(2002). There are several color transformations provided by
\citet{b2} but it is not obvious which set is the most
appropriate. Independent of the transformation used the corrections to
the extinction data are small, and it is safer not to
apply them since the exact values and signs of those corrections
depend on the very fine, unknown details of the photometric system
used. This realization reinforces our decision of using 2MASS
photometry with intrinsic colors by \citet{b15} without any 
wavelength-related adjustments.

\clearpage

\begin{figure}
\includegraphics[width=0.32\textwidth]{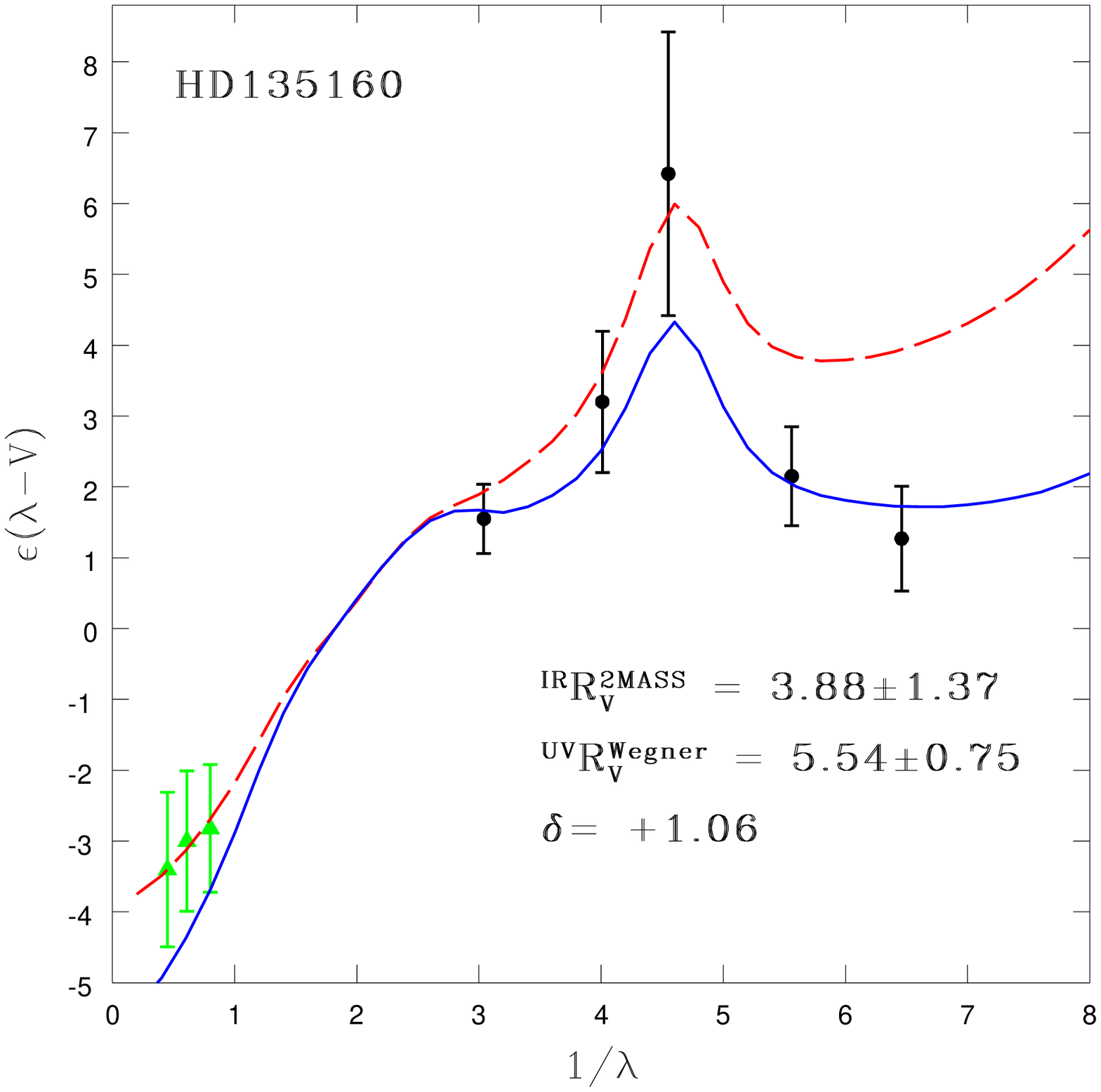}
\includegraphics[width=0.32\textwidth]{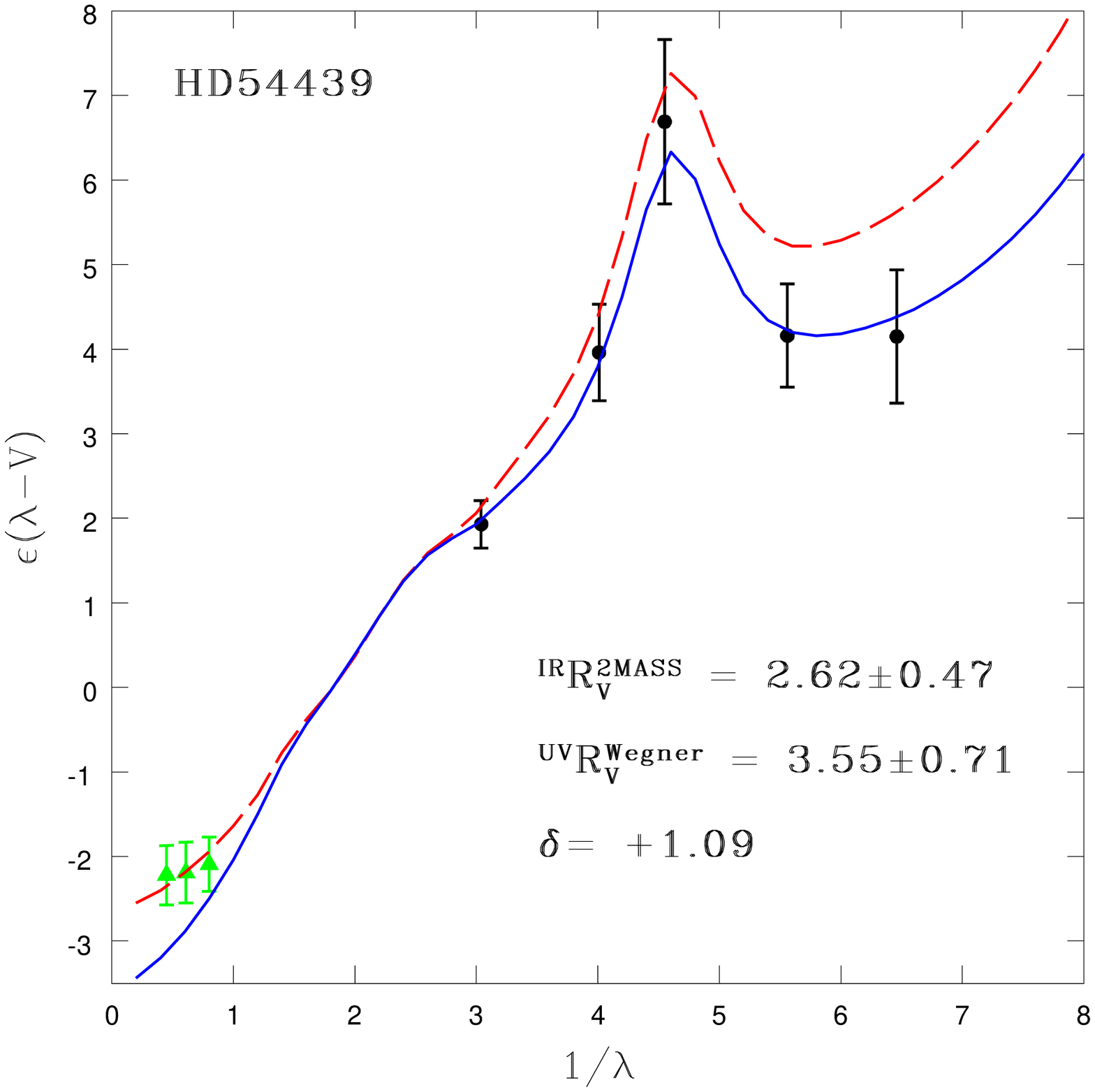}
\includegraphics[width=0.32\textwidth]{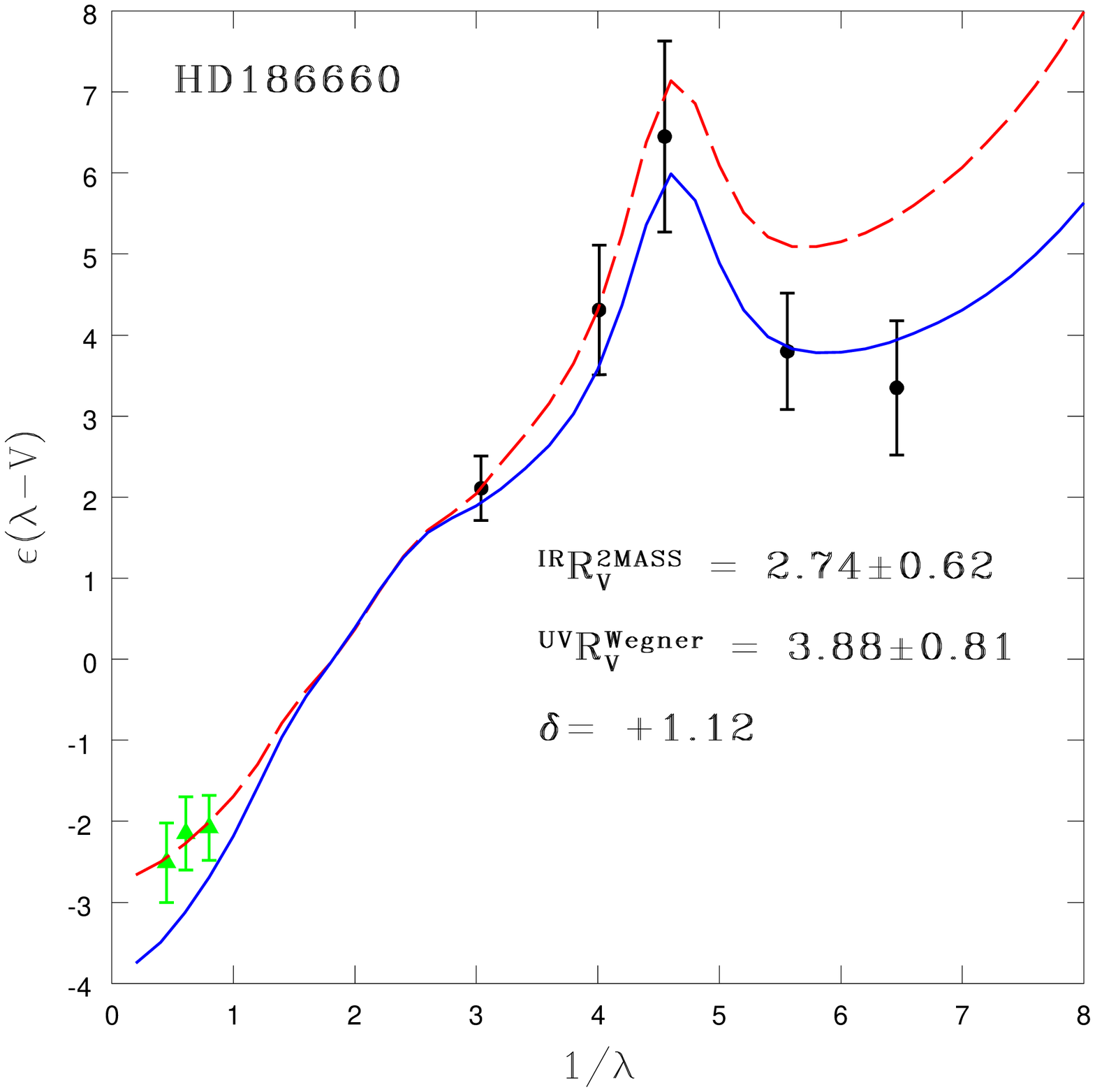}
\includegraphics[width=0.32\textwidth]{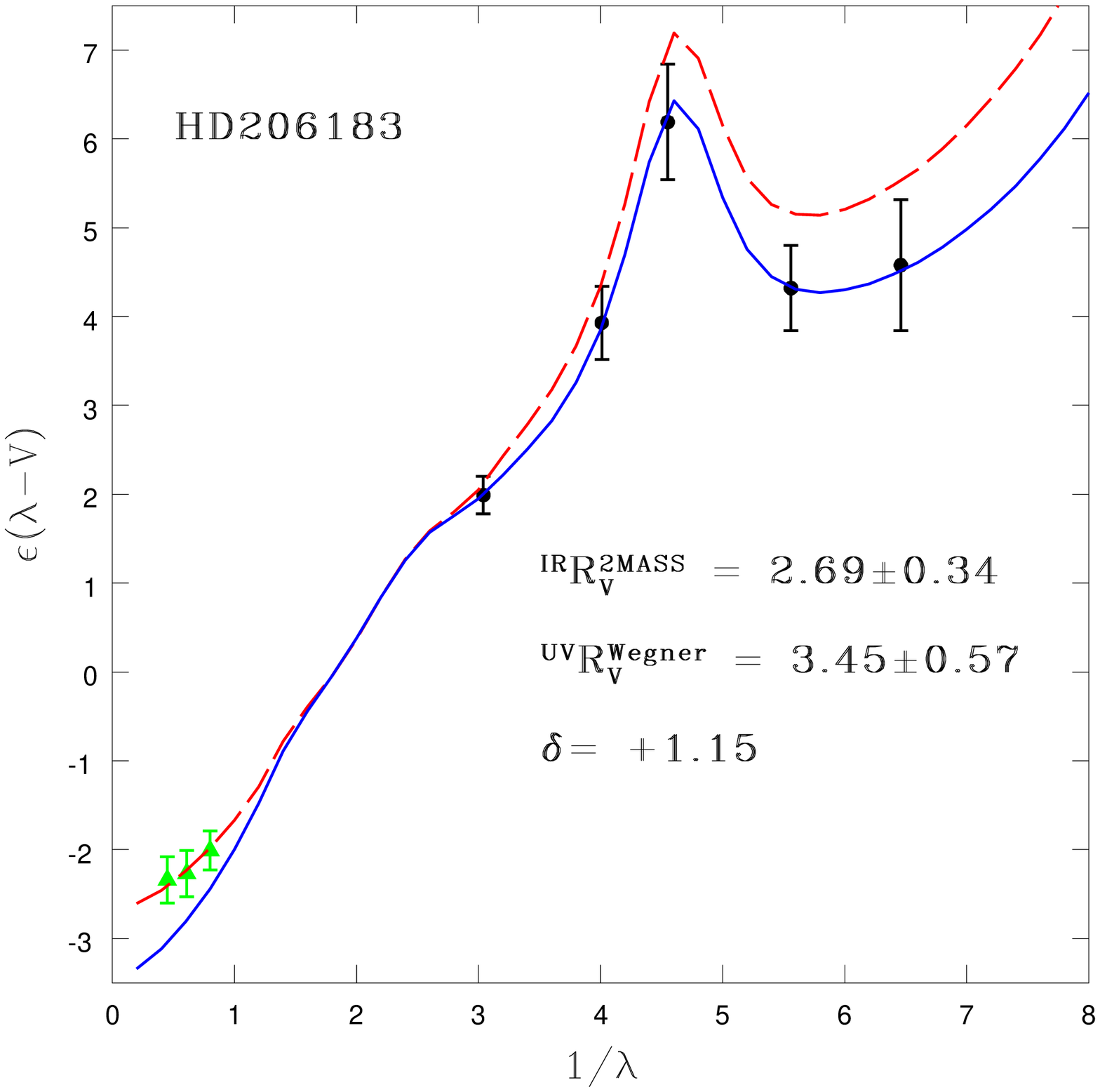}
\includegraphics[width=0.32\textwidth]{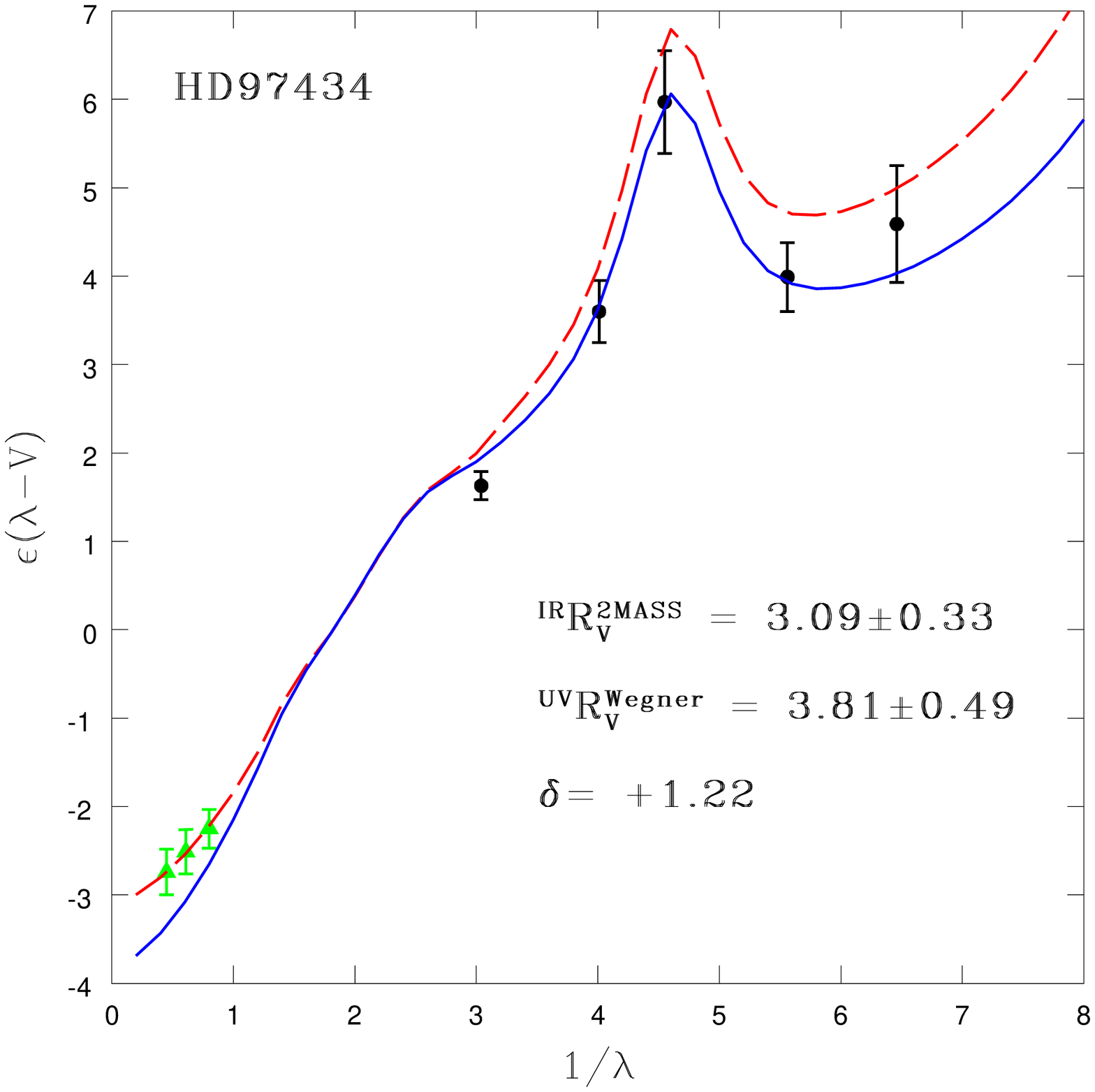}
\includegraphics[width=0.32\textwidth]{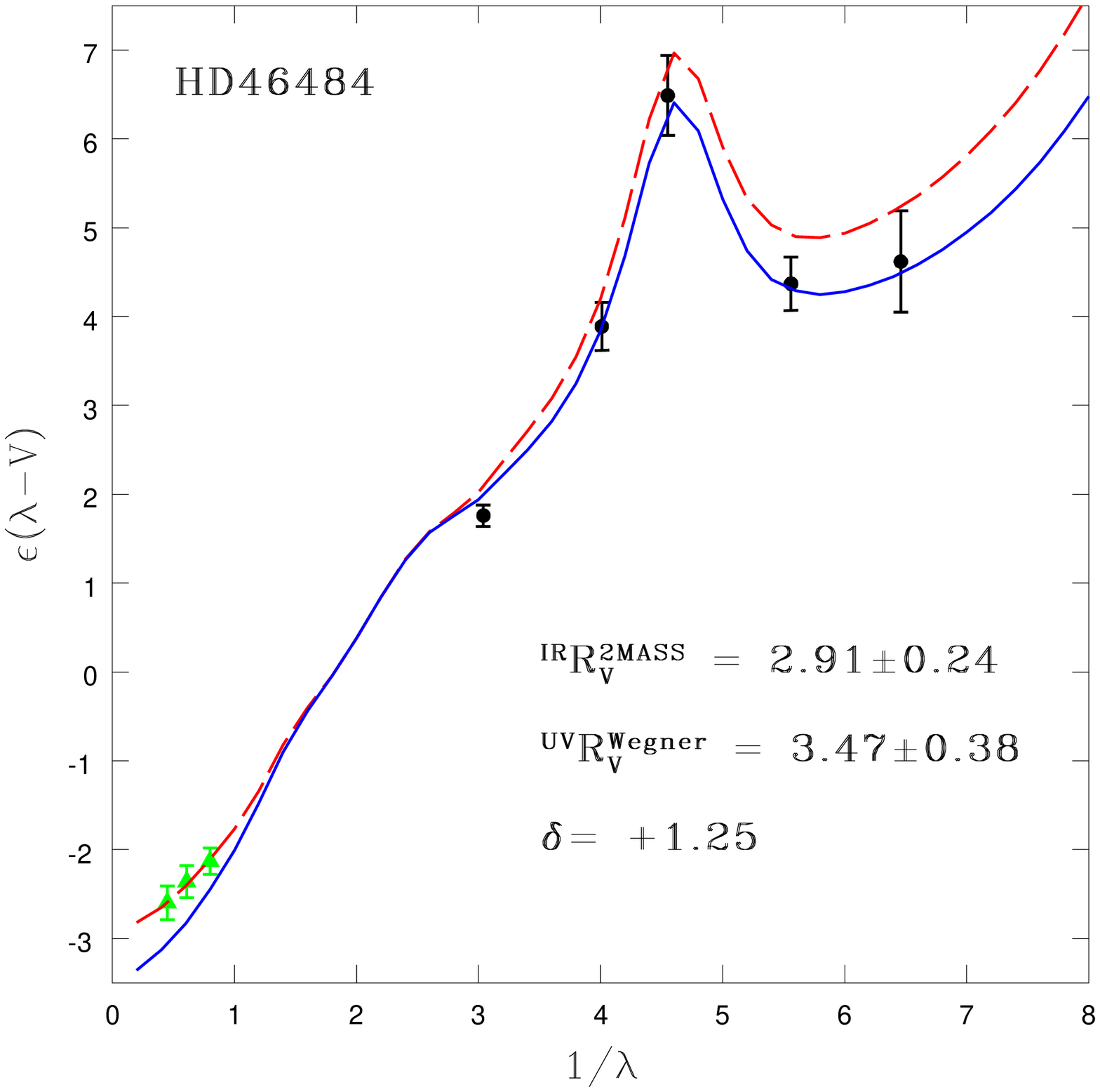}
\includegraphics[width=0.32\textwidth]{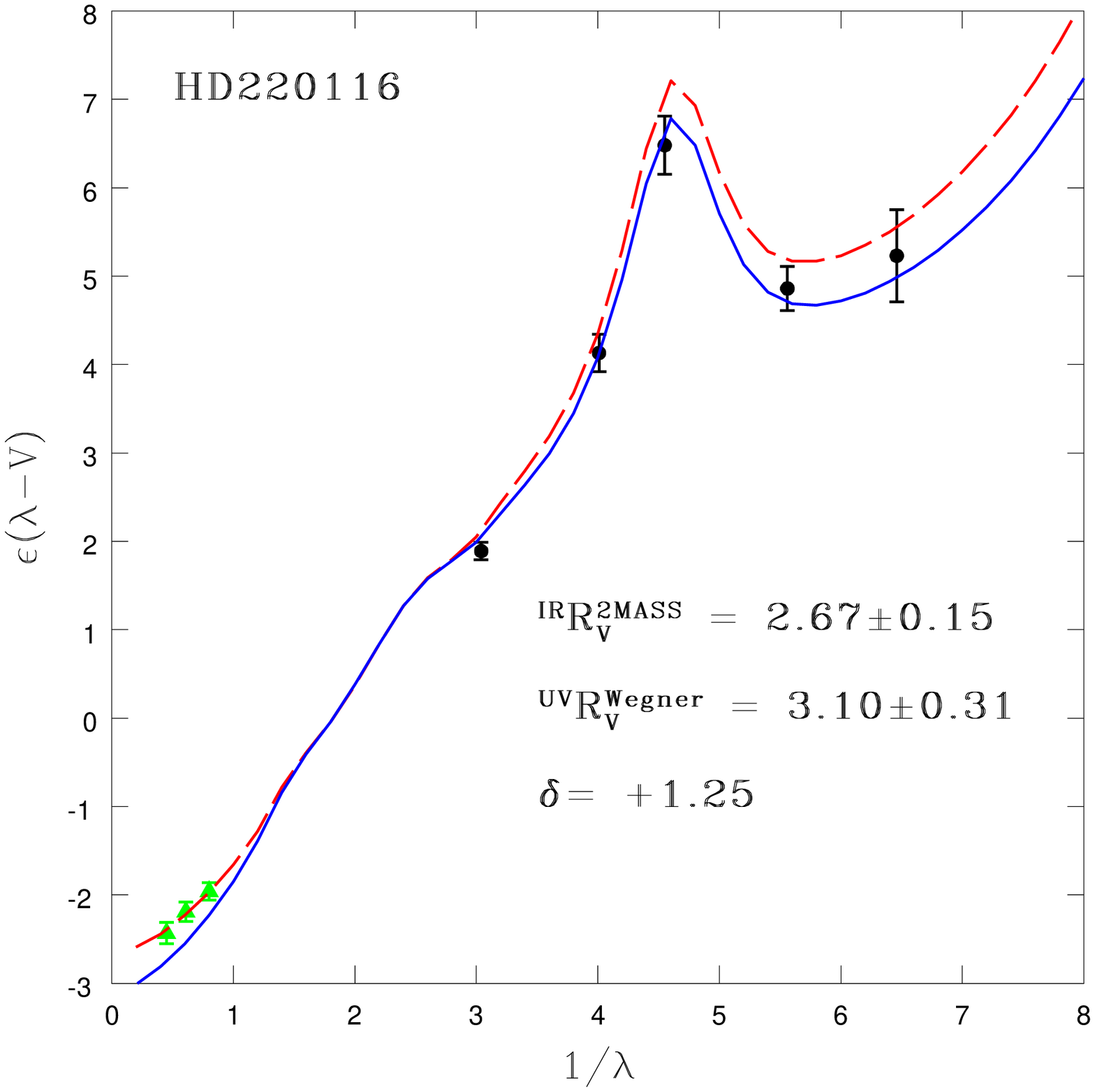}
\includegraphics[width=0.32\textwidth]{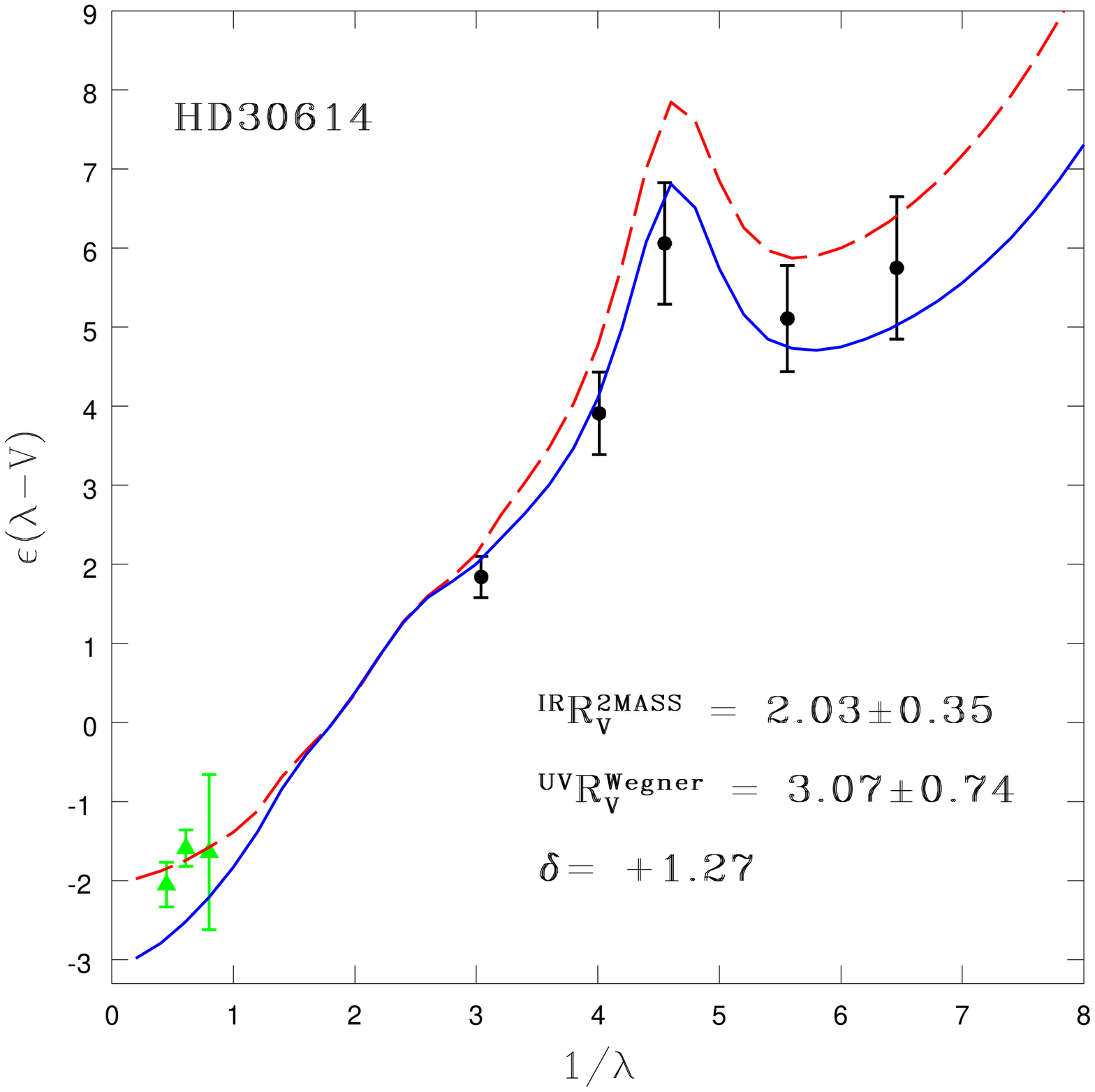}
\includegraphics[width=0.32\textwidth]{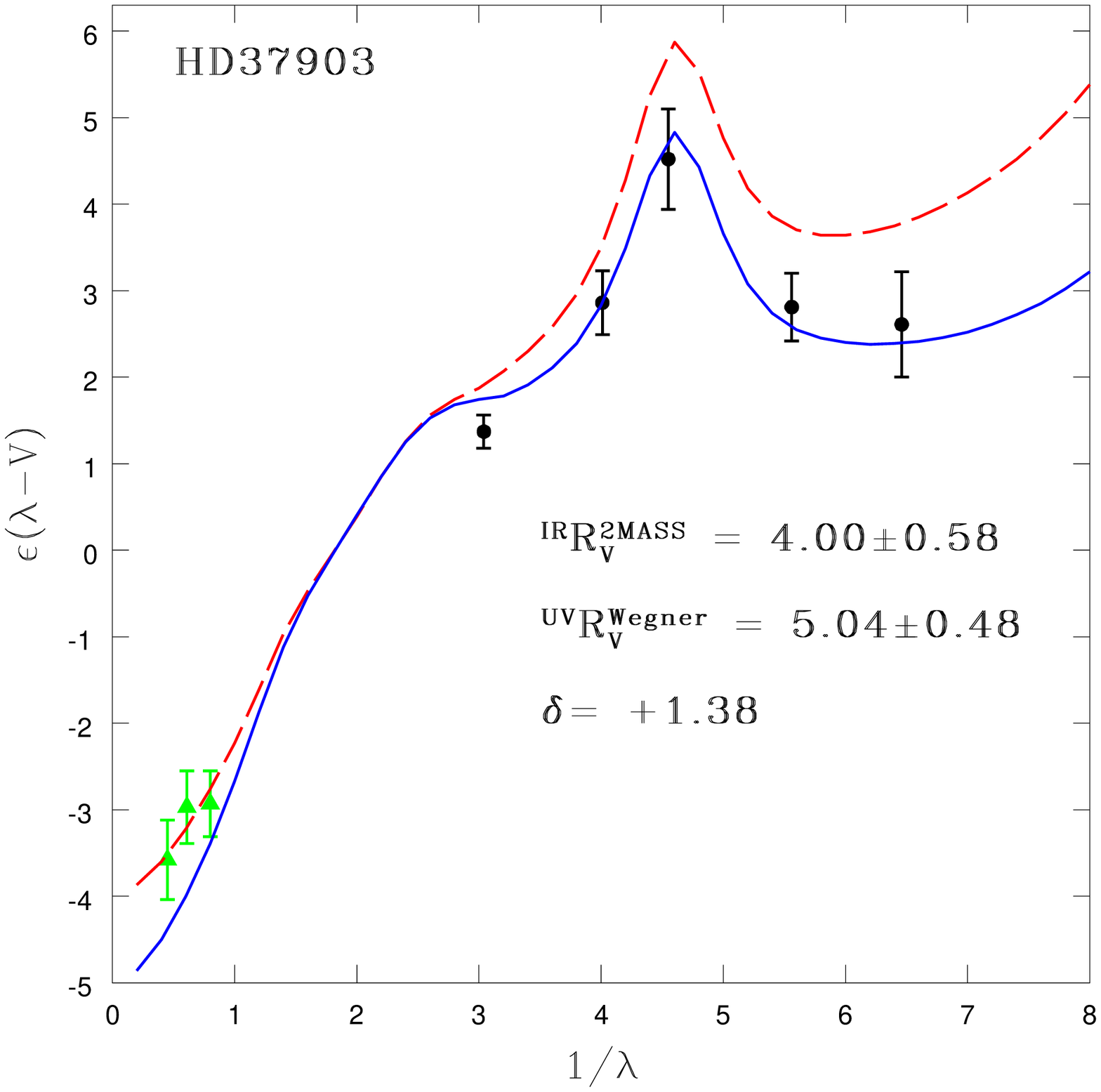}
\includegraphics[width=0.32\textwidth]{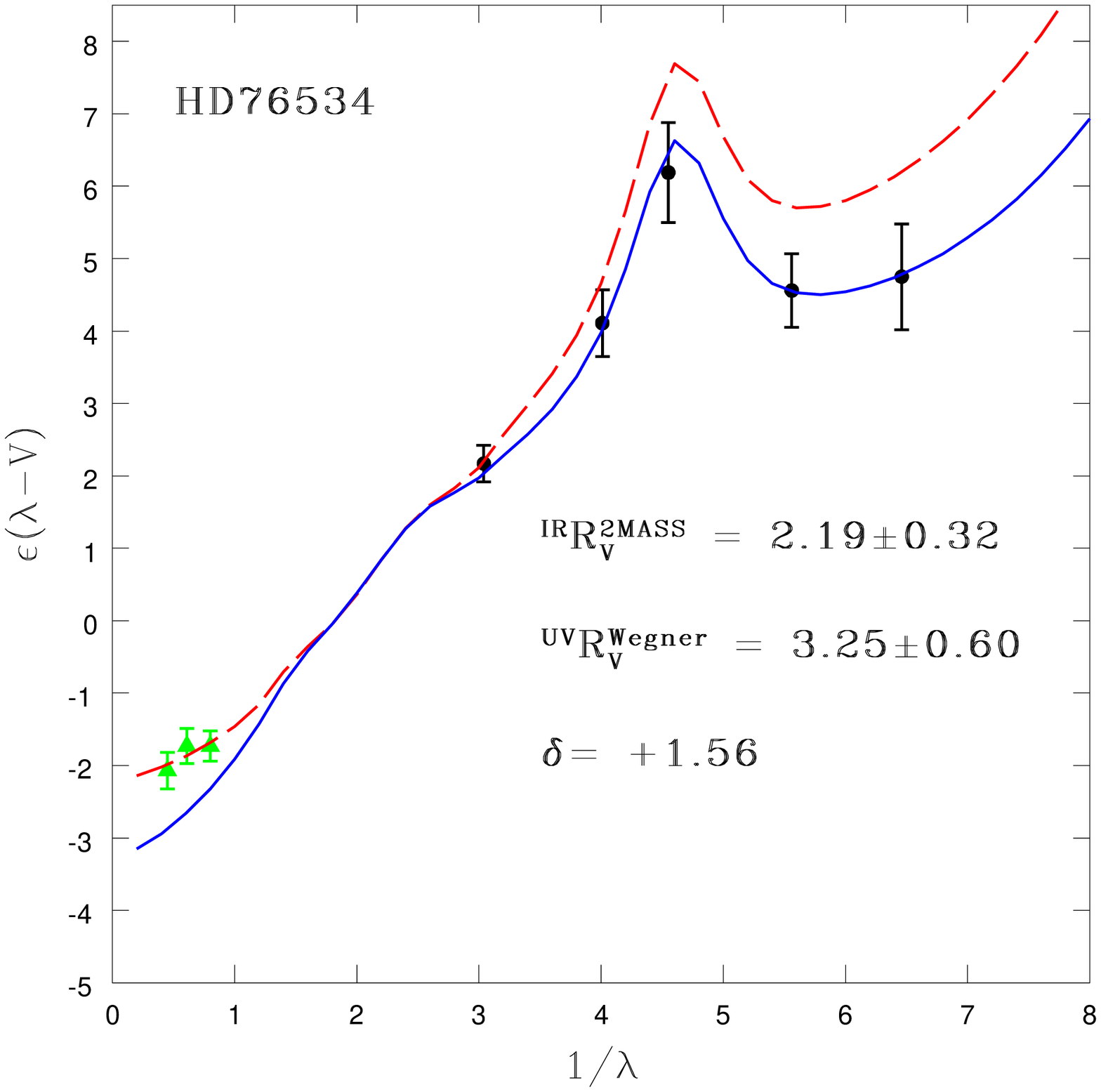}
\includegraphics[width=0.33\textwidth]{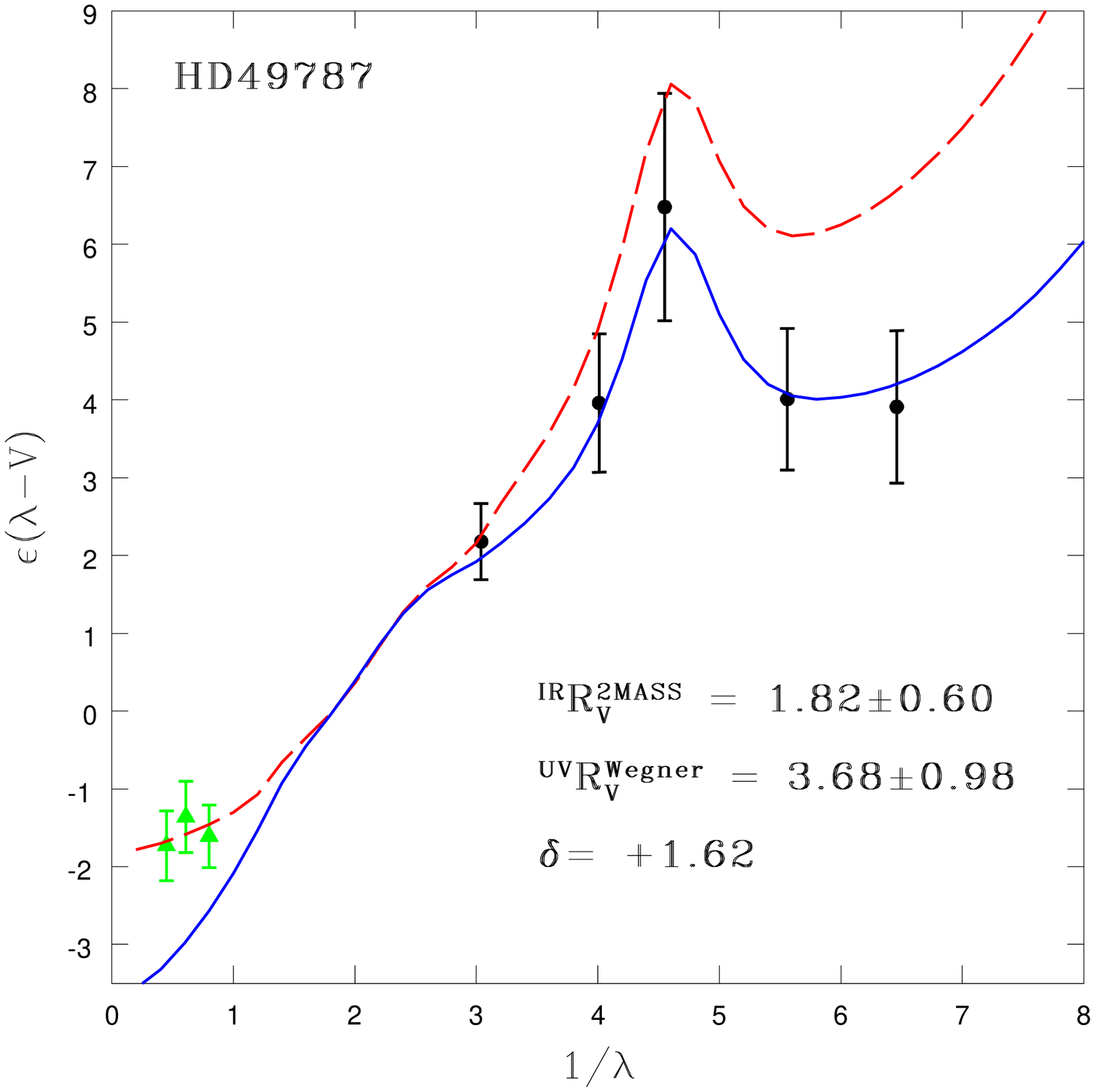}
\includegraphics[width=0.33\textwidth]{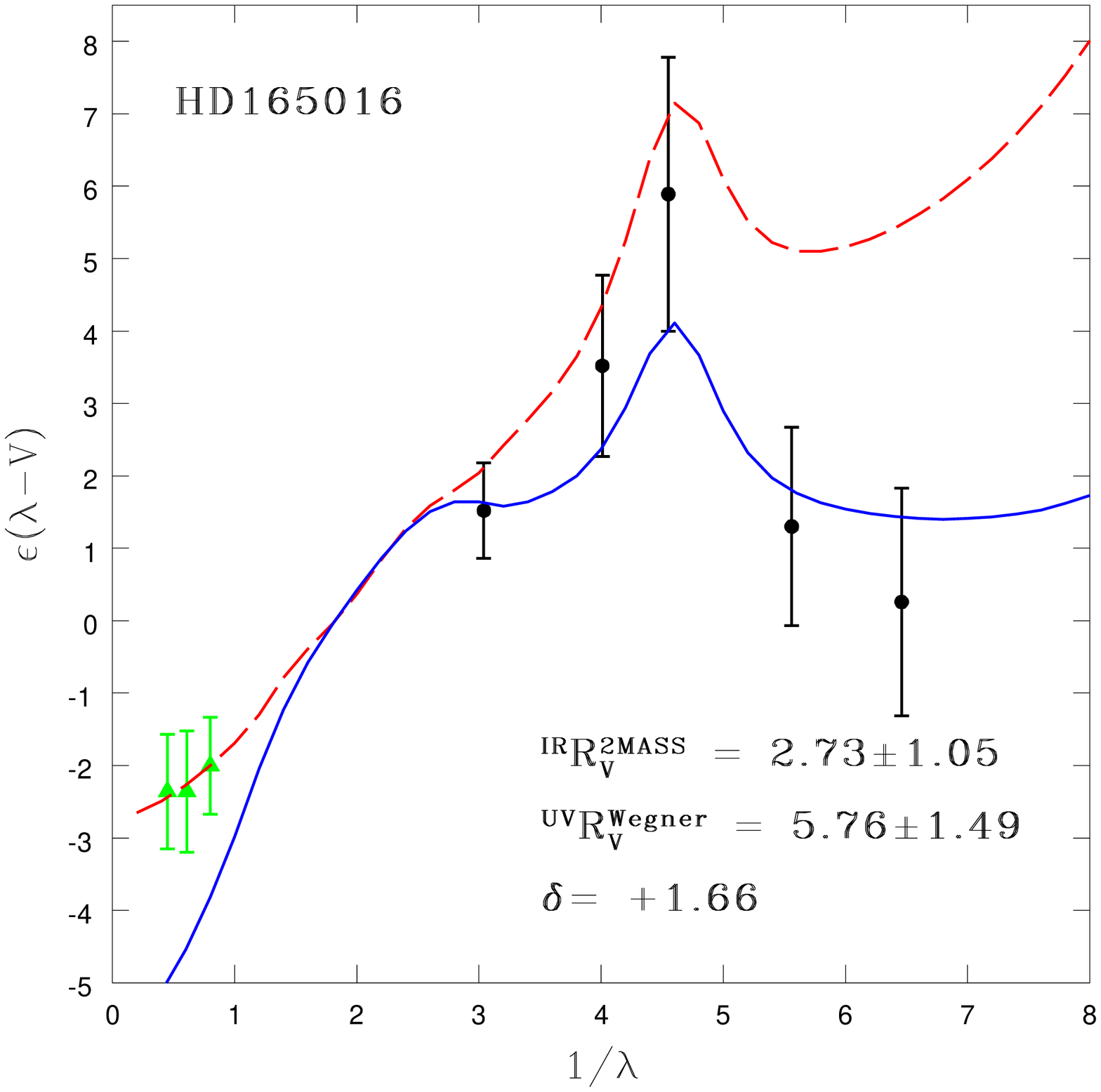}
\end{figure}

\begin{figure}
\includegraphics[width=0.32\textwidth]{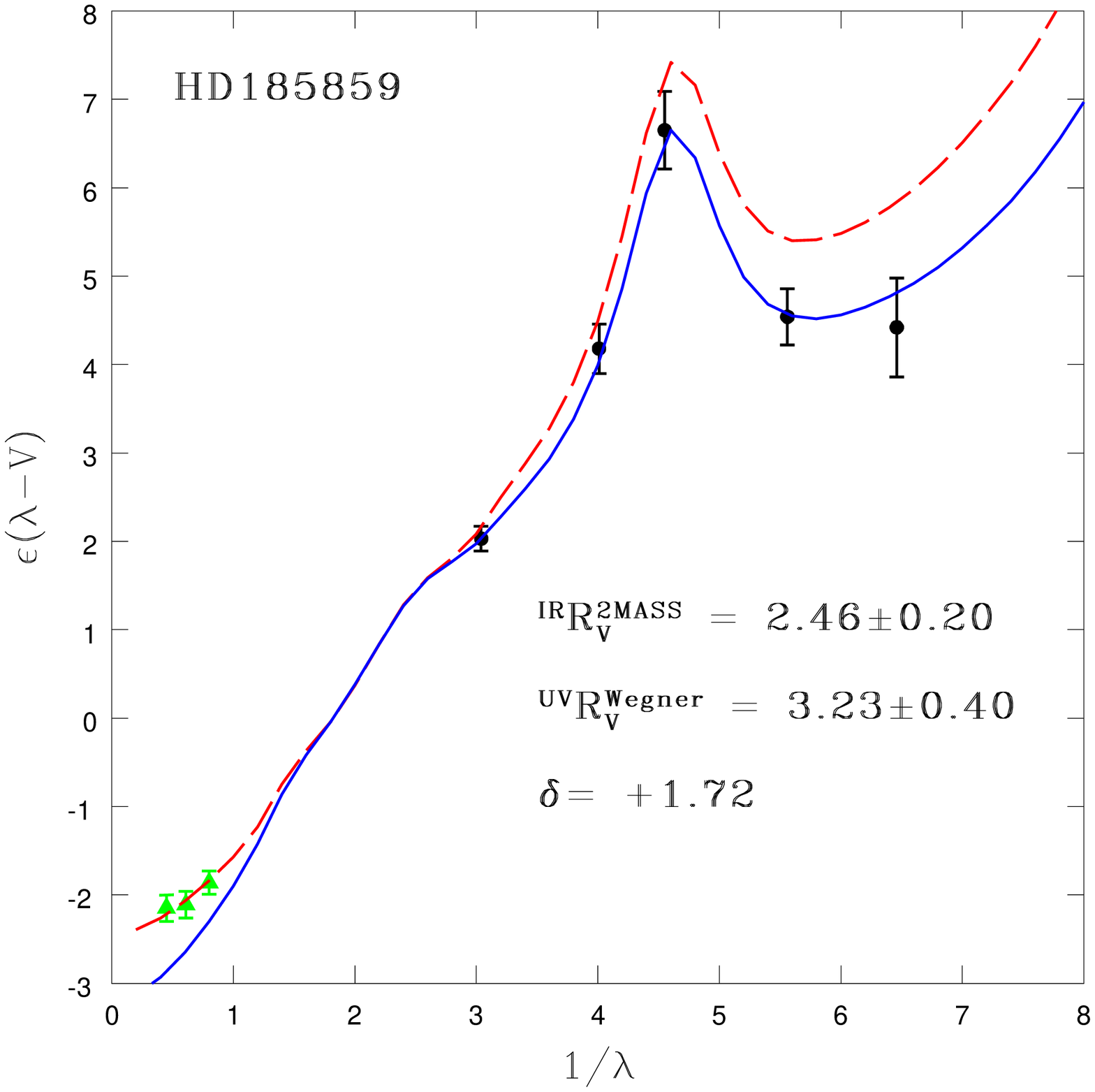}
\includegraphics[width=0.32\textwidth]{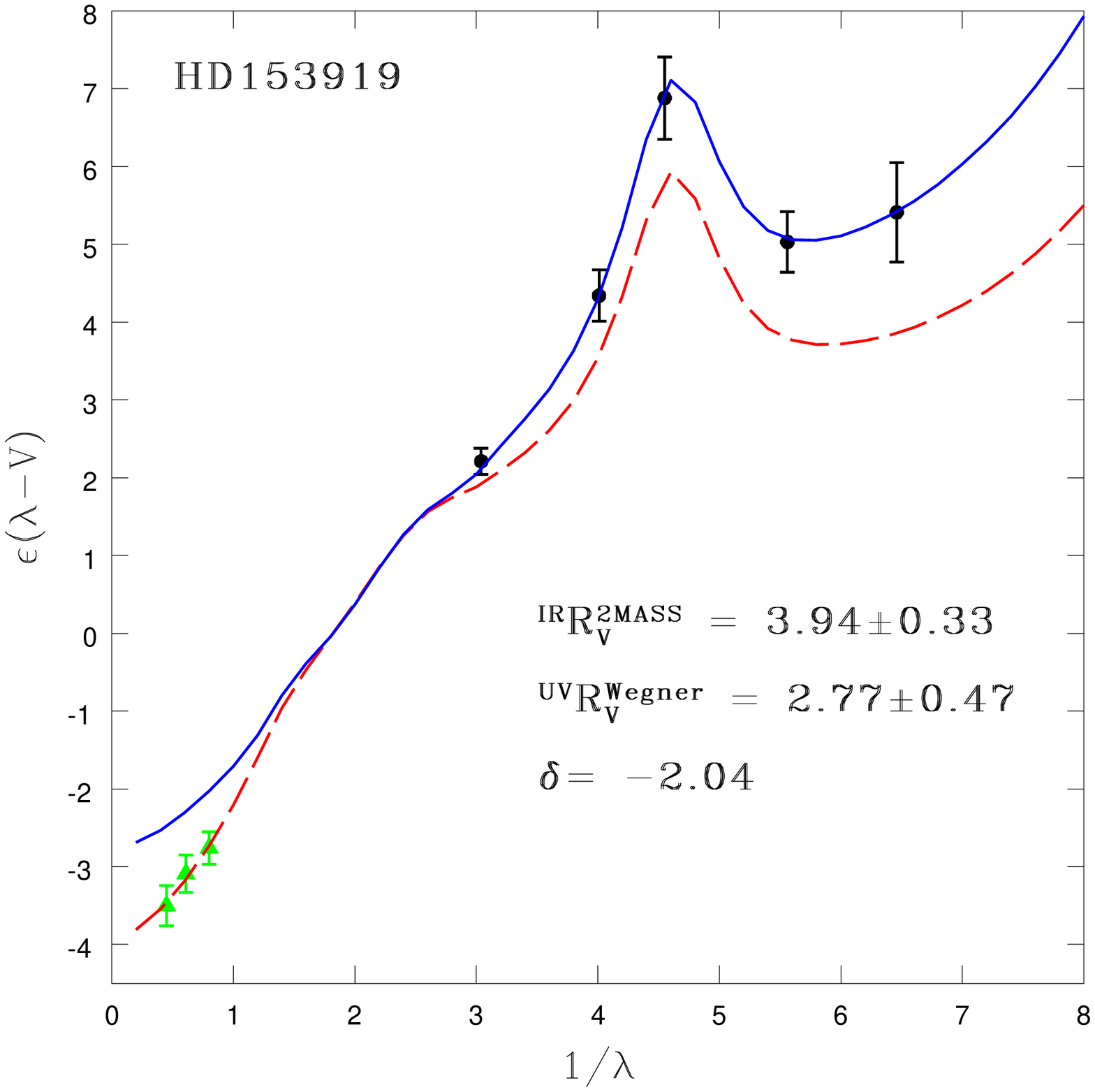}
\includegraphics[width=0.32\textwidth]{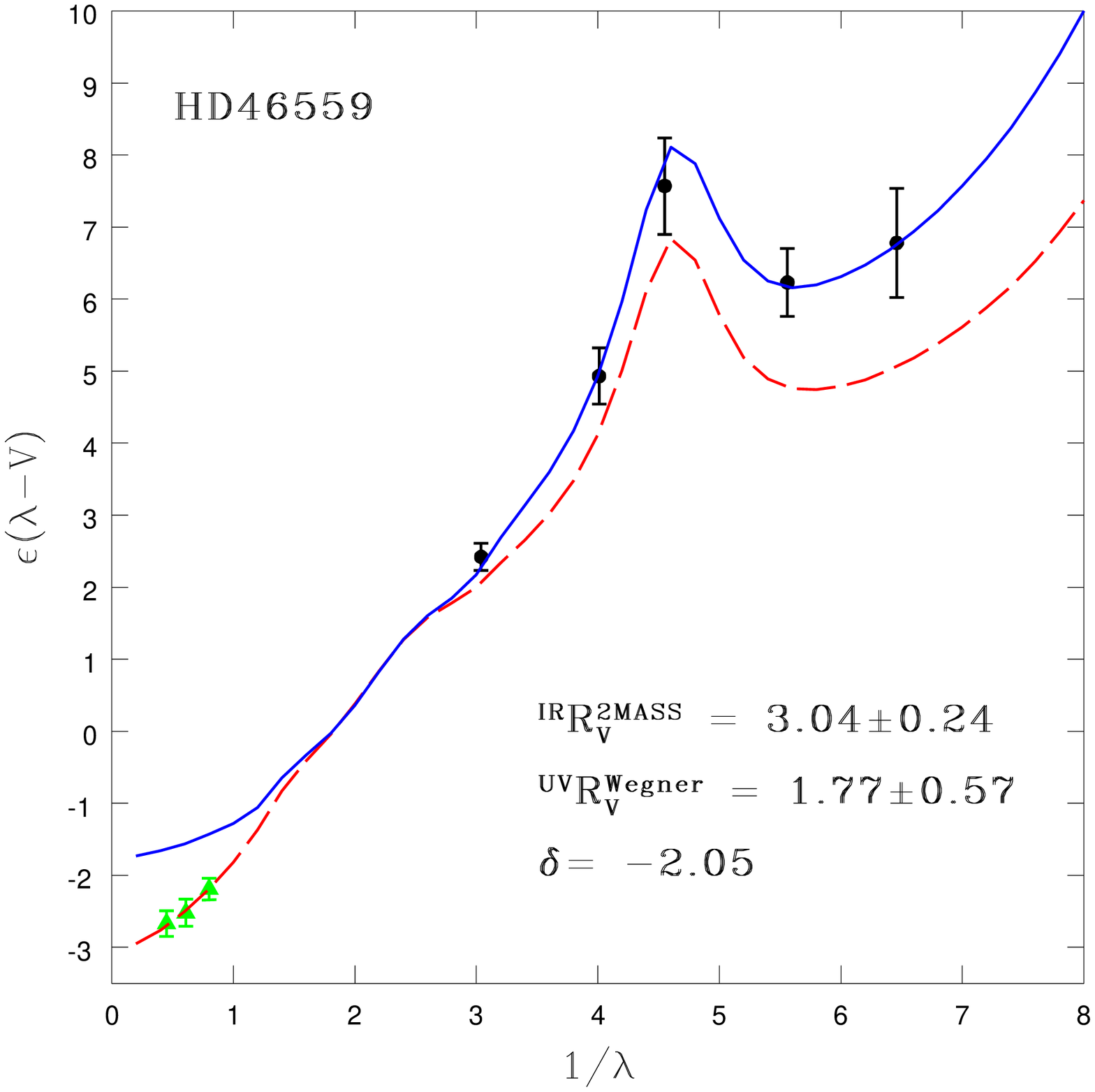}
\includegraphics[width=0.32\textwidth]{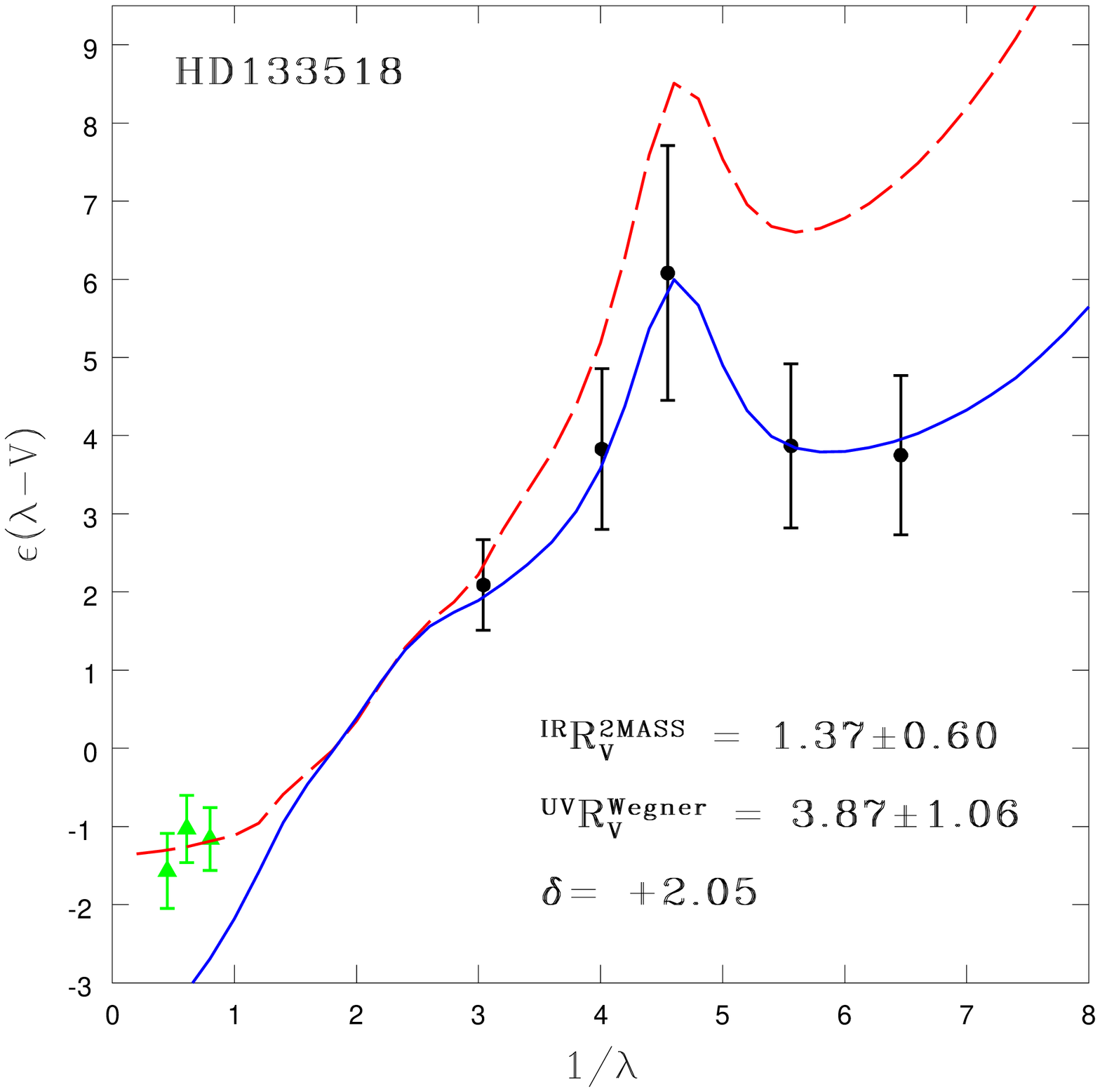}
\includegraphics[width=0.32\textwidth]{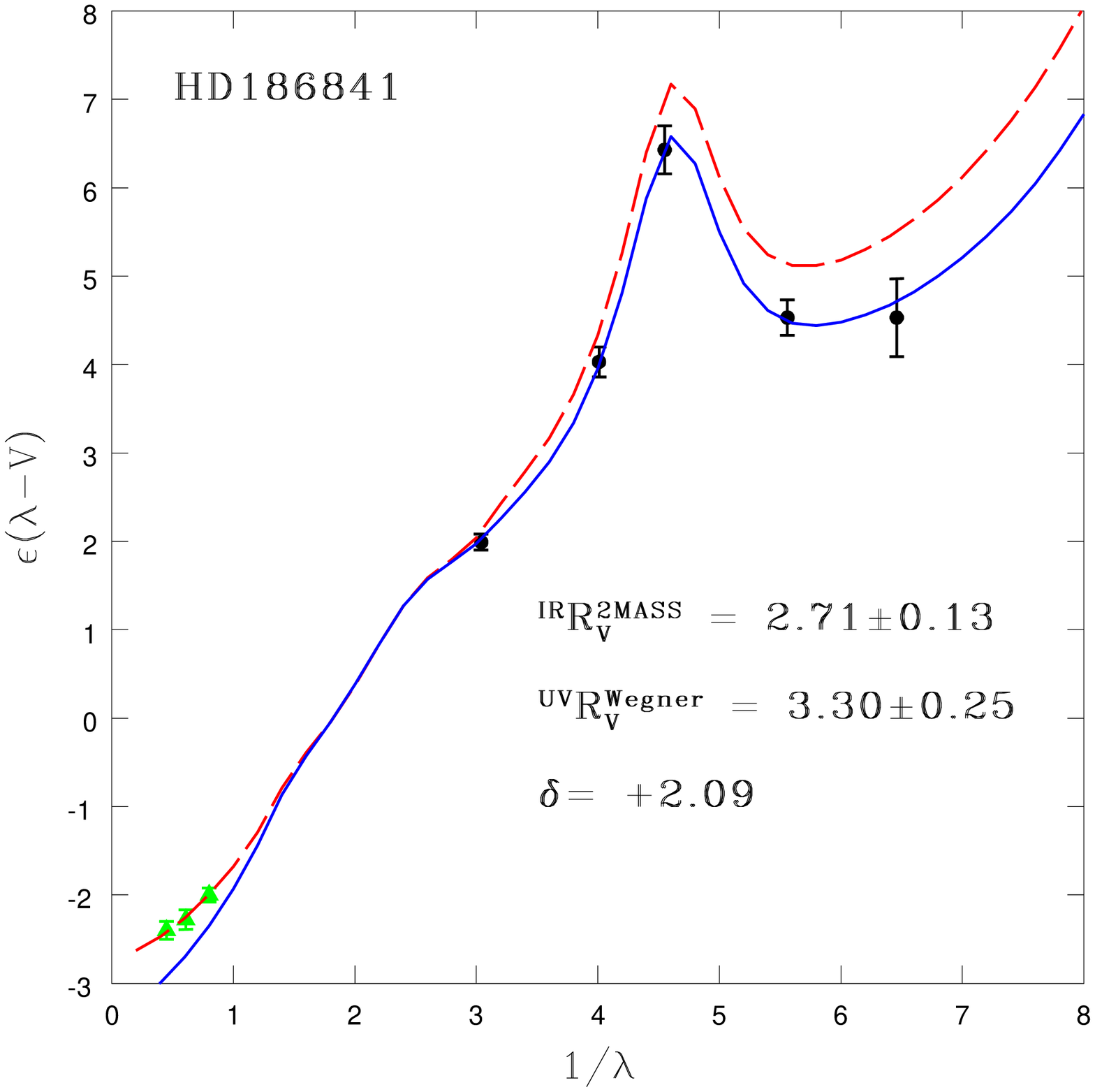}
\includegraphics[width=0.32\textwidth]{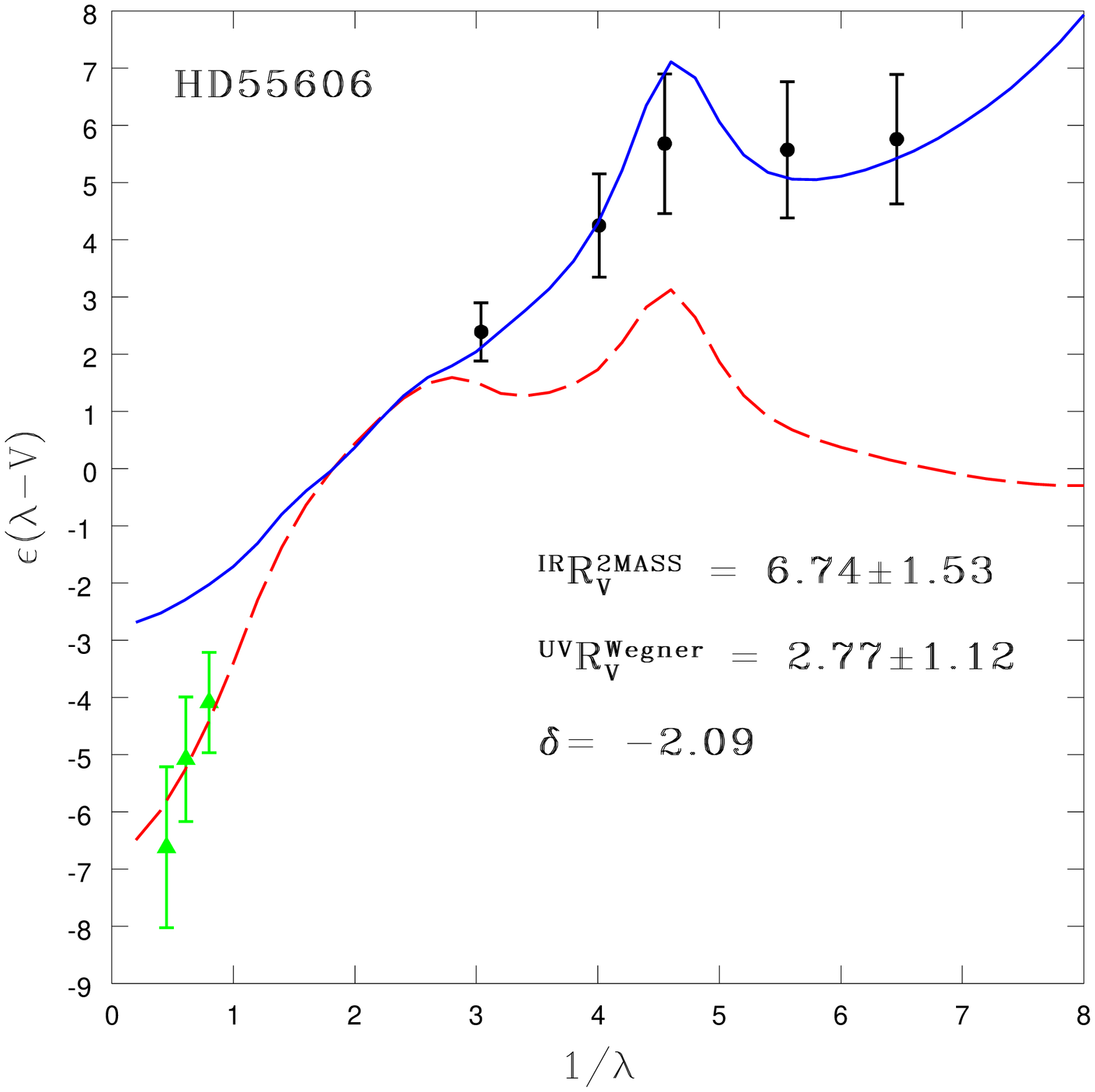}
\includegraphics[width=0.32\textwidth]{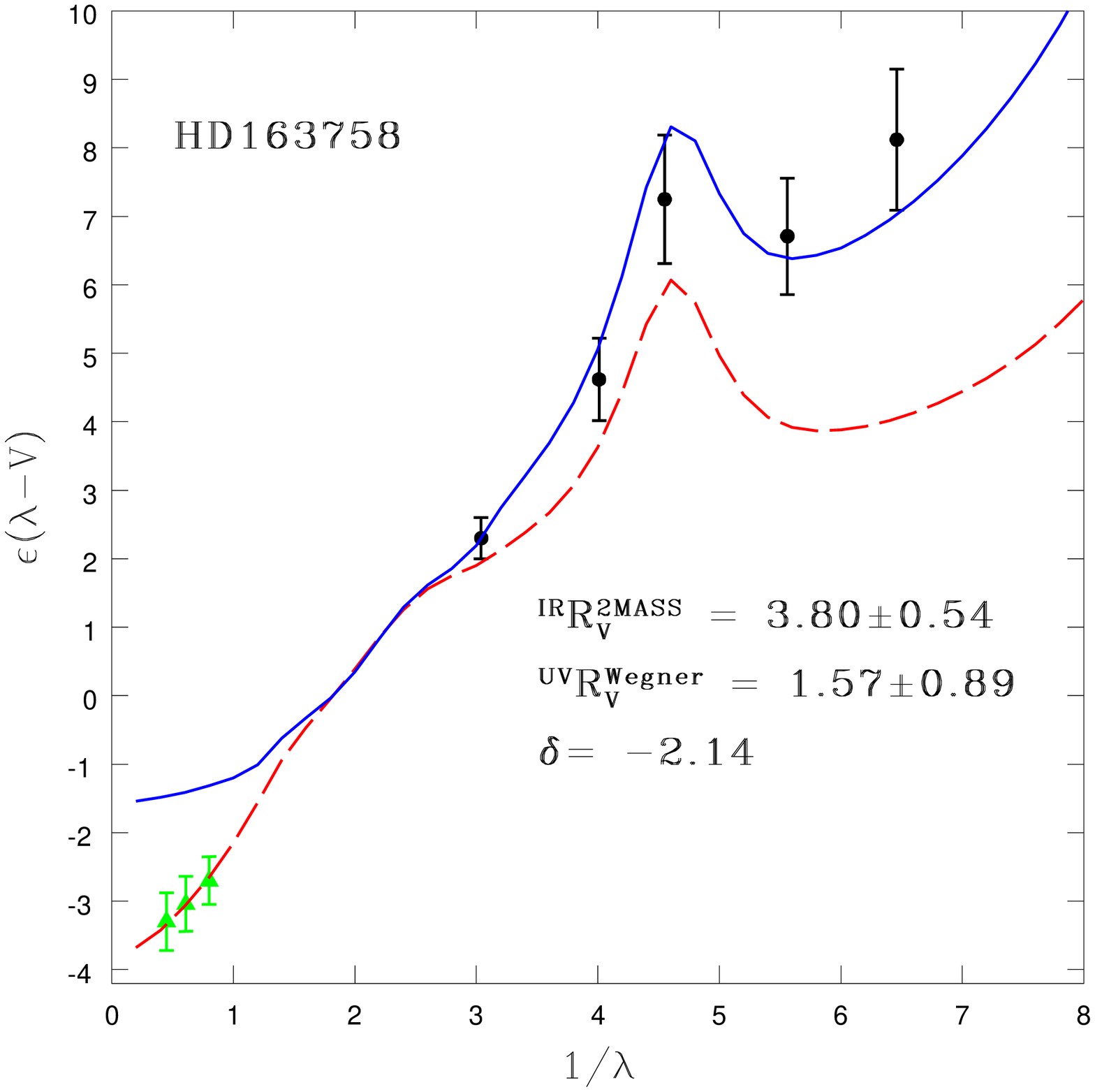}
\includegraphics[width=0.32\textwidth]{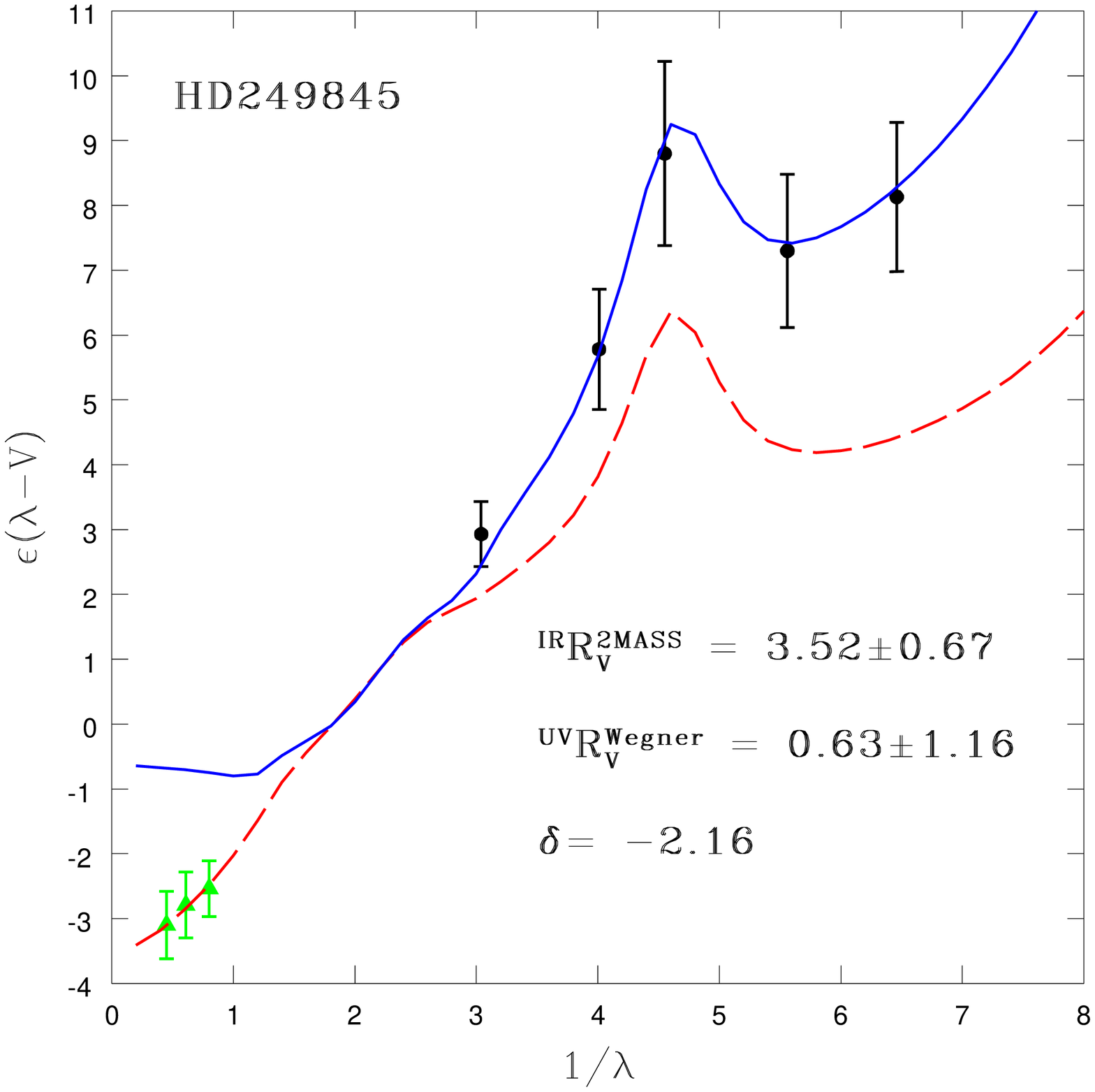}
\includegraphics[width=0.32\textwidth]{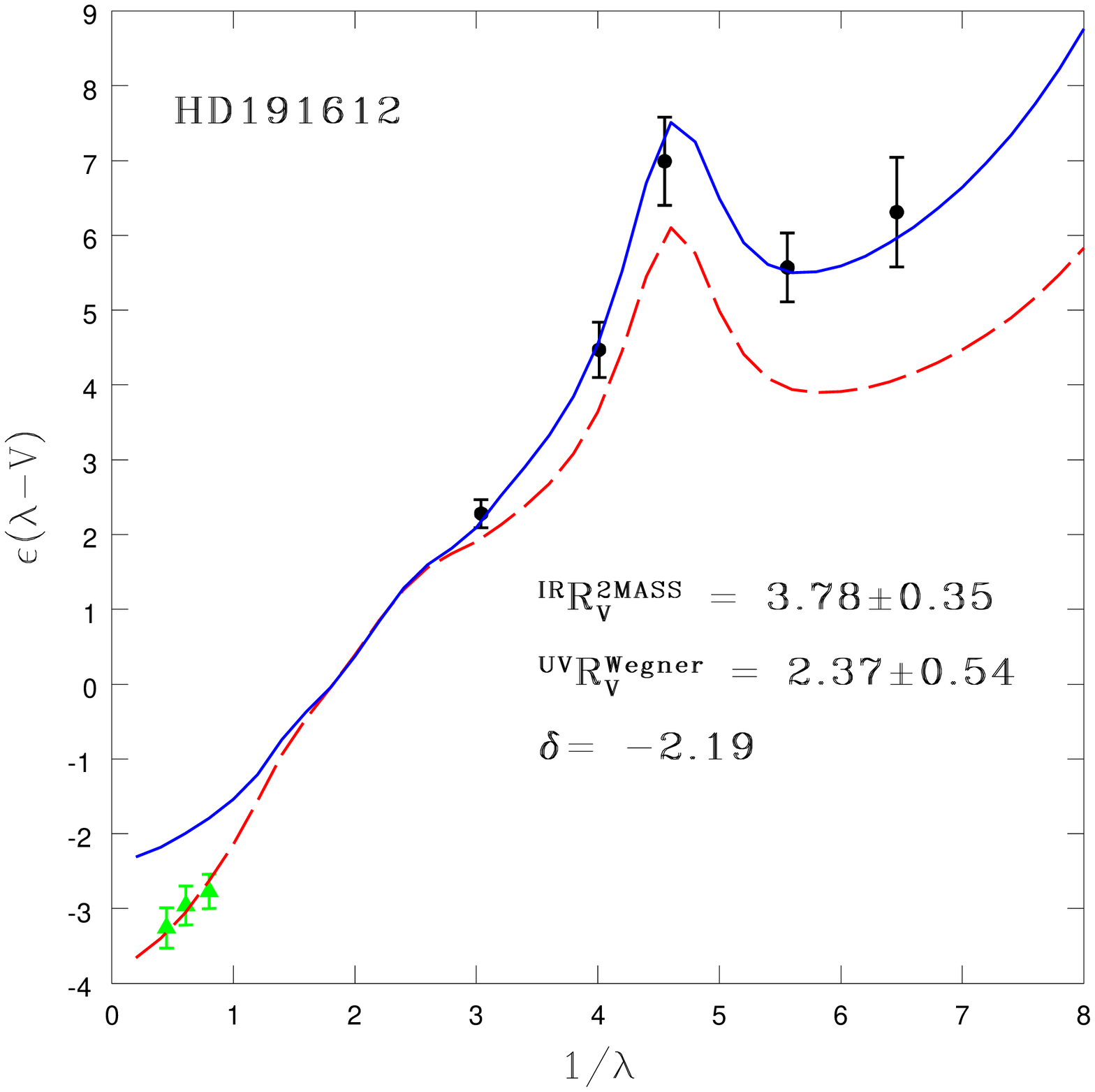}
\includegraphics[width=0.32\textwidth]{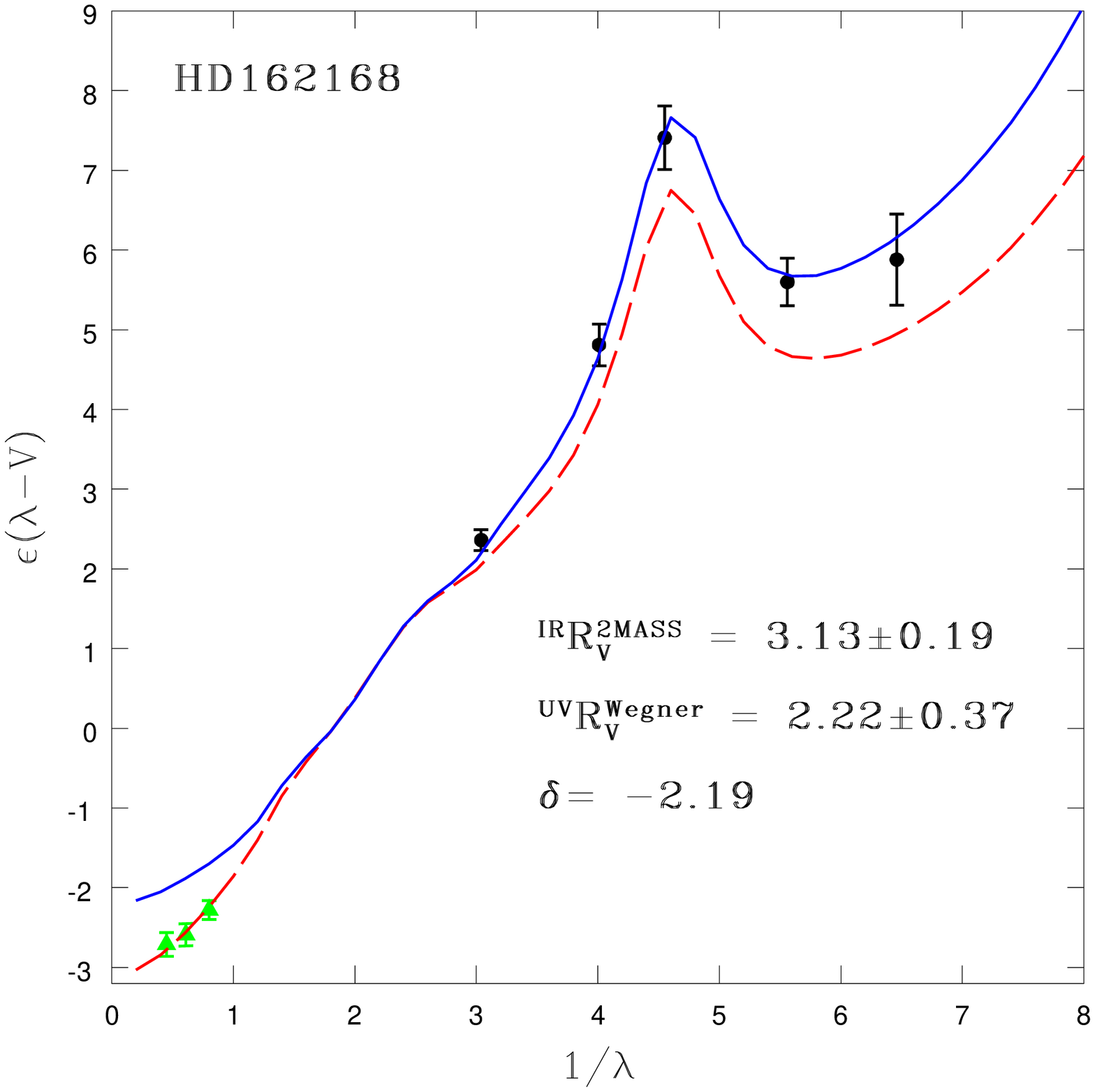}
\includegraphics[width=0.33\textwidth]{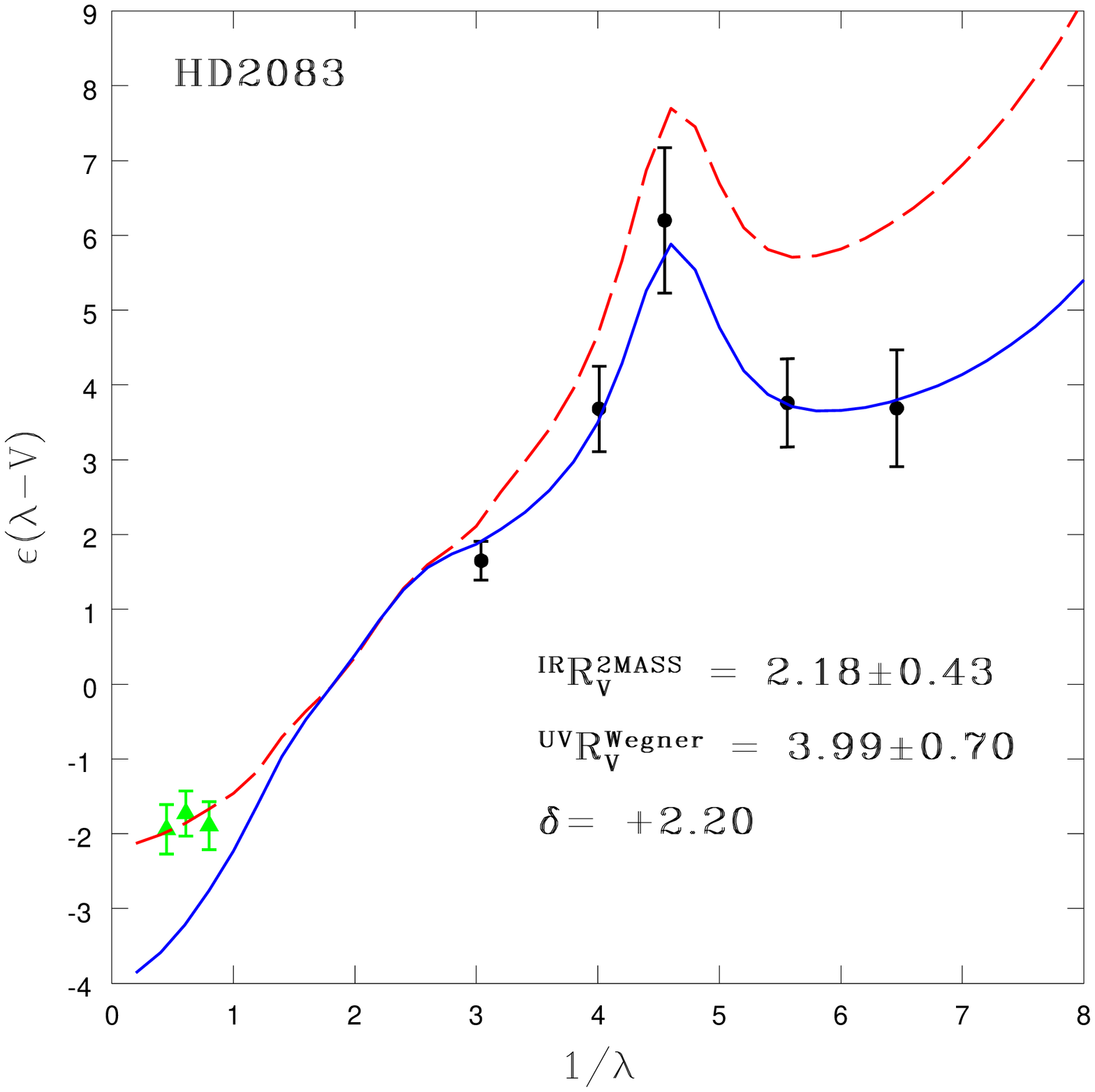}
\includegraphics[width=0.33\textwidth]{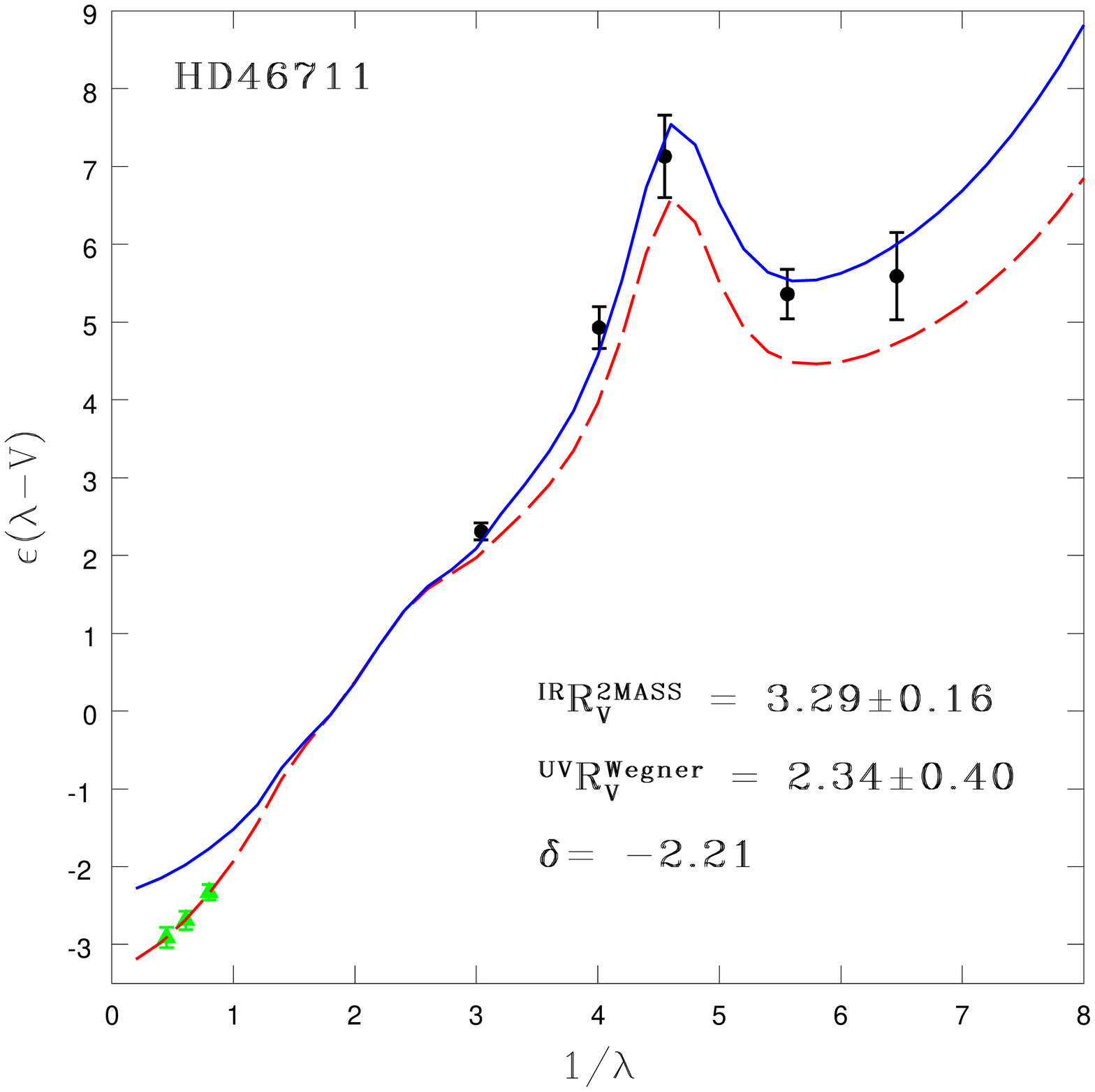}
\end{figure}

\begin{figure}
\includegraphics[width=0.32\textwidth]{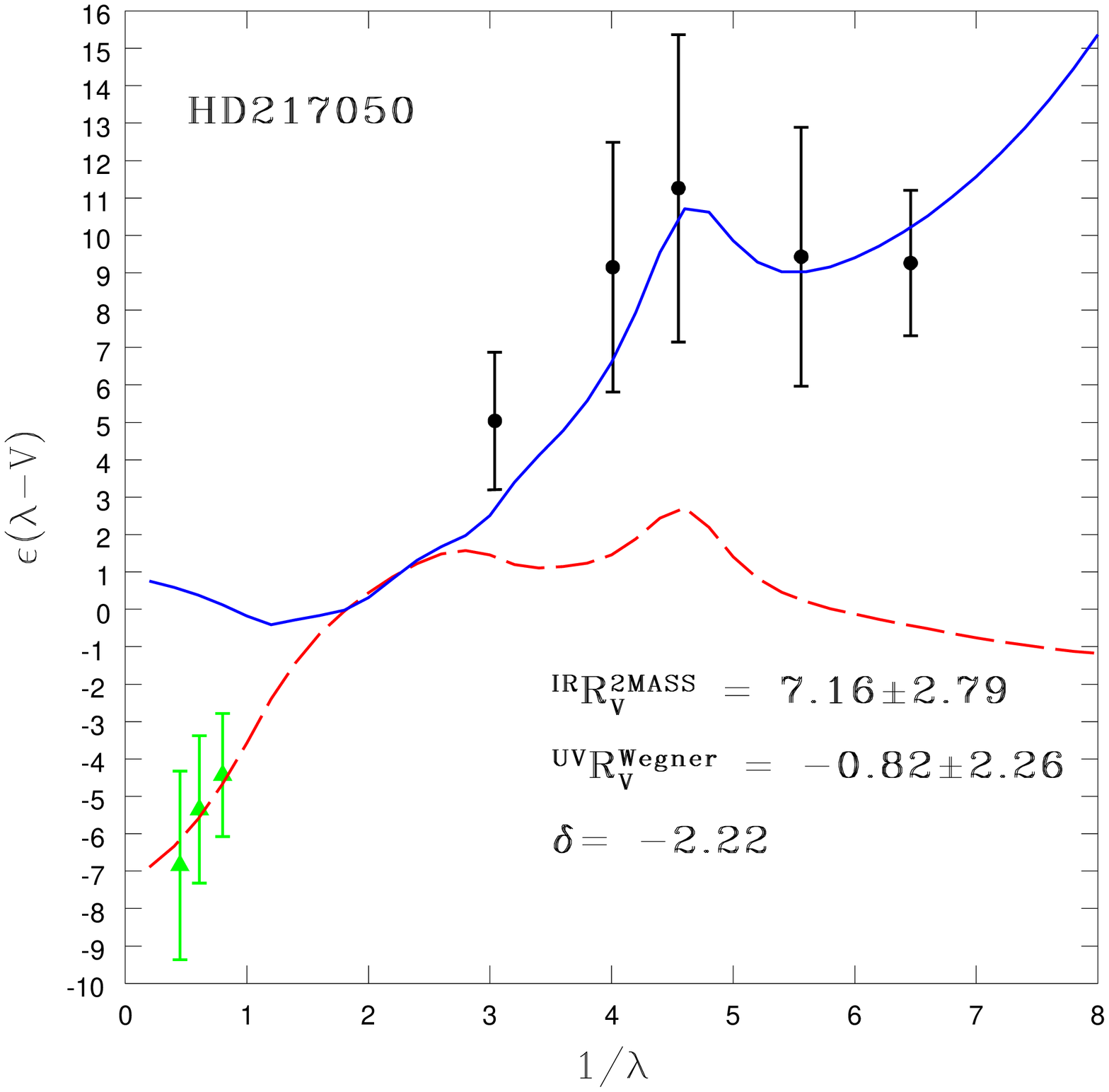}
\includegraphics[width=0.32\textwidth]{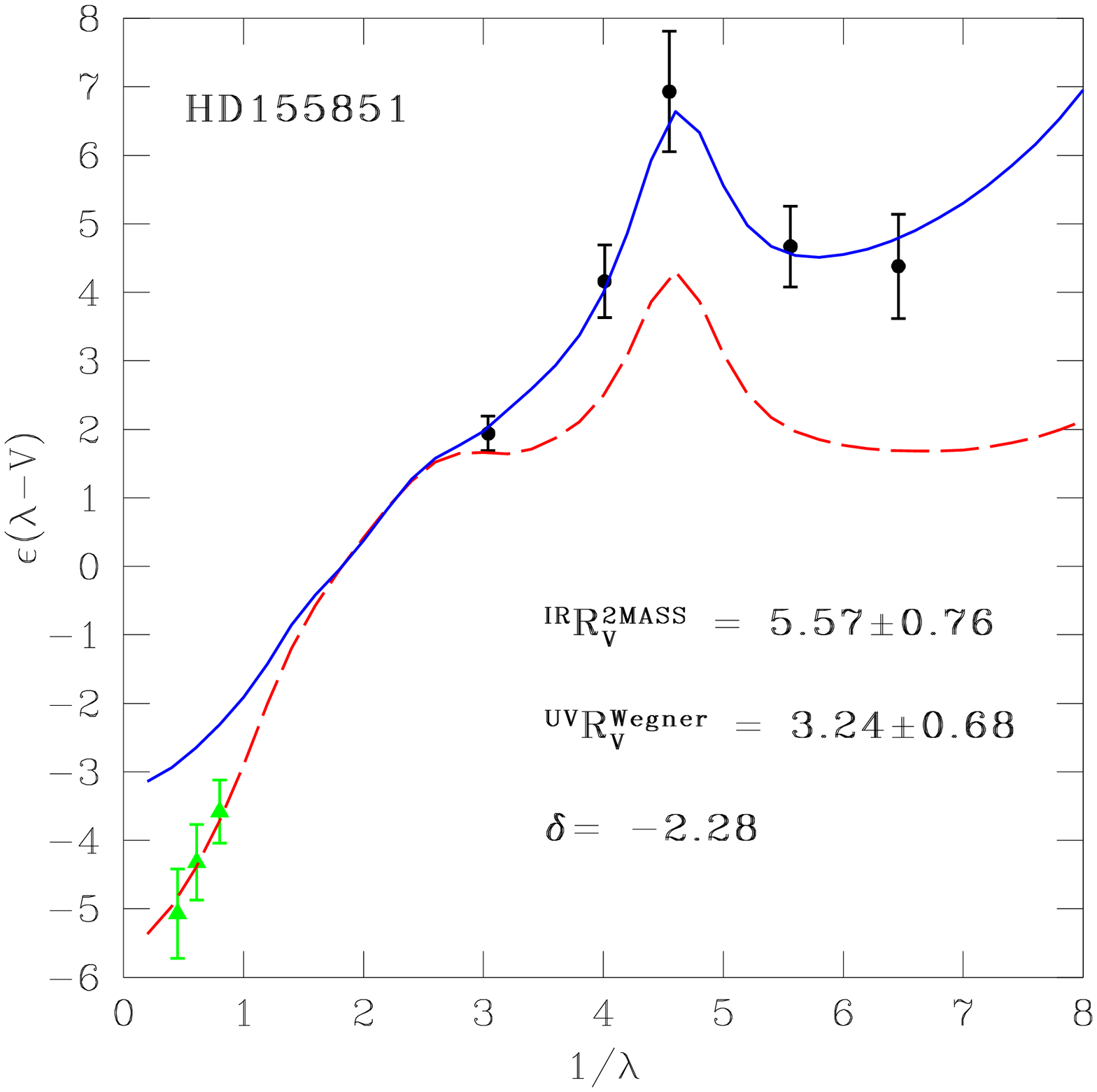}
\includegraphics[width=0.32\textwidth]{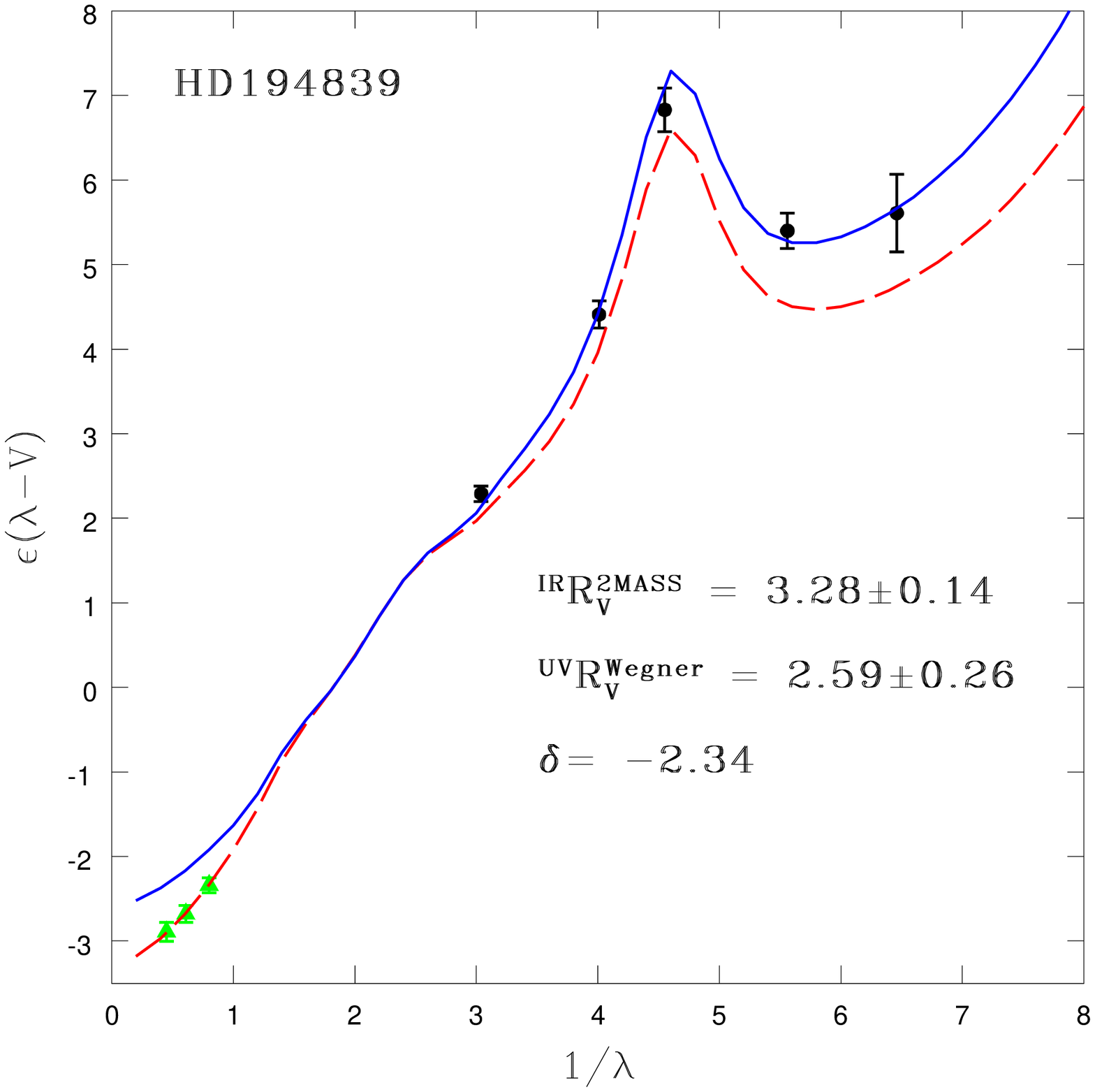}
\includegraphics[width=0.32\textwidth]{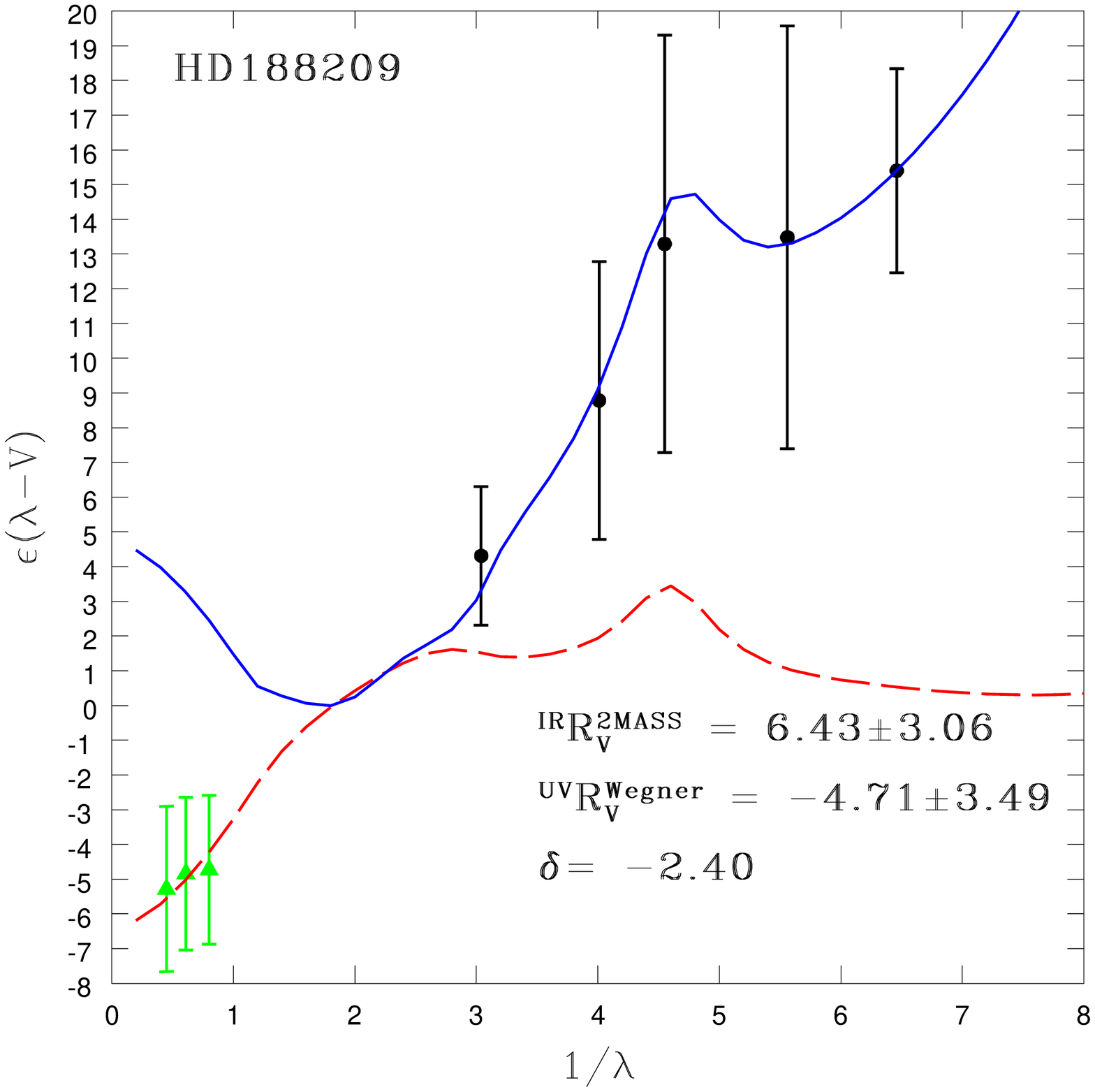}
\includegraphics[width=0.32\textwidth]{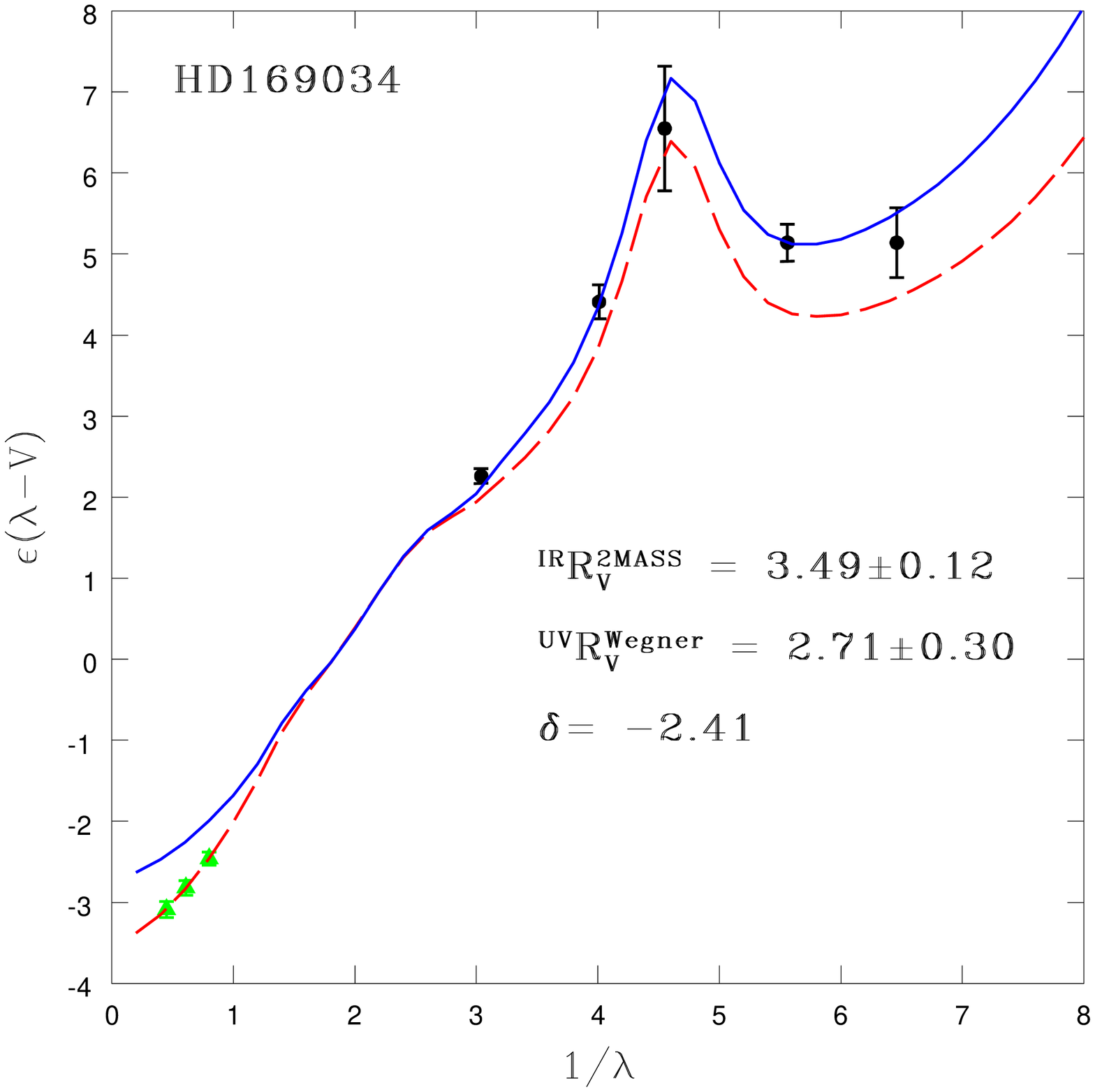}
\includegraphics[width=0.32\textwidth]{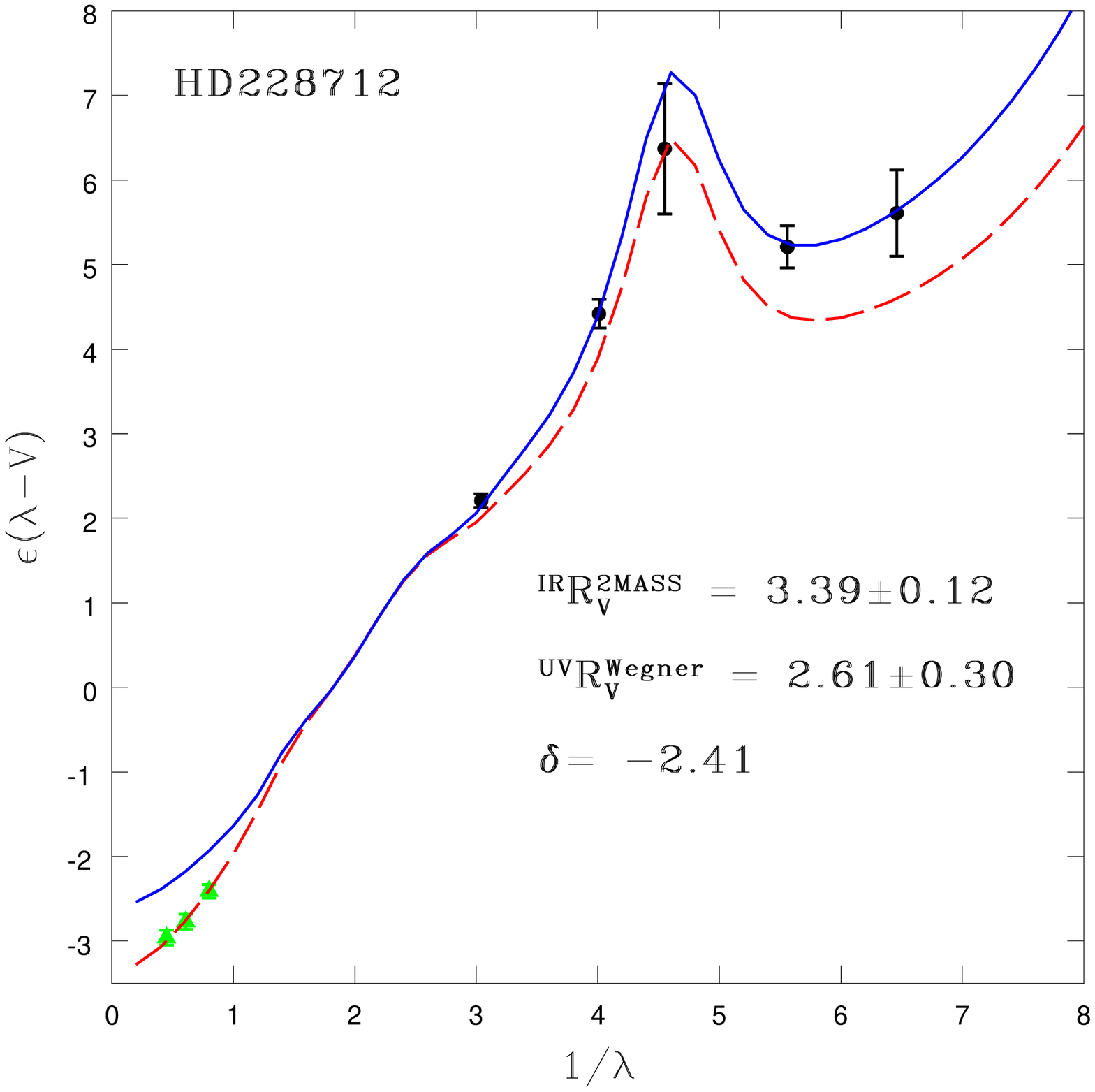}
\includegraphics[width=0.32\textwidth]{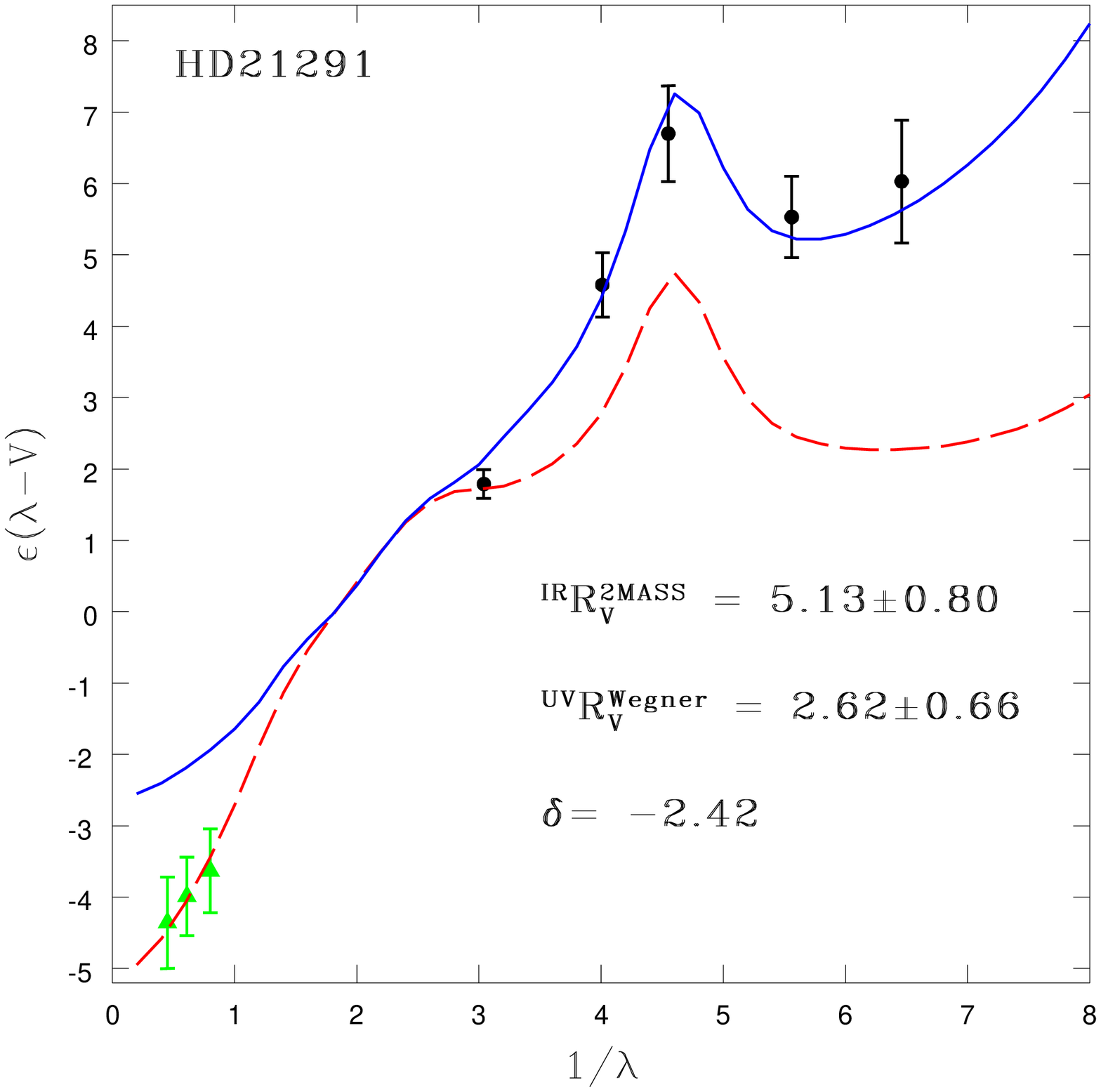}
\includegraphics[width=0.32\textwidth]{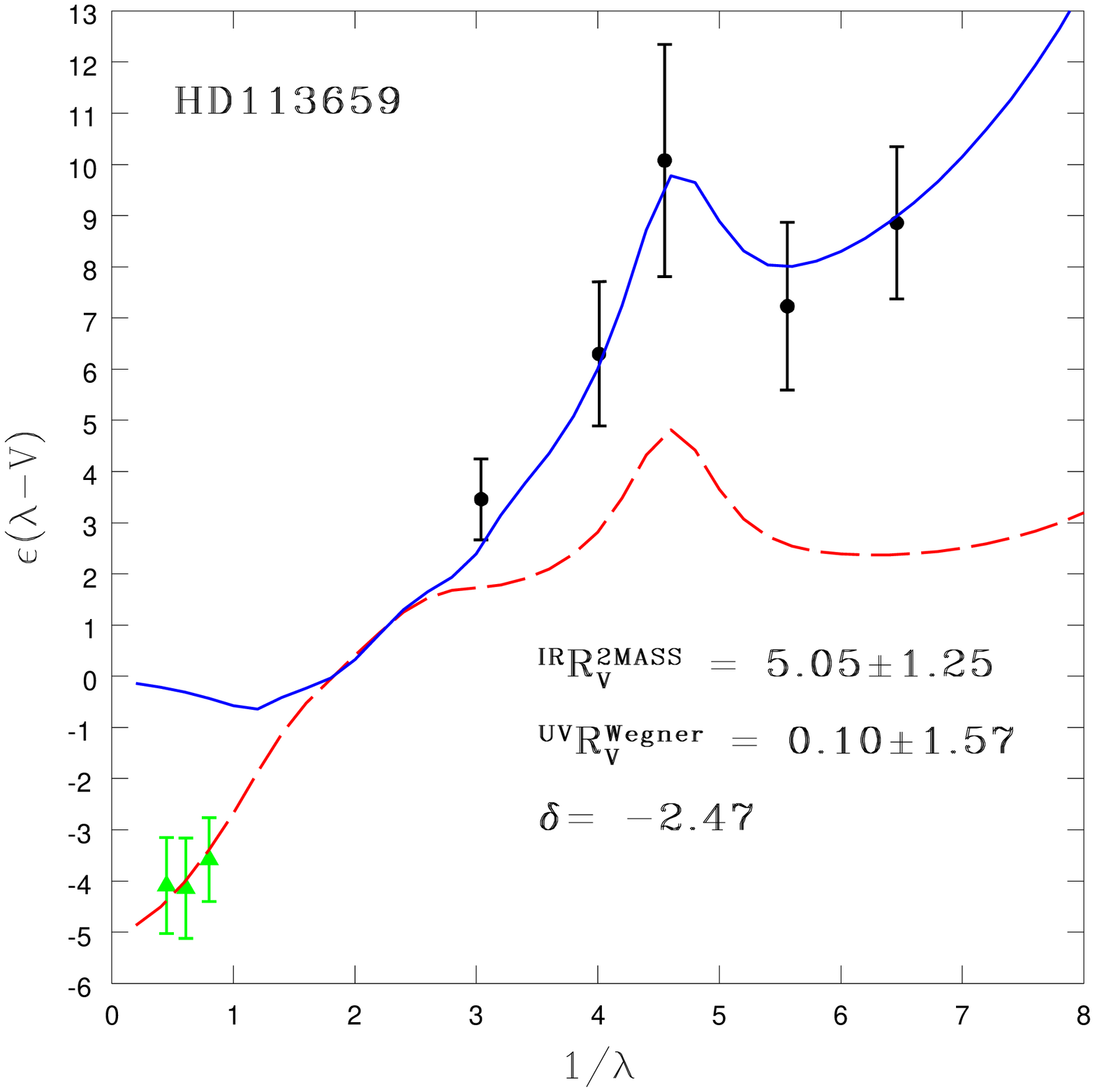}
\includegraphics[width=0.32\textwidth]{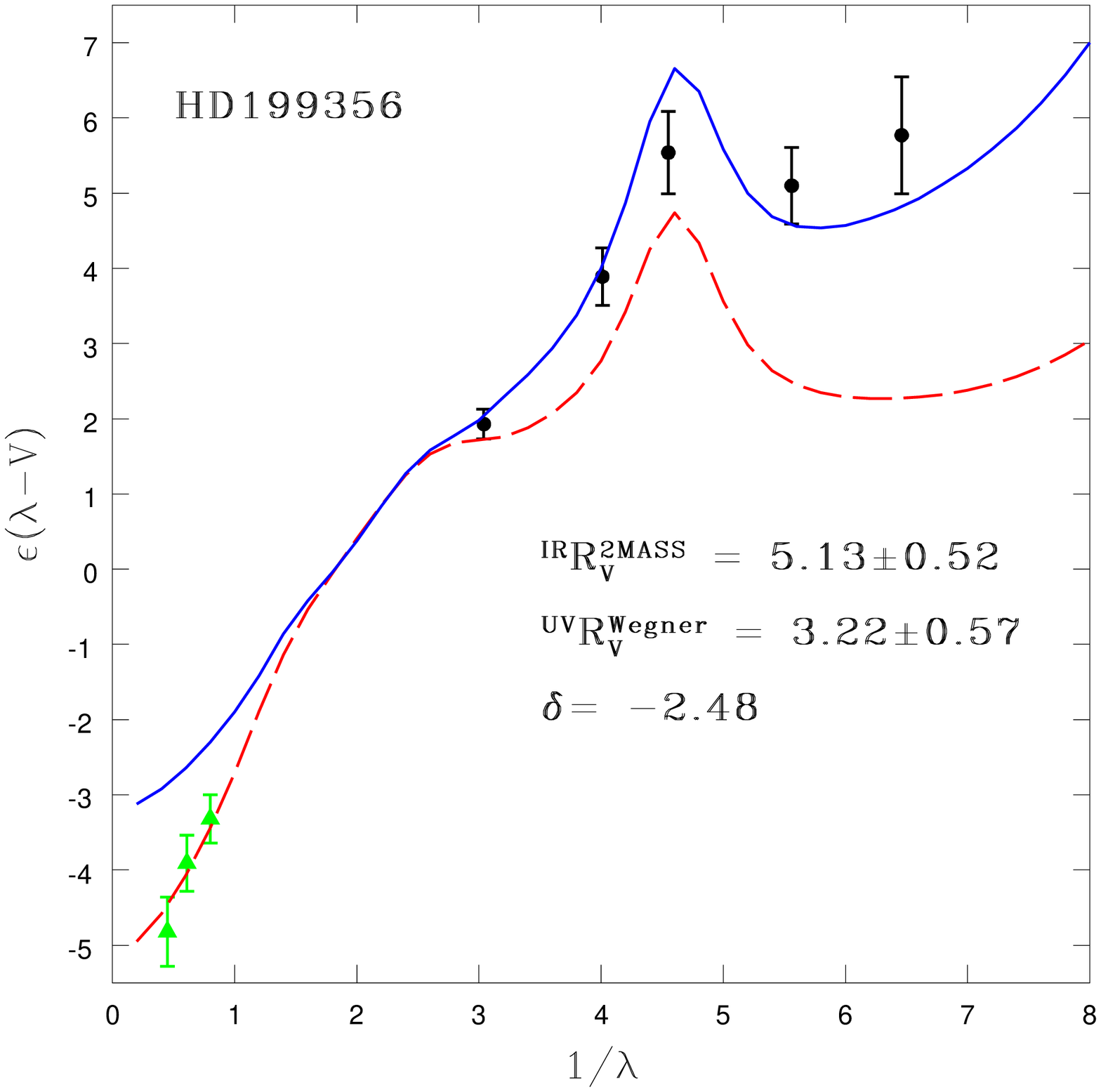}
\includegraphics[width=0.32\textwidth]{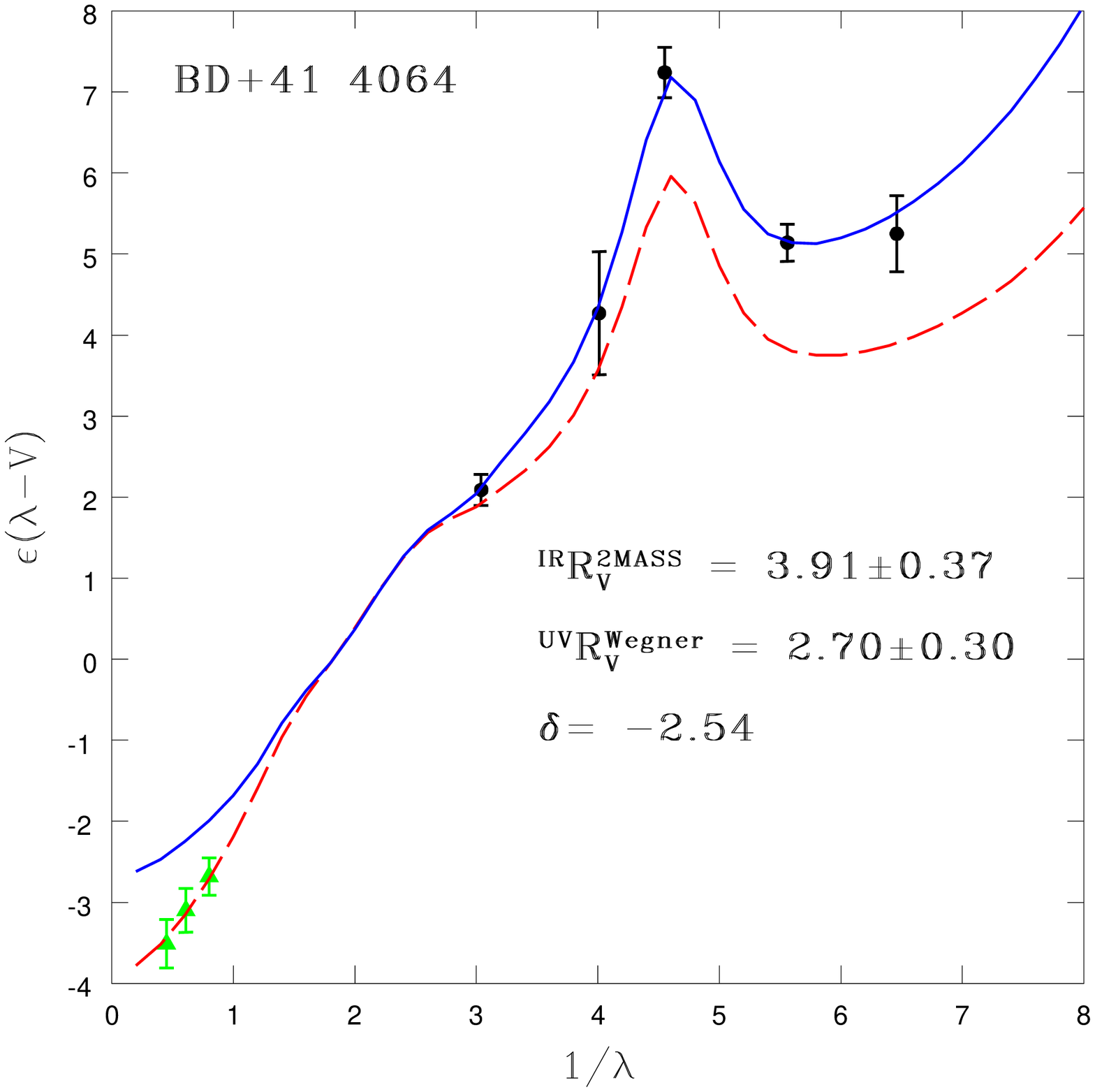}
\includegraphics[width=0.33\textwidth]{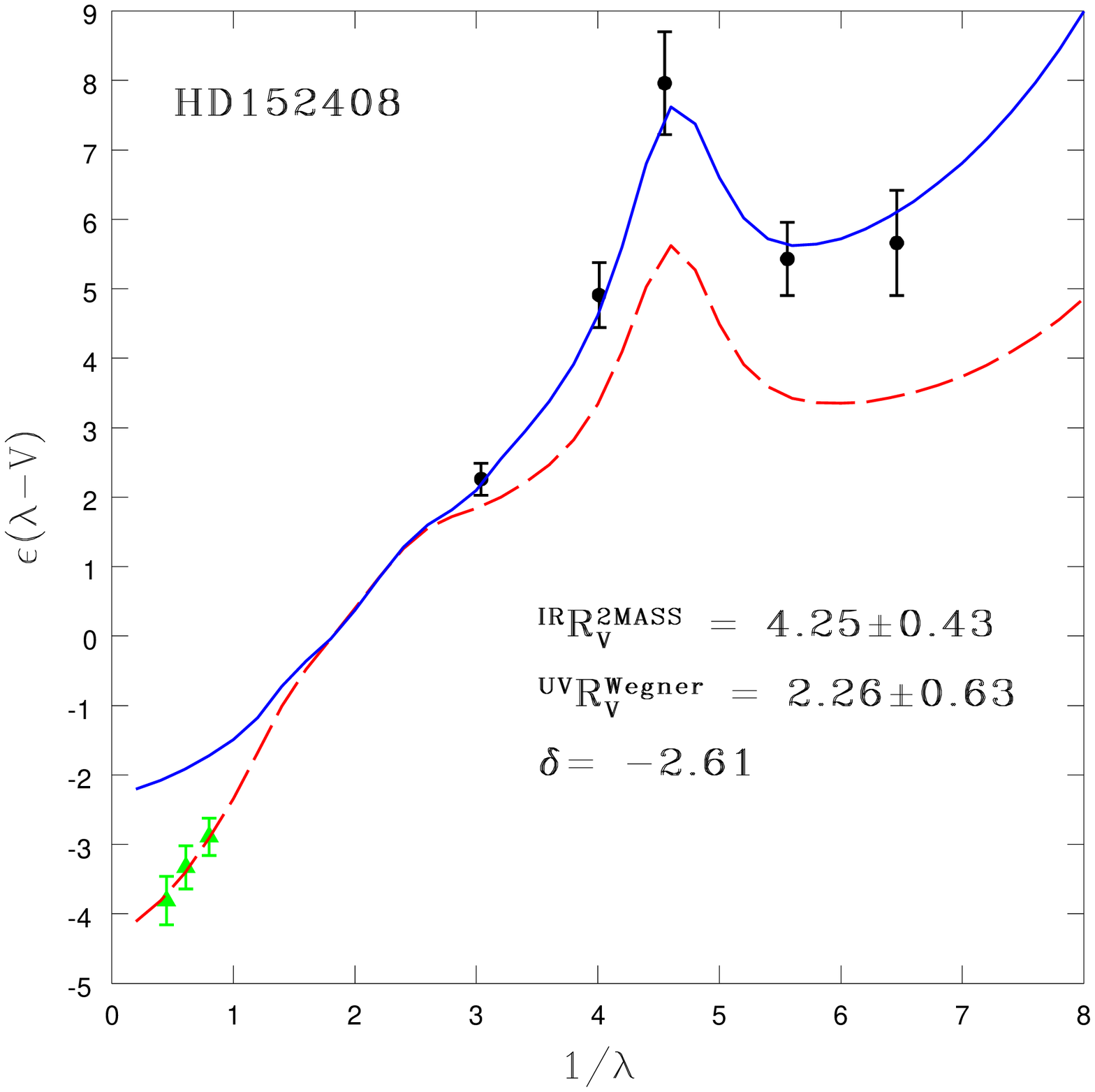}
\includegraphics[width=0.33\textwidth]{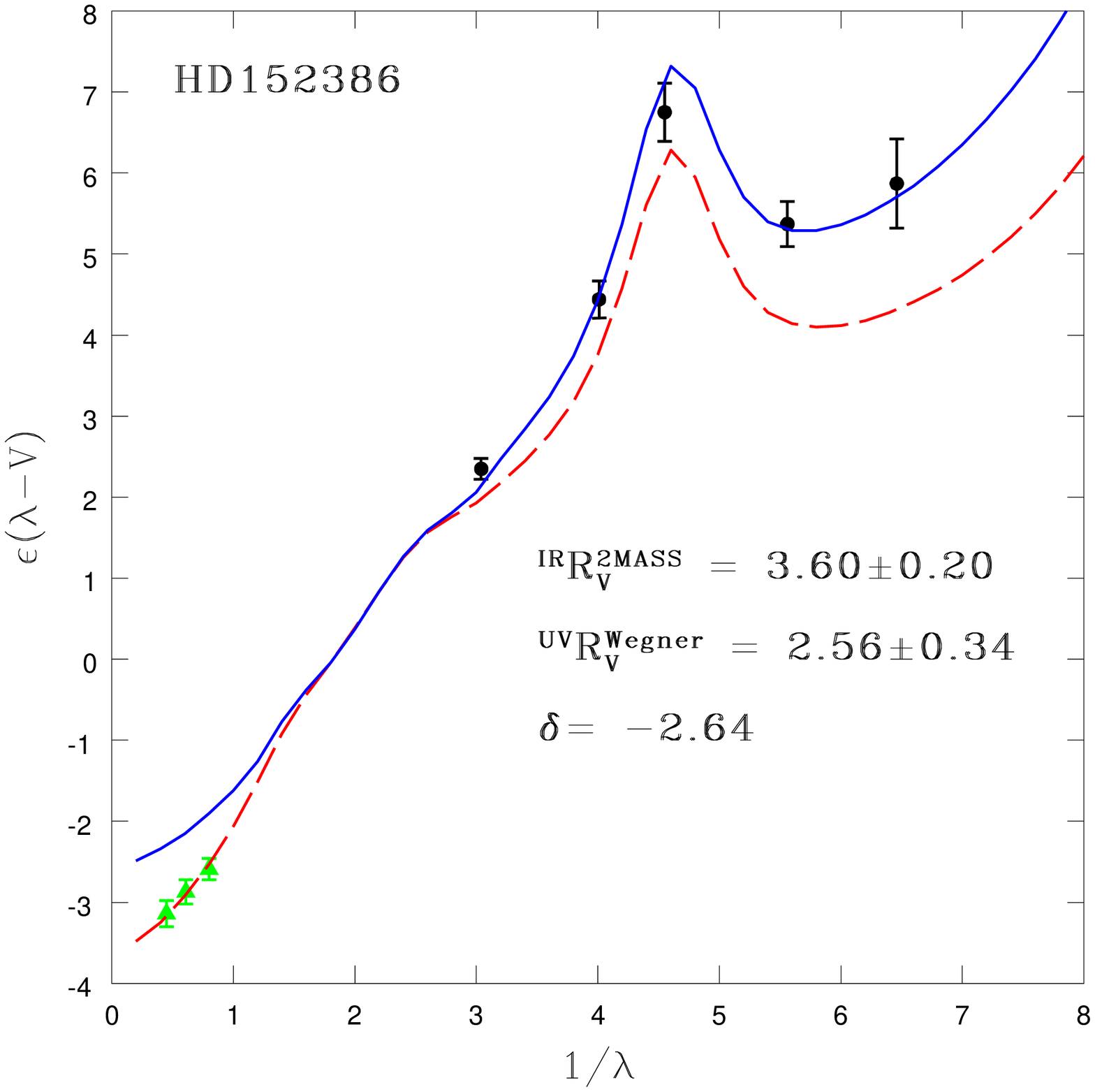}
\end{figure}

\begin{figure}
\includegraphics[width=0.32\textwidth]{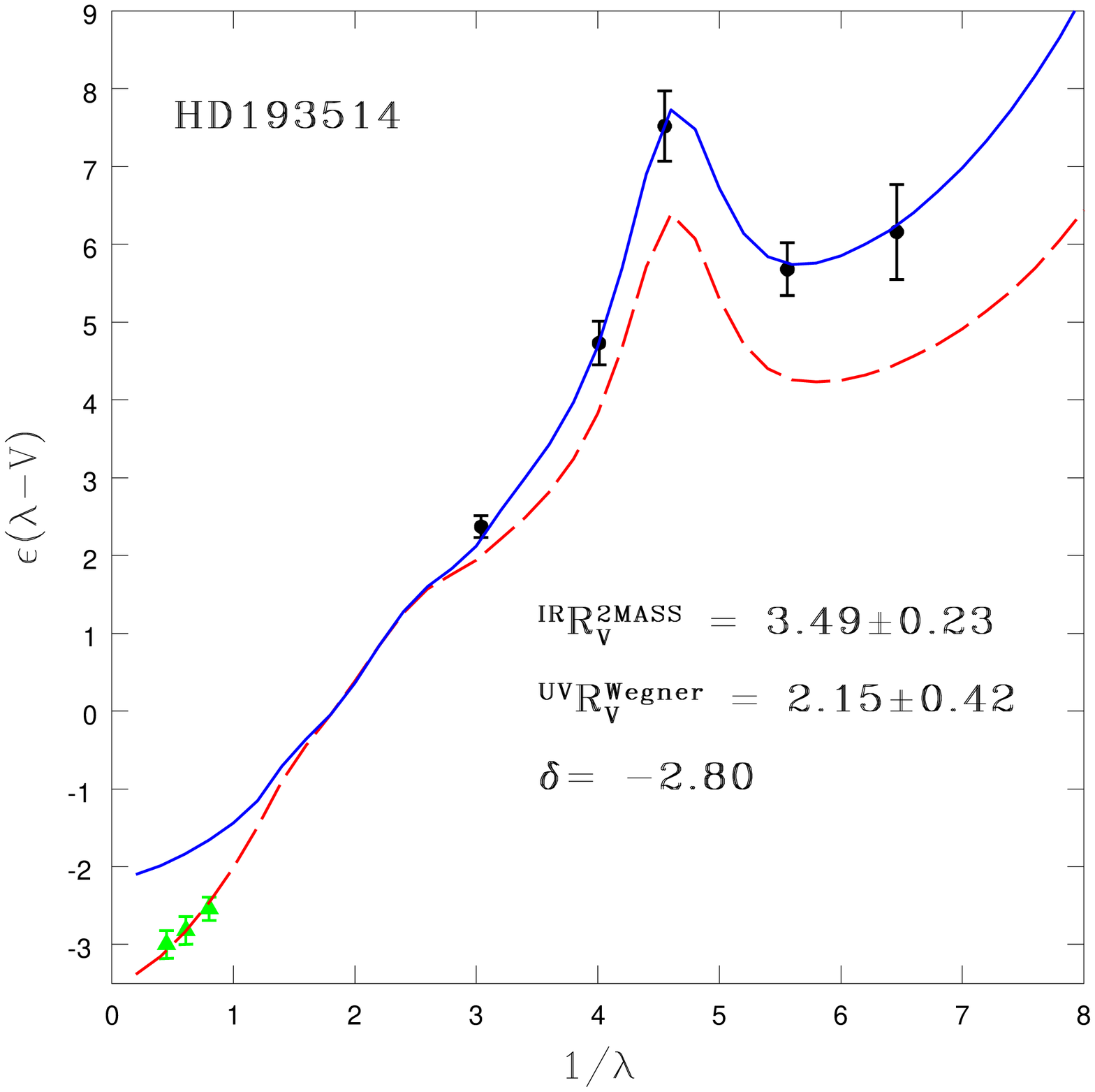}
\includegraphics[width=0.32\textwidth]{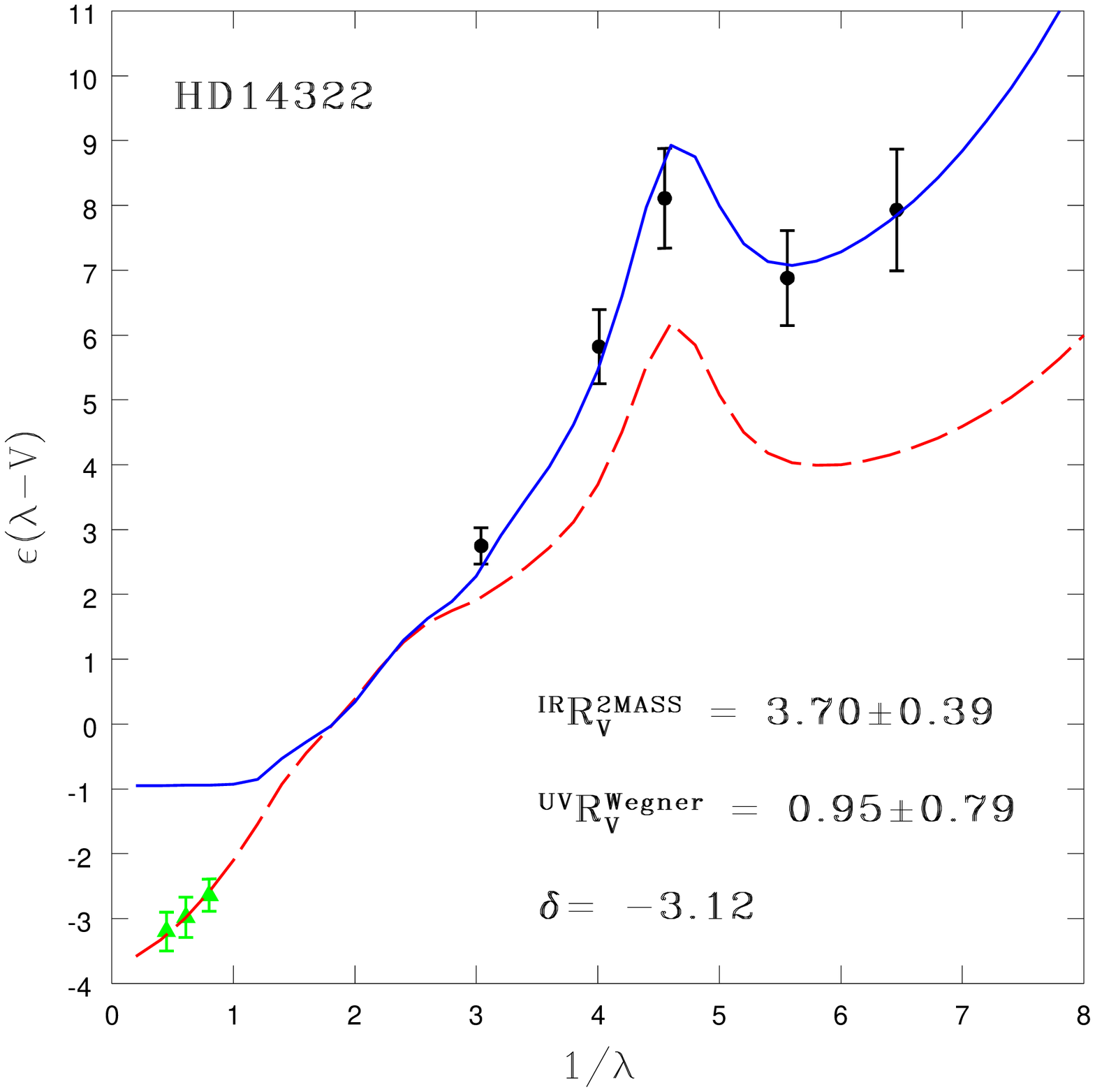}
\includegraphics[width=0.32\textwidth]{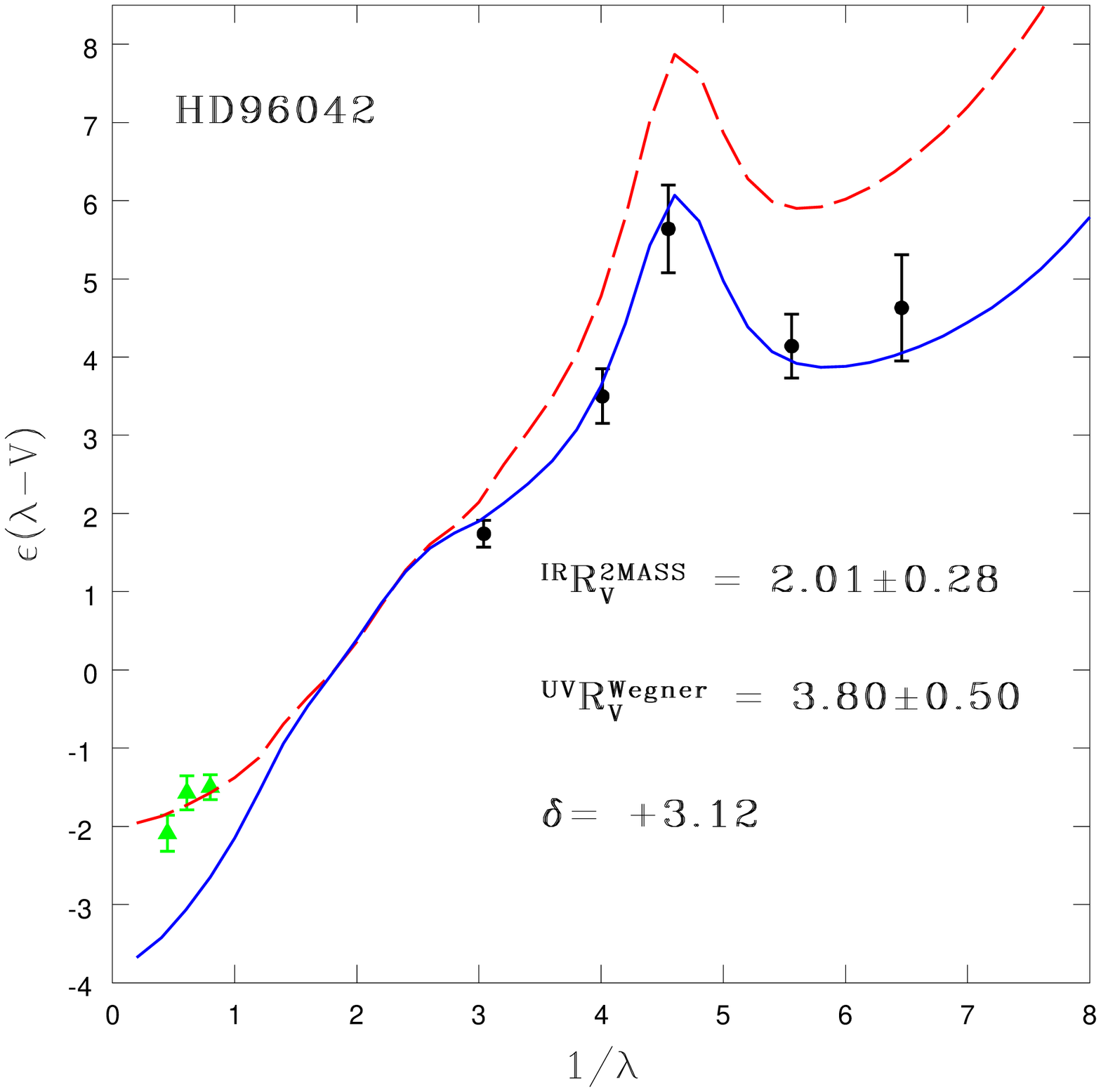}
\includegraphics[width=0.32\textwidth]{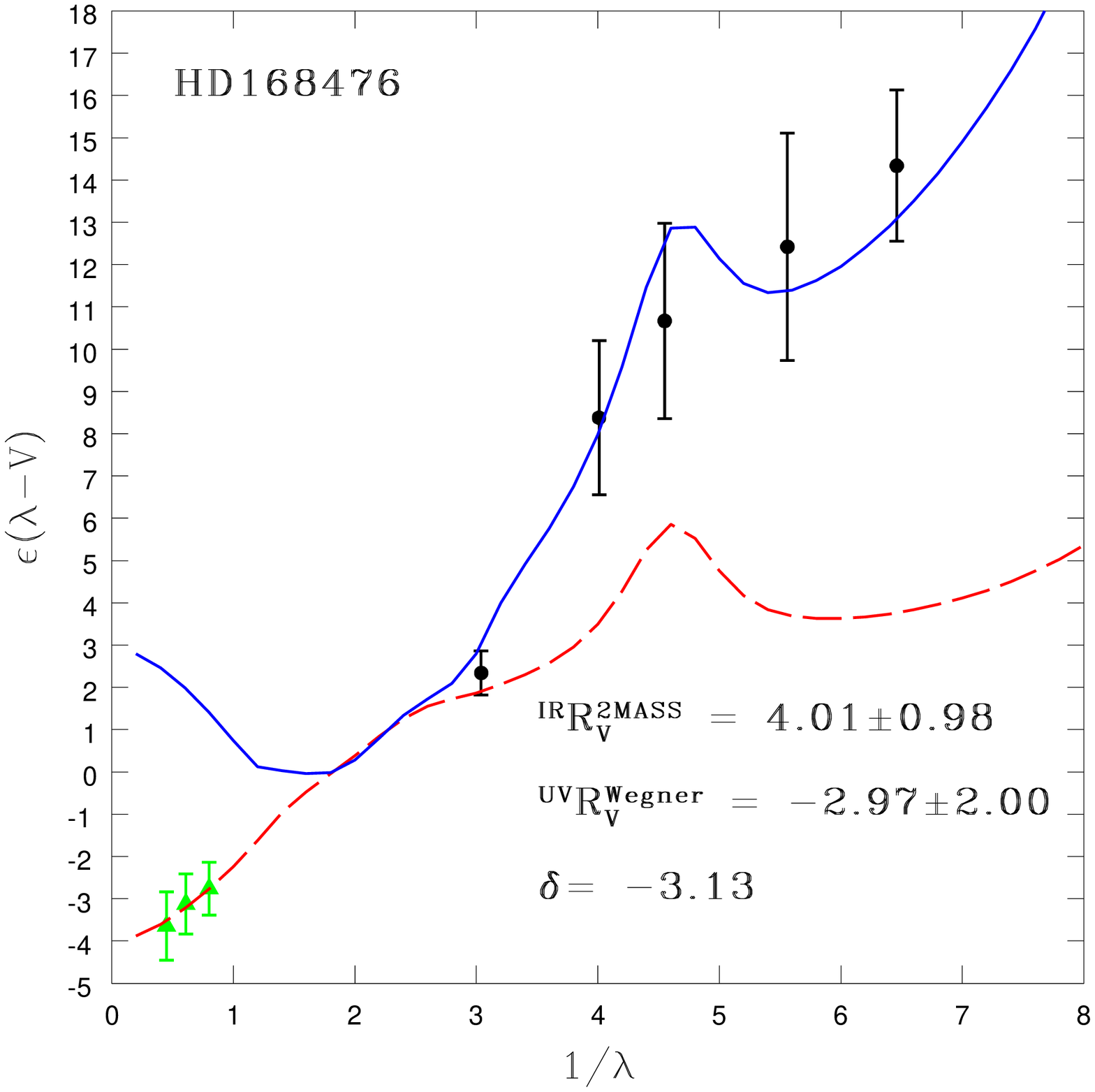}
\includegraphics[width=0.32\textwidth]{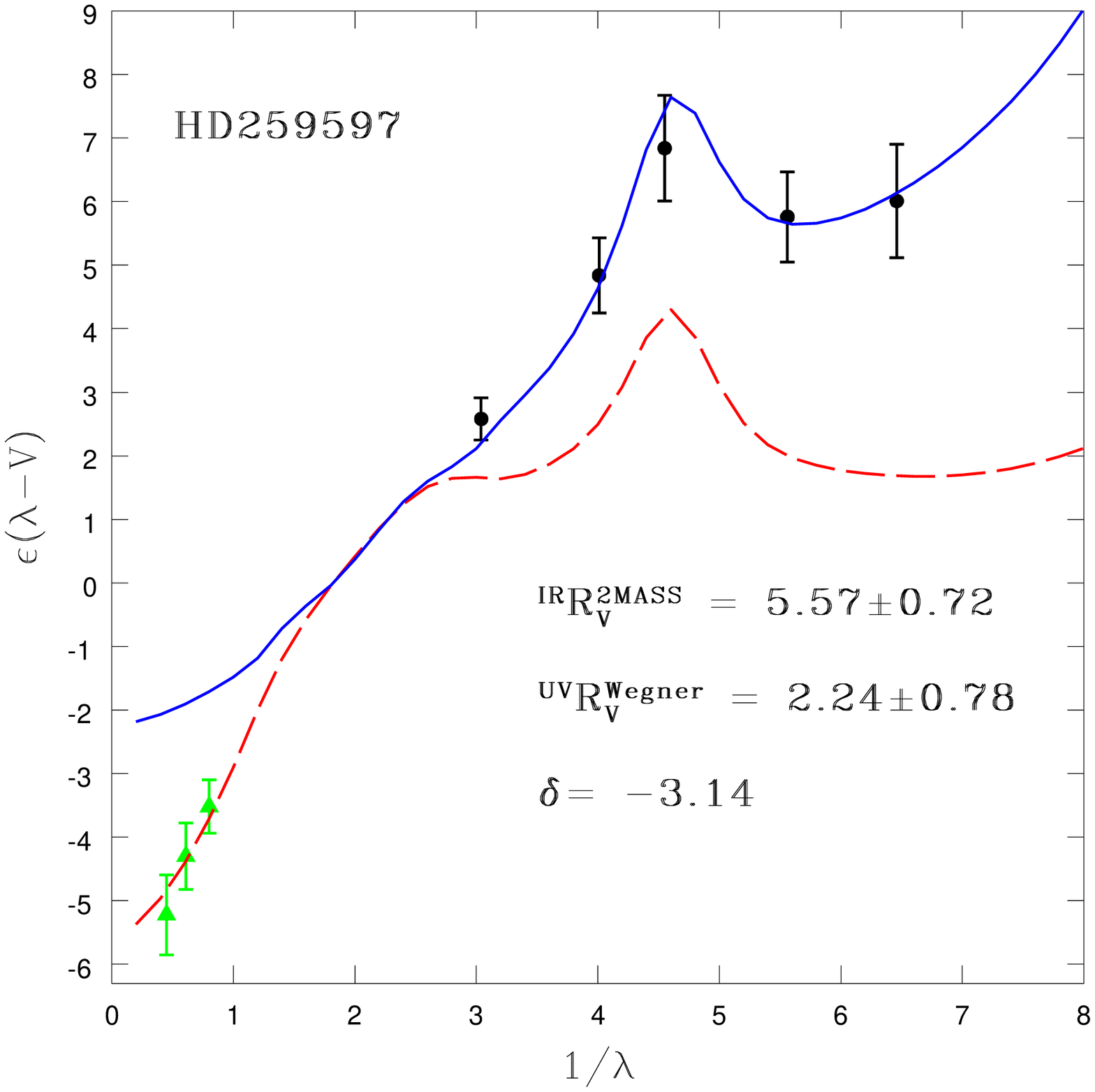}
\includegraphics[width=0.32\textwidth]{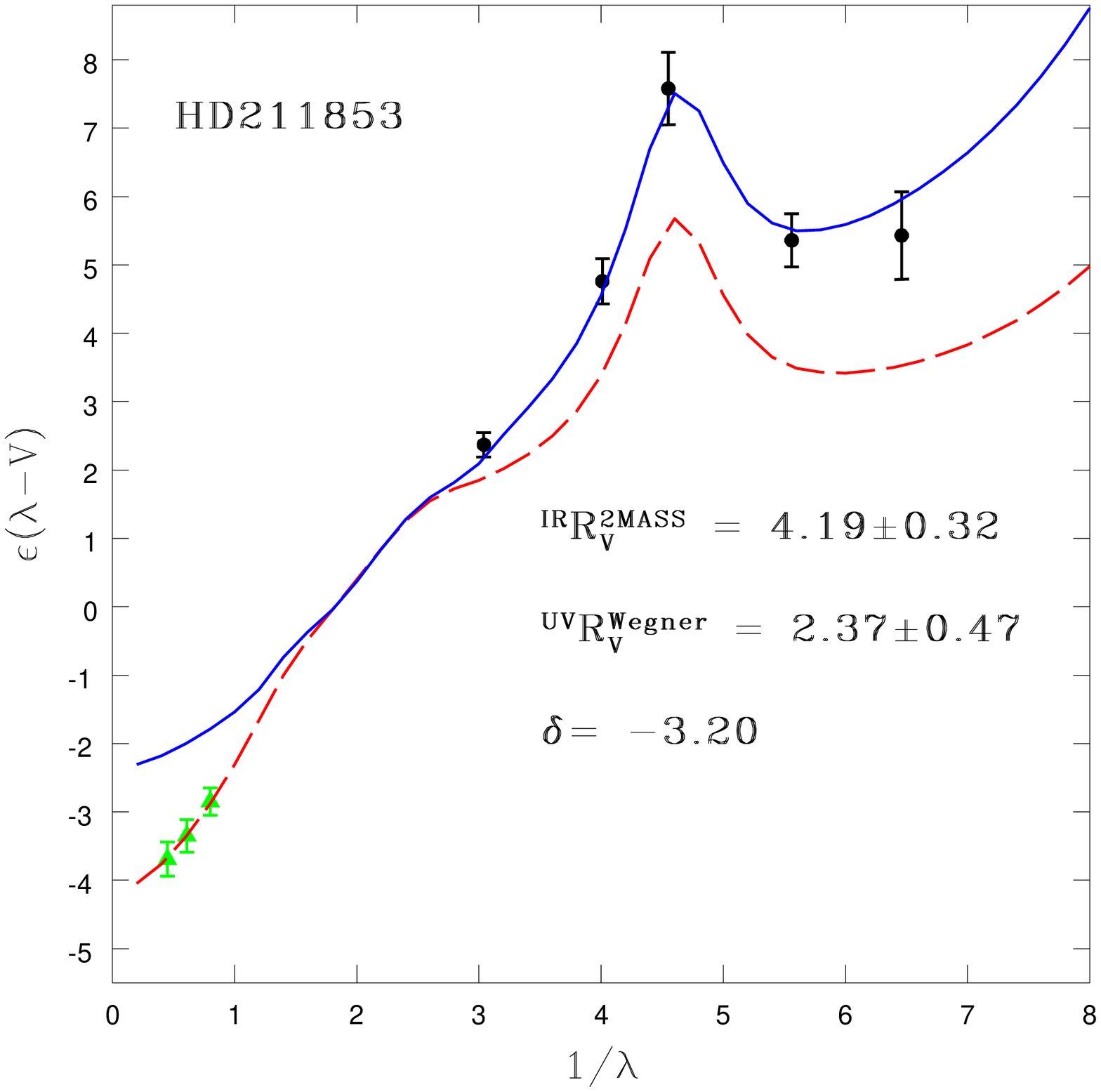}
\includegraphics[width=0.32\textwidth]{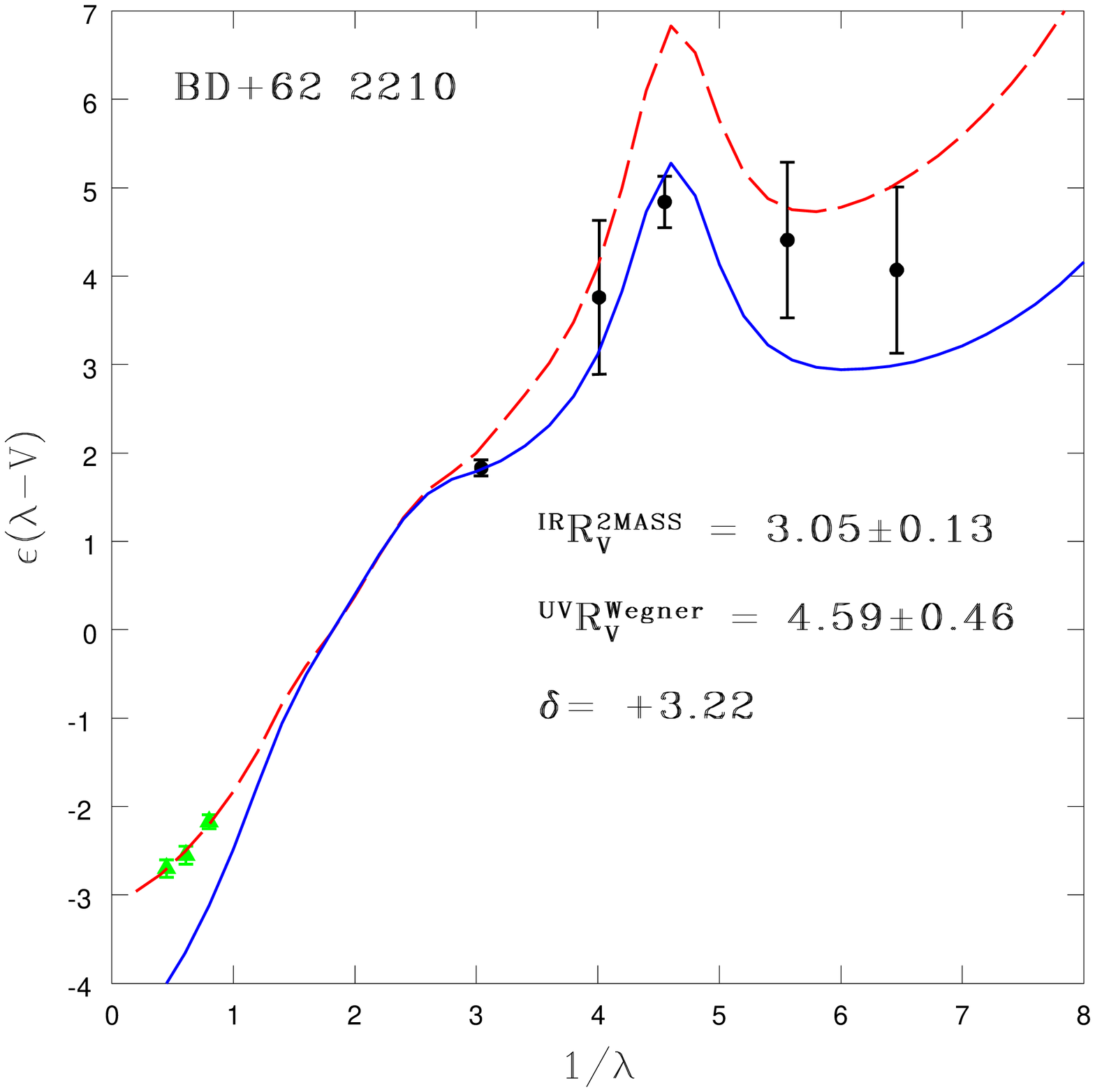}
\includegraphics[width=0.33\textwidth]{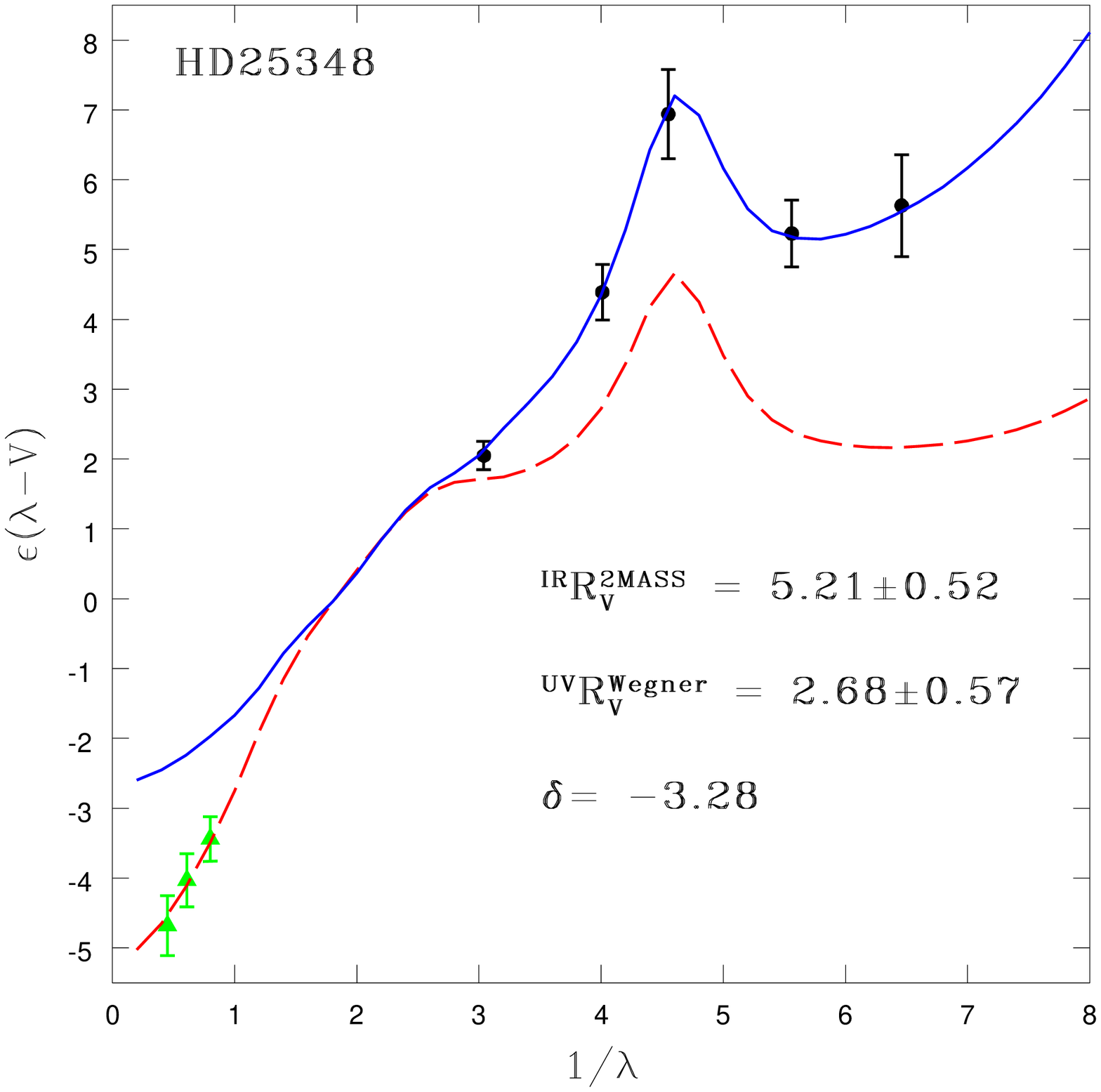}
\includegraphics[width=0.33\textwidth]{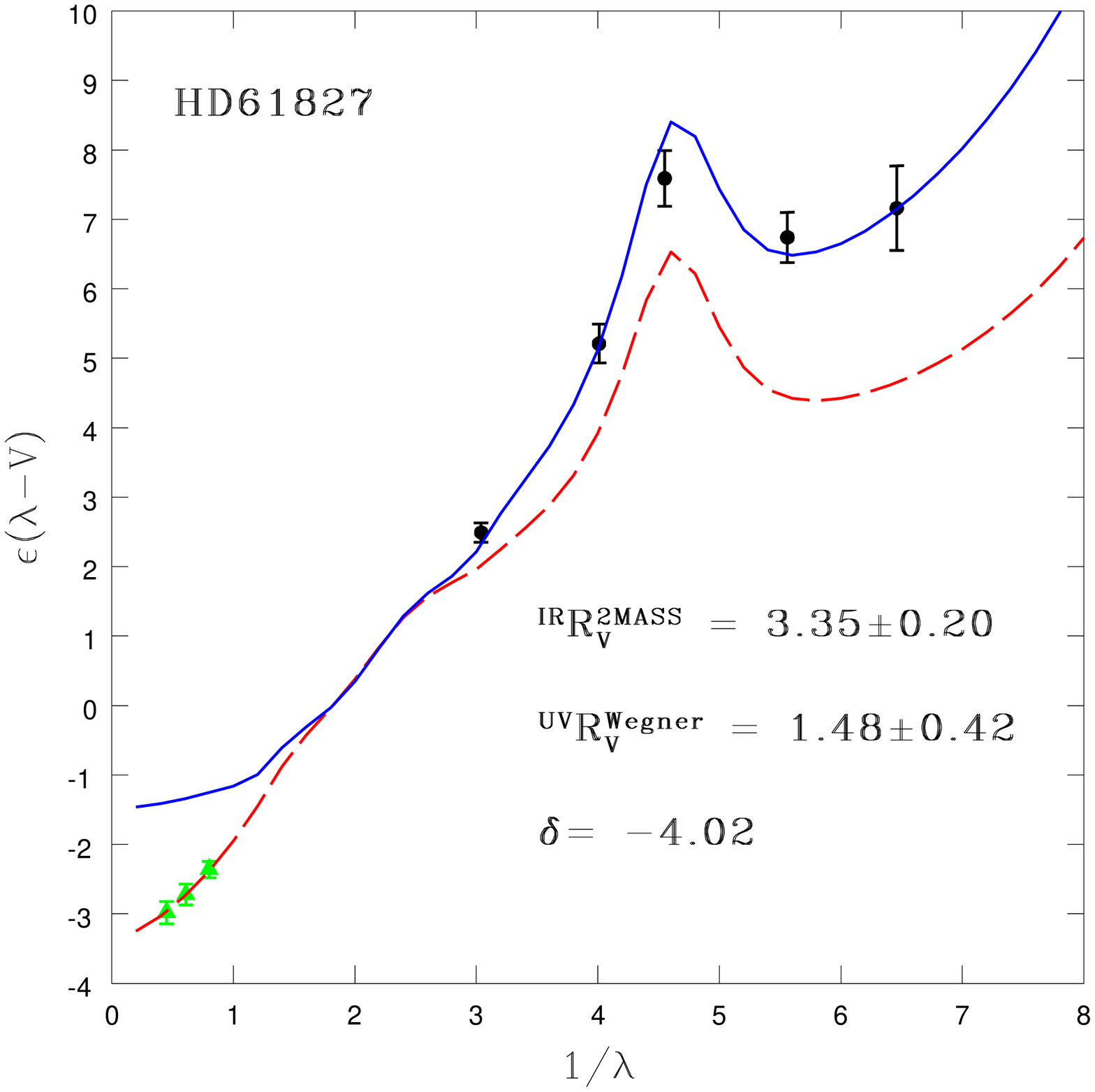}
\caption{Extinction curves for stars listed in Table \ref{table3}. The
dashed line (red) is obtained using CCM law with 
$^{\rm{IR}}\!R_V^{\rm{2MASS}}$; the continuous line (blue)
represents CCM law with $^{\rm{UV}}\!R_V^{\rm{Wegner}}$;
the circular points (black) are the observed UV extinction data taken
from \citet{b16} and the triangular points (green) are 2MASS-based 
extinction data in the $J,H,K_S$ bands.}
\label{figure4}
\end{figure}

\begin{figure}
\begin{center}
\includegraphics[width=80mm]{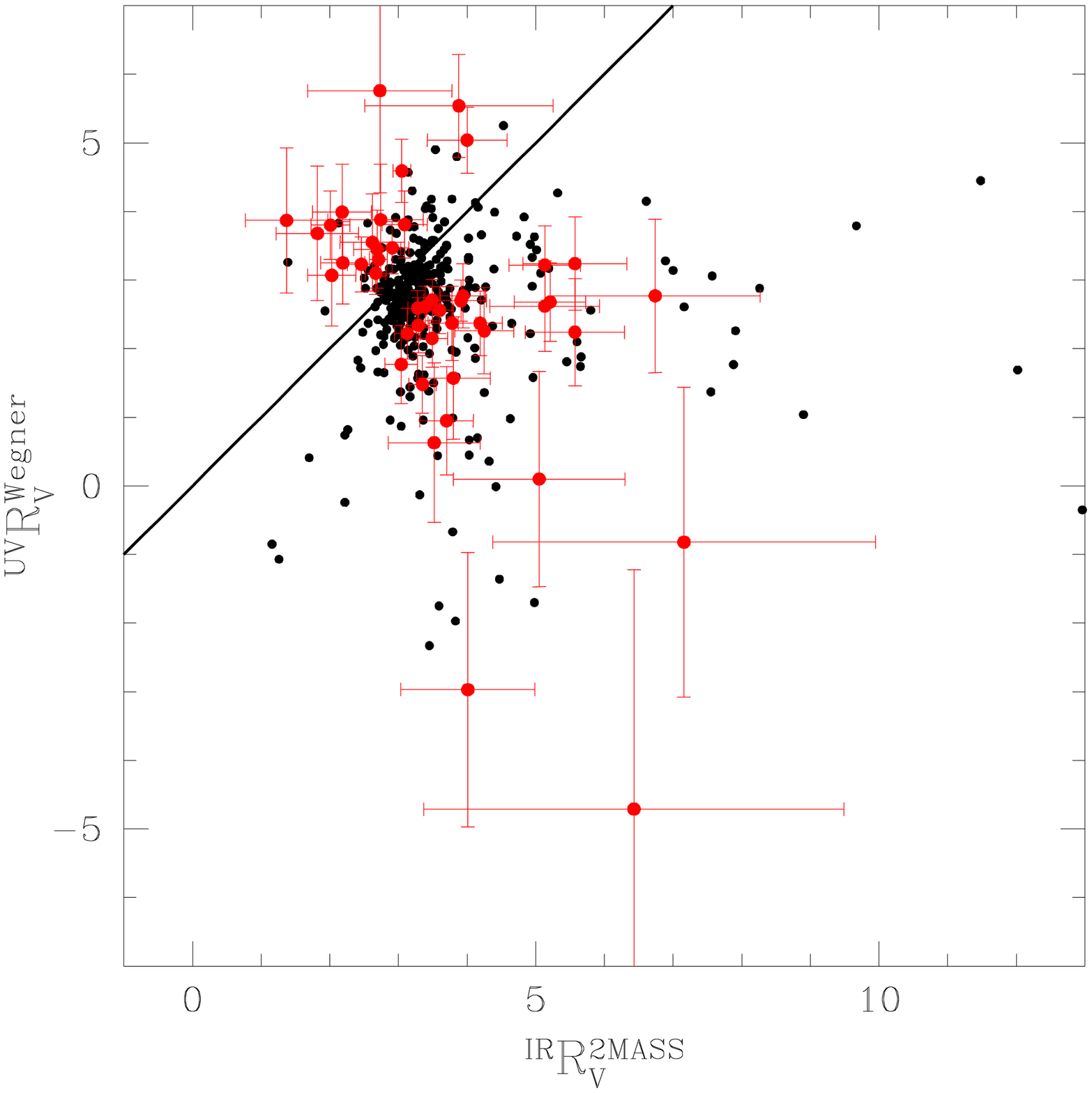}
\includegraphics[width=80mm]{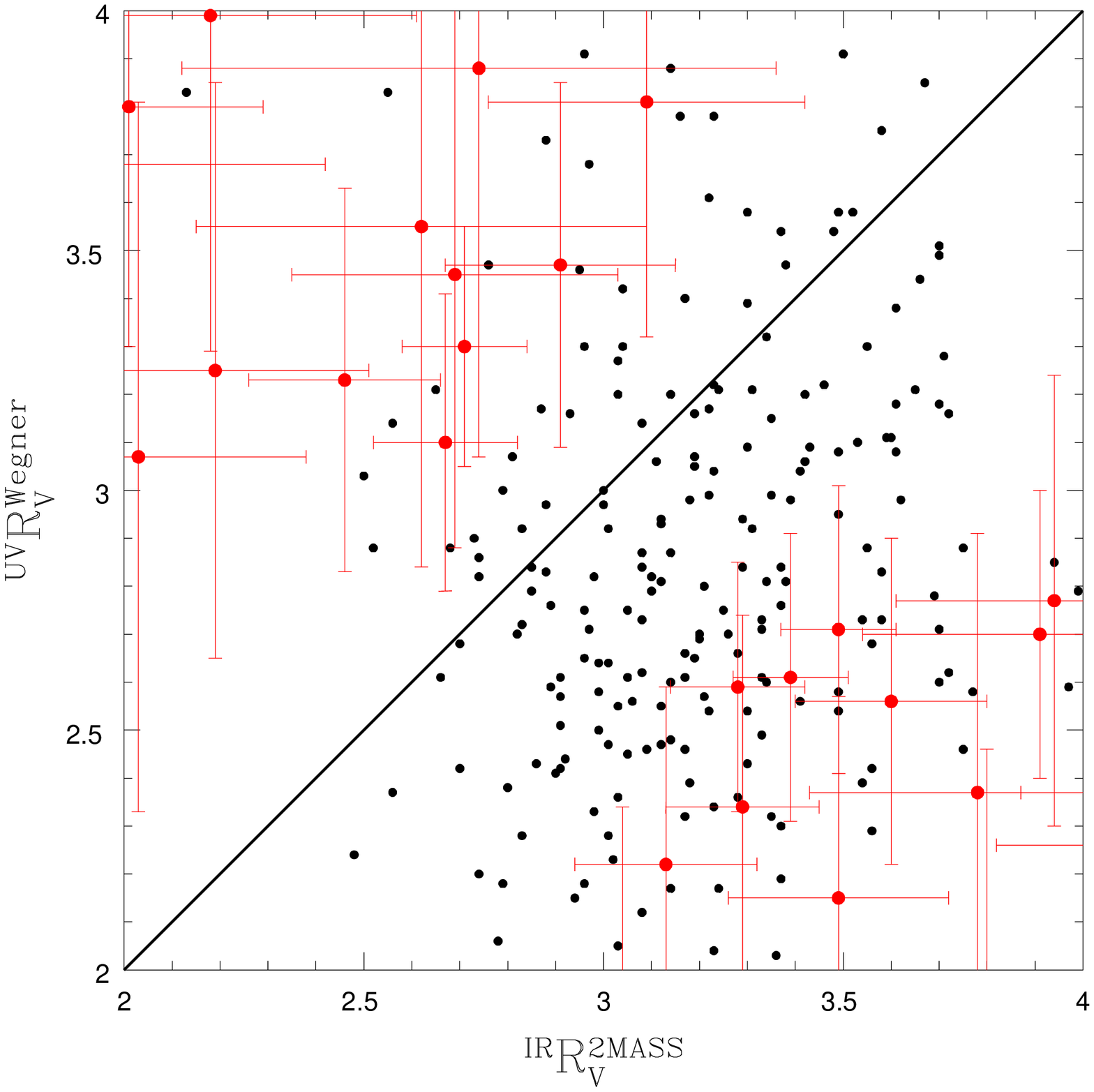}
\caption{The comparison between the $R_V$ values
obtained from NIR and UV extinction data. The points with
error bars (red) are those for which $\delta \leq -2.0$ or $\delta \geq
1.0$. The left panel presents all data points and the right panel
shows the most-crowded region in more
detail.}
\label{figure5}
\end{center}
\end{figure}

\begin{figure}
\begin{center}
\includegraphics[height=10cm,width=8cm]{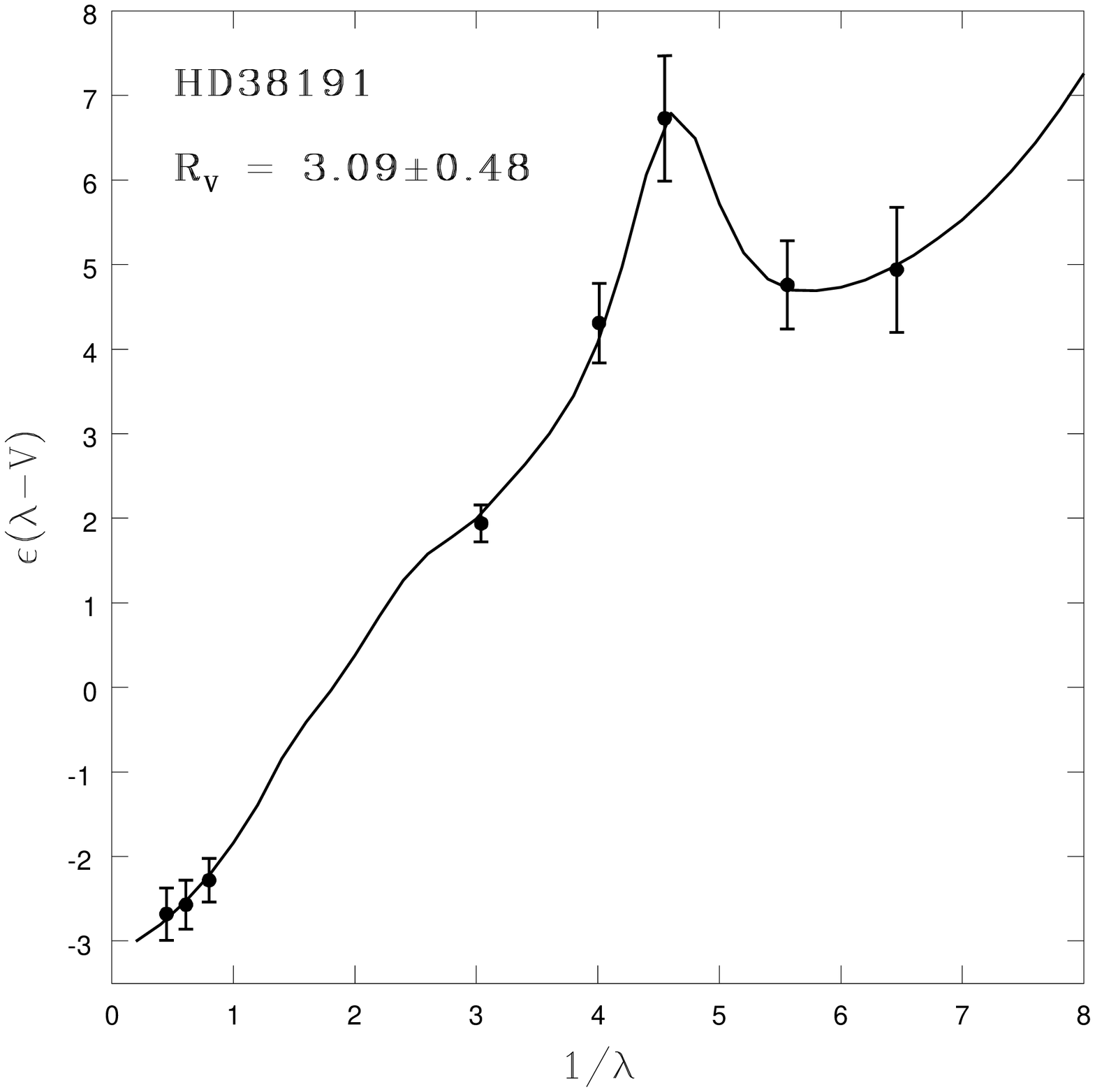}
\includegraphics[height=10cm,width=8cm]{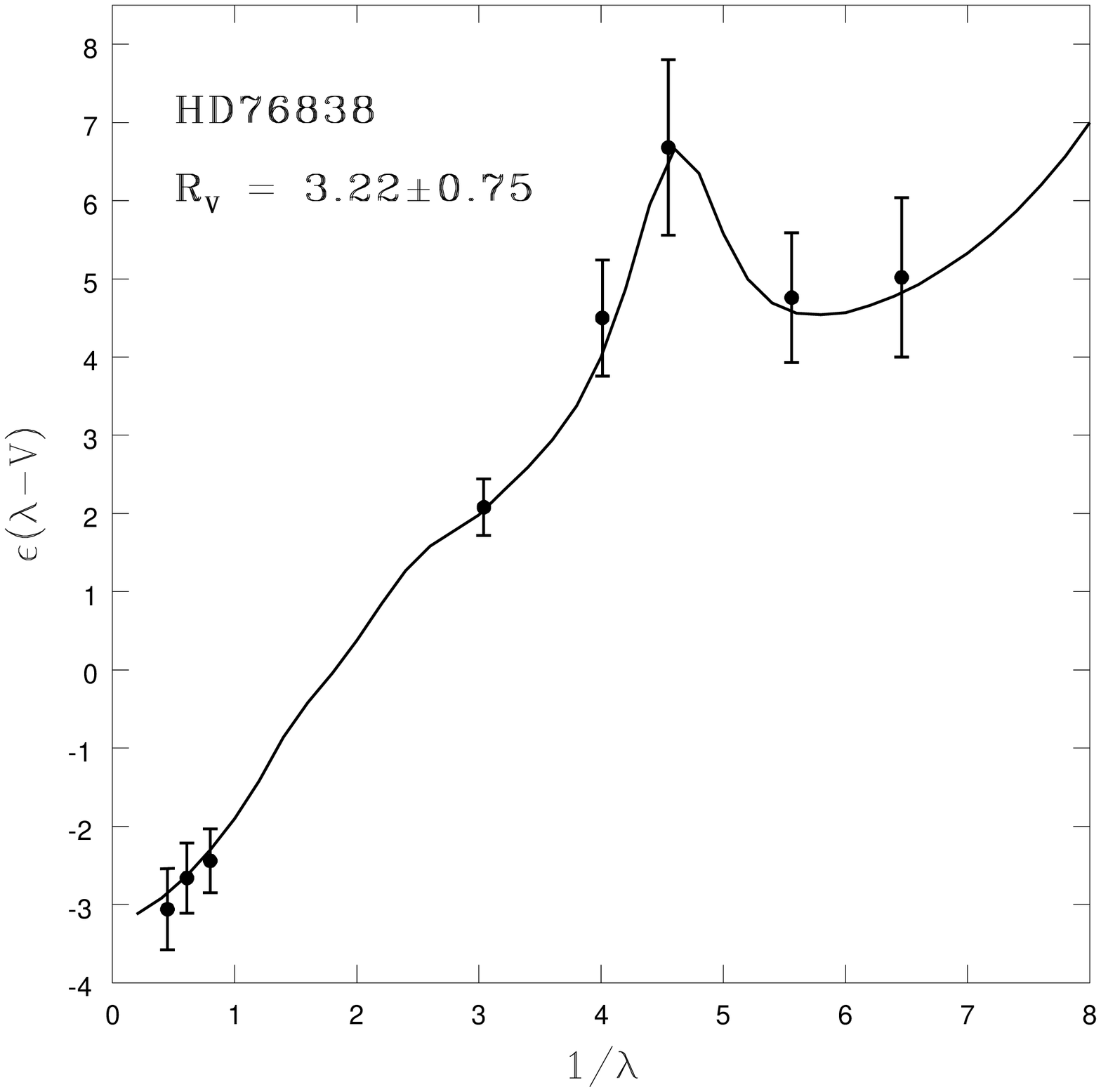}
\caption{Examples of extinction curves for which CCM law works in
the entire spectral range from NIR to UV.}
\label{figure6}
\end{center}
\end{figure}

\begin{deluxetable}{lcccccc}
\tablecolumns{10}
\tabletypesize{\small}
\tablewidth{0pc}
\tablecaption{Lines of sight with highly discrepant 
$^{\rm{IR}}\!R_V^{\rm{2MASS}}$ and
$^{\rm{IR}}\!R_V^{\rm{Wegner}}$}
\tablecomments{
Columns:
[1]  star identification number,
[2]  $^{\rm{IR}}\!R_{V}^{\rm{Wegner}}$ values obtained using optical/NIR data
     from \citet{b16},
[3]  error in $^{\rm{IR}}\!R_{V}^{\rm{Wegner}}$,
[4]  $^{\rm{IR}}\!R_{V}^{\rm{2MASS}}$ values obtained using $J,H,K$
     bands from 2MASS,
[5]  error in $^{\rm{IR}}\!R_{V}^{\rm{2MASS}}$,
[6]  $^{\rm{IR}}\delta \equiv (^{\rm{IR}}\!R_{V}^{\rm{Wegner}}-^{\rm{IR}}\!\!R_{V}^{\rm{2MASS}})/
\sqrt{\sigma^2[^{\rm{IR}}\!R_{V}^{\rm{Wegner}}]+
\sigma^2[^{\rm{IR}}\!R_{V}^{\rm{2MASS}}]}$. 
}
\tablehead{
\colhead{Name} & {$^{\rm{IR}}\!R_{V}^{\rm{Wegner}}$} 
&{$\sigma(^{\rm{IR}}\!R_{V}^{\rm{Wegner}})$} &
{$^{\rm{IR}}\!R_{V}^{\rm{2MASS}}$} 
& {$\sigma(^{\rm{IR}}\!R_{V}^{\rm{2MASS}}$)}
& {$^{\rm{IR}}\delta$}
}
\startdata
 HD    14605  &   4.06  &   0.38 &  0.95  &   0.16  &   7.54\\
 HD    39680  &   6.94  &   0.86 &  3.74  &   0.48  &   3.25\\
 HD    42088  &   3.50  &   0.47 &  1.72  &   0.27  &   3.28\\
 HD    44458  &   6.10  &   1.07 &  2.19  &   0.50  &   3.31\\
 HD    45314  &   3.03  &   0.32 &  4.60  &   0.47  &  $-$2.76\phm{$-$}\\
 HD    46660  &   1.27  &   0.16 &  2.94  &   0.27  &  $-$5.32\phm{$-$}\\
 HD    53367  &   4.29  &   0.28 &  2.40  &   0.18  &   5.68\\
 HD    53975  &   3.21  &   0.82 & $-$2.23\phm{$-$}  &   0.51  &   5.63\\
 HD   102567  &   4.54  &   0.47 &  2.76  &   0.33  &   3.10\\
 HD   155806  &   6.40  &   1.00 &  2.48  &   0.43  &   3.60\\
 HD   164284  &   7.46  &   1.72 &  1.55  &   0.52  &   3.29\\
 HD   170235  &   5.05  &   0.74 &  2.35  &   0.40  &   3.21\\
 HD   178175  &   0.00  &   0.57 &  6.87  &   2.10  &  $-$3.16\phm{$-$}\\
 HD   192639  &   1.80  &   0.16 &  3.30  &   0.25  &  $-$5.05\phm{$-$}\\
 HD   195407  &   4.67  &   0.34 &  2.40  &   0.20  &   5.75\\
 HD   200120  &  $-$0.20\phm{$-$}  &   0.22 &  4.61  &   1.58  &  $-$3.02\phm{$-$}\\
 HD   212044  &   4.46  &   0.70 & $-$0.58\phm{$-$}  &   0.18  &   6.97\\
 BD  $56^{\circ}473$  &   3.38  &   0.35 &  5.02  &   0.49  &  $-$2.72\phm{$-$}\\
 BD  $56^{\circ}586$  &   5.16  &   0.43 &  2.40  &   0.23  &   5.66\\
\enddata
\label{table1}
\end{deluxetable}

\begin{deluxetable}{lcccccc}
\tablecolumns{10}
\tabletypesize{\small}
\tablewidth{0pc}
\tablecaption{$R_V$ values for the poorly fitted extinction curves 
[$^{\rm{IR}}(\chi^2/dof) > 0.28$ in the IR or
$^{\rm{UV}}(\chi^2/dof)>1.6$ in the UV].}
\tablecomments{
Columns:
[1]  star identification number,
[2]  $^{\rm{IR}}\!R_{V}^{\rm{2MASS}}$ values obtained using $J,H,K$
     bands from 2MASS,
[3]  error in $^{\rm{IR}}\!R_{V}^{\rm{2MASS}}$,
[4]  $^{\rm{IR}}(\chi^2/dof)$ for the best CCM fit in IR,
[5]  $^{\rm{UV}}\!R_{V}^{\rm{Wegner}}$ values obtained using UV data
     from \citet{b16},
[6]  error in $^{\rm{UV}}\!R_{V}^{\rm{Wegner}}$,
[7]  $^{\rm{UV}}(\chi^2/dof)$ for the best CCM fit in UV.
}
\tablehead{
\colhead{Name} & {$^{\rm{IR}}\!R_{V}^{\rm{2MASS}}$} 
&{$\sigma(^{\rm{IR}}\!R_{V}^{\rm{2MASS}})$} & {$^{\rm{IR}}(\chi^2/dof)$} &
{$^{\rm{UV}}\!R_{V}^{\rm{Wegner}}$} 
& {$\sigma(^{\rm{UV}}\!R_{V}^{\rm{Wegner}}$)}
& {$^{\rm{UV}}(\chi^2/dof)$}
}
\startdata
HD      7902  &   3.27  &   0.32 &  0.08  &   2.62  &   0.56 &  2.66\\ 
HD     14134  &   2.98  &   0.26 &  0.06  &   4.34  &   0.43 &  8.90 \\
HD     14357  &   2.68  &   0.27 &  0.01  &   3.28  &   0.54 &  2.88 \\
HD     14422  &   2.34  &   0.17 &  0.85  &   3.23  &   0.35 &  0.33 \\
HD     17145  &   3.12  &   0.17 &  0.00  &   2.64  &   0.45 &  2.33 \\
HD     32343  &   4.81  &   2.07 &  0.91  &   1.96  &   1.61 &  0.34 \\
HD     37061  &   5.03  &   0.48 &  0.01  &   5.86  &   0.29 &  4.24 \\
HD     45910  &   5.99  &   0.55 &  0.08  &   4.44  &   0.38 &  6.31 \\
HD     46380  &   4.11  &   0.33 &  0.08  &   3.82  &   0.37 &  2.46 \\
HD     46867  &   2.99  &   0.33 &  0.00  &   5.04  &   0.41 & 18.69 \\
HD     50064  &   4.03  &   0.23 &  0.18  &   3.27  &   0.38 &  7.87 \\
HD     50820  &   5.11  &   0.49 &  0.29  &   6.18  &   0.15 & 46.08 \\
HD     59094  &   4.31  &   0.54 &  0.01  &   3.90  &   0.57 &  3.26 \\
HD     63462  &   5.75  &   1.65 &  0.21  &   4.95  &   0.93 &  2.26 \\
HD     73882  &   3.78  &   0.28 &  0.07  &   3.24  &   0.38 &  3.54 \\
HD     76868  &   6.16  &   0.69 &  0.42  &   6.06  &   0.40 &  9.17 \\
HD     93205  &   4.32  &   0.55 &  0.10  &   5.55  &   0.51 & 32.75 \\
HD     97966  &   0.39  &   0.17 &  1.15  &   3.11  &   0.67 &  0.83 \\
HD    101205  &   3.24  &   0.45 &  0.01  &  $-$2.17\phm{$-$}  &
1.16 &  7.84 \\
HD    110432  &   3.84  &   0.78 &  0.30  &   4.44  &   0.38 &  1.87 \\
HD    147648  &   3.88  &   0.22 &  0.03  &   4.32  &   0.35 &  4.91 \\
HD    147889  &   4.21  &   0.19 &  0.07  &   5.55  &   0.17 & 94.33 \\
HD    147933  &   5.91  &   1.08 &  0.01  &   5.50  &   0.28 &  9.46 \\
HD    148184  &   5.78  &   0.94 &  0.01  &   4.49  &   0.42 &  2.48 \\
HD    149038  &   3.01  &   0.53 &  0.42  &   2.69  &   0.80 &  0.12 \\
HD    152235  &   3.11  &   0.23 &  0.56  &   2.69  &   0.38 &  0.51 \\
HD    152560  &   3.42  &   0.44 &  0.04  &  $-$1.05\phm{$-$}  &
1.02 &  7.15 \\
HD    164492  &   4.59  &   0.69 &  0.02  &   5.04  &   0.60 & 11.22 \\
HD    168076  &   3.75  &   0.23 &  0.03  &   4.28  &   0.28 & 12.33 \\
HD    168112  &   3.24  &   0.16 &  0.25  &   3.49  &   0.29 & 31.42 \\
HD    168137  &   3.55  &   0.31 &  0.00  &   6.03  &   0.37 &  6.21 \\
HD    169454  &   3.43  &   0.35 &  0.05  &   2.83  &   0.26 &  4.86 \\
HD    172252  &   3.14  &   0.17 &  0.00  &   3.55  &   0.27 &  4.62 \\
HD    173438  &   2.95  &   0.14 &  0.00  &   2.47  &   0.30 &  1.72 \\
HD    177291  &   5.46  &   0.46 &  0.20  &   4.58  &   0.39 &  5.13 \\
HD    184943  &   2.96  &   0.19 &  0.00  &   3.69  &   0.32 &  1.95 \\
HD    186745  &   2.85  &   0.15 &  1.00  &   2.99  &   0.32 &  0.68 \\
HD    190918  &   3.80  &   0.43 &  0.02  &   3.78  &   0.57 & 15.75 \\
HD    190944  &   4.37  &   0.44 &  0.09  &   4.21  &   0.46 &  3.81 \\
HD    194279  &   3.35  &   0.18 &  0.03  &   3.23  &   0.26 &  2.36 \\
HD    198931  &   3.70  &   0.20 &  0.15  &   3.63  &   0.27 &  4.91 \\
HD    199478  &   2.47  &   0.26 &  0.29  &   2.94  &   0.55 &  0.77 \\
HD    200775  &   5.25  &   0.41 &  2.73  &   4.21  &   0.34 &  3.68 \\
HD    204827  &   2.56  &   0.12 &  0.11  &   2.36  &   0.28 &  5.67 \\
HD    206165  &   2.64  &   0.46 &  0.61  &   1.45  &   0.79 &  0.33 \\
HD    206267  &   3.05  &   0.28 &  0.33  &   2.70  &   0.51 &  0.24 \\
HD    206773  &   4.36  &   0.40 &  0.25  &   4.07  &   0.44 &  2.13 \\
HD    208501  &   2.82  &   0.49 &  0.01  &   2.88  &   0.40 &  2.73 \\
HD    209975  &   3.02  &   0.48 &  0.40  &   1.85  &   0.87 &  0.08 \\
HD    210839  &   2.99  &   0.32 &  1.18  &   1.89  &   0.58 &  0.46 \\
HD    217086  &   3.15  &   0.16 &  0.29  &   3.08  &   0.28 &  0.93 \\
HD    226868  &   3.36  &   0.15 &  0.07  &   3.00  &   0.27 &  4.10 \\
HD    228779  &   2.87  &   0.09 &  0.14  &   2.48  &   0.52 &  2.73 \\
HD    236689  &   3.05  &   0.29 &  0.57  &   2.58  &   0.52 &  0.03 \\
HD    236923  &   2.95  &   0.22 &  0.07  &   2.60  &   0.46 &  1.61 \\
HD    254577  &   3.08  &   0.15 &  0.01  &   4.01  &   0.29 & 45.97 \\
HD    262013  &   6.51  &   5.60 &  0.00  &  $-$4.57\phm{$-$}  &
4.10 & 31.01 \\
BD   $-12^{\circ}5008$ & 3.17 &    0.12 &  0.01  &   3.21  &   0.20 &  4.77 \\
BD   $40^{\circ}4220$ & 3.26  &   0.08  & 0.16   &  3.65   &  0.51 &  5.14 \\
BD   $40^{\circ}4227$ & 3.22  &   0.10  & 0.00   &  3.08   &  0.18 &  2.09 
\enddata
\label{table2}
\end{deluxetable}

\begin{deluxetable}{lccccc}
\tablecolumns{10}
\tabletypesize{\small}
\tablewidth{0pc}
\tablecaption{$R_V$ values for the outlier extinction curves with $\delta \leq
-2.0$ and $\delta \geq 1.0$.}
\tablecomments{
Columns:
[1]  star identification number,
[2]  $^{\rm{IR}}\!R_{V}^{\rm{2MASS}}$ values obtained using $J,H,K$
     bands from 2MASS,
[3]  error in $^{\rm{IR}}\!R_{V}^{\rm{2MASS}}$,
[4]  $^{\rm{UV}}\!R_{V}^{\rm{Wegner}}$ values obtained using UV data
     from \citet{b16},
[5]  error in $^{\rm{UV}}\!R_{V}^{\rm{Wegner}}$,
[6]  $\delta$ given by equation (\ref{deviation}) using 2MASS data in
     IR and Wegner's (2002) data in UV.
}
\tablehead{
\colhead{Name} & {$^{\rm{IR}}\!R_{V}^{\rm{2MASS}}$} 
&{$\sigma(^{\rm{IR}}\!R_{V}^{\rm{2MASS}})$} & 
{$^{\rm{UV}}\!R_{V}^{\rm{Wegner}}$} 
& {$\sigma(^{\rm{UV}}\!R_{V}^{\rm{Wegner}}$)}
& {$\delta$} 
}
\startdata
HD      2083  & 2.18 & 0.43 &  3.99 & 0.70  &   2.20 \\
HD     14322  & 3.70 & 0.39 &  0.95 & 0.79  &  $-$3.12\phm{$-$} \\
HD     21291  & 5.13 & 0.80 &  2.62 & 0.66  &  $-$2.42\phm{$-$} \\
HD     25348  & 5.21 & 0.52 &  2.68 & 0.57  &  $-$3.28\phm{$-$} \\
HD     30614  & 2.03 & 0.35 &  3.07 & 0.74  &   1.27 \\
HD     37903  & 4.00 & 0.58 &  5.04 & 0.48  &   1.38 \\
HD     46484  & 2.91 & 0.24 &  3.47 & 0.38  &   1.25 \\
HD     46559  & 3.04 & 0.24 &  1.77 & 0.57  &  $-$2.05\phm{$-$} \\
HD     46711  & 3.29 & 0.16 &  2.34 & 0.40  &  $-$2.21\phm{$-$} \\
HD     49787  & 1.82 & 0.60 &  3.68 & 0.98  &   1.62 \\
HD     54439  & 2.62 & 0.47 &  3.55 & 0.71  &   1.09 \\
HD     55606  & 6.74 & 1.53 &  2.77 & 1.12  &  $-$2.09\phm{$-$} \\
HD     61827  & 3.35 & 0.20 &  1.48 & 0.42  &  $-$4.02\phm{$-$} \\
HD     76534  & 2.19 & 0.32 &  3.25 & 0.60  &   1.56 \\
HD     96042  & 2.01 & 0.28 &  3.80 & 0.50  &   3.12 \\
HD     97434  & 3.09 & 0.33 &  3.81 & 0.49  &   1.22 \\
HD    113659  & 5.05 & 1.25 &  0.10 & 1.57  &  $-$2.47\phm{$-$} \\
HD    133518  & 1.37 & 0.60 &  3.87 & 1.06  &   2.05 \\
HD    135160  & 3.88 & 1.37 &  5.54 & 0.75  &   1.06 \\
HD    152386  & 3.60 & 0.20 &  2.56 & 0.34  &  $-$2.64\phm{$-$} \\
HD    152408  & 4.25 & 0.43 &  2.26 & 0.63  &  $-$2.61\phm{$-$} \\
HD    153919  & 3.94 & 0.33 &  2.77 & 0.47  &  $-$2.04\phm{$-$} \\
HD    155851  & 5.57 & 0.76 &  3.24 & 0.68  &  $-$2.28\phm{$-$} \\
HD    162168  & 3.13 & 0.19 &  2.22 & 0.37  &  $-$2.19\phm{$-$} \\
HD    163758  & 3.80 & 0.54 &  1.57 & 0.89  &  $-$2.14\phm{$-$} \\
HD    165016  & 2.73 & 1.05 &  5.76 & 1.49  &   1.66 \\
HD    168476  & 4.01 & 0.98 & $-$2.97\phm{$-$} & 2.00  &  $-$3.13\phm{$-$} \\
HD    169034  & 3.49 & 0.12 &  2.71 & 0.30  &  $-$2.41\phm{$-$} \\
HD    185859  & 2.46 & 0.20 &  3.23 & 0.40  &   1.72 \\
HD    186660  & 2.74 & 0.62 &  3.88 & 0.81  &   1.12 \\
HD    186841  & 2.71 & 0.13 &  3.30 & 0.25  &   2.09 \\
HD    188209  & 6.43 & 3.06 & $-$4.71\phm{$-$} & 3.49  &  $-$2.40\phm{$-$} \\
HD    191612  & 3.78 & 0.35 &  2.37 & 0.54  &  $-$2.19\phm{$-$} \\
HD    193514  & 3.49 & 0.23 &  2.15 & 0.42  &  $-$2.80\phm{$-$} \\
HD    194839  & 3.28 & 0.14 &  2.59 & 0.26  &  $-$2.34\phm{$-$} \\
HD    199356  & 5.13 & 0.52 &  3.22 & 0.57  &  $-$2.48\phm{$-$} \\
HD    206183  & 2.69 & 0.34 &  3.45 & 0.57  &   1.15 \\
HD    211853  & 4.19 & 0.32 &  2.37 & 0.47  &  $-$3.20\phm{$-$} \\
HD    217050  & 7.16 & 2.79 & $-$0.82\phm{$-$} & 2.26  &
$-$2.22\phm{$-$} \\
HD    220116  & 2.67 & 0.15 &  3.10 & 0.31  &   1.25 \\
HD    228712  & 3.39 & 0.12 &  2.61 & 0.30  &  $-$2.41\phm{$-$} \\
HD    249845  & 3.52 & 0.67 &  0.63 & 1.16  &  $-$2.16\phm{$-$} \\
HD    259597  & 5.57 & 0.72 &  2.24 & 0.78  &  $-$3.14\phm{$-$} \\
BD+   $41^{\circ}4064$ &  3.91 & 0.37 &  2.70 & 0.30  &  $-$2.54\phm{$-$}\\ 
BD+   $62^{\circ}2210$ &  3.05 & 0.13 &  4.59 & 0.46  &   3.22 \\
\enddata
\label{table3}
\end{deluxetable}

\begin{deluxetable}{lccccccccc}
\tablecolumns{10}
\tabletypesize{\small}
\tablewidth{0pc}
\tablecaption{Catalog of $R_V$ and $A_V$ values.}
\tablecomments{
Columns:
[1]  star identification number,
[2],[3]  star coordinates,
[4]  $E(B-V)$ taken from \citet{b16},
[5]  $R_V$ value obtained using NIR data from 2MASS and UV data from 
	\citet{b16},
[6]  error in $R_V$,
[7]  $\chi^2$ per degree of freedom derived from equation (\ref{chiweighted}),
[8]  $A_V$ obtained from columns four and five,
[9]  error in $A_V$.
}
\tablehead{
\colhead{Name} & {RA(J2000)} &{DEC(J2000)} & {$E(B-V)$} & {$R_{V}$} &
{$\sigma(R_{V})$}  & {$\chi^2/dof$} & {$A_{V}$} & {$\sigma(A_V)$}
}
\startdata
HD108 &   00 06 03.37 &  63 40 46.8 &  0.480  &   3.28  &
0.43  &   0.16  &   1.58  &   0.34\\
HD1544   &   00 20 05.55 &  62  03 58.7 &  0.370  &   2.99  &
0.49  &   0.02  &   1.11  &   0.30\\
HD2905   &   00 33 00.00 &  62 55 54.2 &  0.300  &   1.30  &
1.10  &   0.01  &   0.39  &   0.38\\
HD3901   &   00 42 03.90 &  50 30 45.1 &  0.085  &   2.36  &
2.24  &   0.11  &   0.20  &   0.28\\
HD4180   &   00 44 43.51 &  48 17  3.7 &  0.130  &   1.21  &
2.26  &   0.03  &   0.16  &   0.34\\
HD4841   &   00 51 25.93 &  63 46 52.1 &  0.650  &   3.14  &
0.29  &   0.00  &   2.04  &   0.31\\
HD6811   &   01 09 30.14 &  47 14 30.3 &  0.060  &   0.70  &
3.75  &   0.01  &   0.04  &   0.25\\
HD9311   &   01 33 14.01 &  60 41 11.2 &  0.360  &   2.99  &
0.49  &   0.00  &   1.07  &   0.29\\
HD10516  &   01 43 39.65 &  50 41 19.3 &  0.200  &   4.40  &
1.09  &   0.00  &   0.88  &   0.39\\
HD12867  &   02 07 53.68 &  57 42 45.5 &  0.380  &   2.84  &
0.45  &   0.02  &   1.08  &   0.29\\
HD13267 &  02 11 29.20 &  57 38 44.0  &   0.420 &    3.08  &   0.41  &
0.06  &   1.30  &   0.30\\
HD13900 &  02 17 15.56 &  56 53 52.9  &   0.380 &    2.89  &   0.47  &
0.04  &   1.10  &   0.30\\
HD13969 &  02 17 49.85 &  57  5 25.7  &   0.540 &    2.75  &   0.32  &
0.01  &   1.48  &   0.28\\
HD14092 &  02 18 41.89 &  56 45 40.7  &   0.460 &    2.90  &   0.40  &
0.06  &   1.33  &   0.30\\
HD14250 &  02 20 15.73 &  57  5 55.0  &   0.550 &    2.80  &   0.33  &
0.04  &   1.54  &   0.29\\
\enddata
\label{table4}
\end{deluxetable}

\end{document}